\documentclass[11pt]{article}
\pdfoutput=1
\usepackage{jheppub}
\usepackage[T1]{fontenc}
\usepackage{cancel,tensor,enumitem,fix-cm}
\usepackage{epsfig}
\usepackage{graphicx}
\usepackage[english]{babel}
\usepackage{amsmath}
\usepackage{textcomp}
\usepackage{multirow}
\usepackage{booktabs}
\usepackage{tikz}
\usepackage{overpic}
\usepackage{bm}
\usepackage{physics}
\usepackage[lofdepth,lotdepth]{subfig}

\newcommand{\be}{\begin{equation}}
\newcommand{\ee}{\end{equation}}
\newcommand{\bea}{\begin{eqnarray}}
\newcommand{\eea}{\end{eqnarray}}

\def\({\left(}
\def\){\right)}
\def\[{\left[}
\def\]{\right]}

\def\p{\partial}
\newcommand{\f}[2]{\frac{#1}{#2}}

\subheader{\begin{flushright}
\texttt{YITP-20-19, UTTG-04-2020, BRX-TH-6666}
\end{flushright}}

\title{Bit threads, Einstein's equations and bulk locality}

\author[a]{Cesar A. Ag\'on,}
\author[b]{Elena C\'aceres}
\author[c,d]{and Juan F. Pedraza}
\affiliation[a]{C.~N. Yang Institute for Theoretical Physics, State University of New York,\\
Stony Brook, NY 11794, USA}
\affiliation[b]{Theory Group, Department of Physics, University of Texas, Austin, TX 78712, USA}
\affiliation[c]{Department of Physics and Astronomy, University College London, London WC1E 6BT, UK}
\affiliation[d]{Martin Fisher School of Physics, Brandeis University, Waltham MA 02453, USA}
\emailAdd{cesar.agon@stonybrook.edu}
\emailAdd{elenac@utexas.edu}
\emailAdd{j.pedraza@ucl.ac.uk}

\abstract{In the context of holography, entanglement entropy can be studied either by i)
extremal surfaces or ii) bit threads, i.e., divergenceless vector fields with a norm bound set by the Planck length.
In this paper we develop
a new method for metric reconstruction based on the latter approach and show the advantages over existing ones.
We start by studying general linear perturbations
around the vacuum state. Generic thread configurations turn out to encode the information about the metric in a highly nonlocal way,
however, we show that for boundary regions with a local modular Hamiltonian there is always a canonical choice for the perturbed thread configurations that exploits bulk locality. To do so, we express the bit thread formalism in terms of differential forms so that it becomes manifestly background independent.
We show that the Iyer-Wald formalism provides a natural candidate for a canonical local perturbation, which can be used to recast the problem of metric reconstruction in terms of the inversion of a particular linear differential operator. We examine in detail the inversion problem for the case of spherical regions and give explicit expressions for the inverse operator in this case. Going beyond linear order, we argue that the operator that must be inverted naturally increases in order. However, the inversion can be done recursively at different orders in the perturbation.
 Finally, we comment on an alternative way of reconstructing the metric non-perturbatively by phrasing the inversion problem as a particular optimization problem.}

\begin{document}
\maketitle
\flushbottom

\section{Introduction}

\subsection{Motivation}
Recent progress in the joint program on quantum information and holography has uncovered striking connections between entanglement and spacetime. Arguably, the most exciting discovery in this context, and the one which ignited most of the research in this field, was the proposal of Ryu and Takayanagi that relates the entanglement entropy of a region $A$ in the boundary to the area of a minimal codimension-two bulk surface $\gamma_A$ \cite{Ryu:2006bv},
\be\label{RyuT}
S_A={\frac1{4G_N}}\underset{\gamma_A\sim A}{\min}\left[\text{area}(\gamma_A)\right]\,.
\ee
This formula was further generalized to a fully covariant setting in \cite{Hubeny:2007xt}
and  proved formally in \cite{Lewkowycz:2013nqa,Dong:2016hjy} using the known entries of the holographic dictionary. The RT prescription (\ref{RyuT}) and its covariant version generalize in an elegant way the well-known Bekenstein-Hawking formula for black hole entropy and provide a natural way to interpret it directly in terms of a microscopic CFT description. Given its elegance and simplicity, entanglement entropy became a robust tool to investigate fundamental aspects in holography, ranging from the problem of bulk reconstruction \cite{Balasubramanian:2013rqa,Balasubramanian:2013lsa,Myers:2014jia,Headrick:2014eia,Czech:2014ppa,Czech:2015qta,Dong:2016eik,Czech:2016xec,Cao:2016mst,Espindola:2017jil,Espindola:2018ozt,Roy:2018ehv,Faulkner:2018faa,Czech:2019hdd,Balasubramanian:2018uus,Bao:2019bib,Cao:2020uvb}, to the emergence and dynamics of spacetime \cite{Lashkari:2013koa,Faulkner:2013ica,Swingle:2014uza,Caceres:2016xjz,Czech:2016tqr,Faulkner:2017tkh,Haehl:2017sot,Rosso:2020zkk}.

Recently, Freedman and Headrick proposed an alternative way to compute entanglement entropy that does not rely on bulk surfaces, but instead, is phrased in terms of a specific flow maximization problem \cite{Freedman:2016zud}. More specifically, the new prescription states that
\be\label{BitThreadReform}
S_A=\frac1{4G_N} \max_{v \in {\cal F}}\int_{A}\sqrt{h} \, n_\mu v^\mu\,, \qquad {\cal F}\equiv\{v \, \vert\, \nabla_\mu v^\mu=0,\, \abs{v}\leq 1\}\,,
\ee
and can be shown to be equivalent to the RT formula through the continuous version of the max flow-min cut theorem of network theory. The maximization above is an example of a convex optimization program and, hence, the equivalence between (\ref{RyuT}) and (\ref{BitThreadReform}) can also be proved using techniques borrowed from convex optimization \cite{Headrick:2017ucz}. Soon after this paper appeared, it was realized that various geometric problems could likewise be translated to the realm of convex optimization leading to interesting new results \cite{Headrick:2018dlw,Headrick:2018ncs}. The connection with convex optimization has also helped uncover various properties of entanglement entropy from the bit thread perspective \cite{Cui:2018dyq}, as well as some generalizations and applications to other entanglement related quantities \cite{Chen:2018ywy,Harper:2018sdd,Du:2019emy,Bao:2019wcf,Harper:2019lff,Agon:2019qgh,Du:2019vwh}. A complementary approach that departs from the realm of convex optimization was put forward in \cite{Hubeny:2018bri,Agon:2018lwq} and studies aspects of bit threads and entanglement by considering explicit constructions of max flows. This is the line of work that we will mostly follow in this paper.

There is one crucial distinction between the two prescriptions to compute entanglement entropy that we believe deserves further investigation: while the minimal surface $\gamma_A$ is in most cases unique, the solution to the max flow problem $v$ is highly degenerate. More specifically, it can be shown that $v$ is uniquely determined only at the bulk bottle-neck $\gamma_A$, but is highly non-unique away from it. This non-uniqueness raises the question:
\vspace{2mm}
\noindent\fbox{
    \parbox{0.975\textwidth}{
    \centering
    \emph{Out of the infinitely many thread configurations that could be associated with a boundary region, is there any meaningful separation or classification that could be associated with states of special ``entanglement classes'' in the dual field theory?}
    }%
}
\vspace{2mm}
$\quad$\\
Intuitively, it would seem that this large degeneracy could indeed be associated to a choice of microstate (or a particular class of microstates) that give rise to the same amount of entanglement between the region $A$ and its complement,\footnote{The standard lore asserts that states with (semi)-classical bulk duals can only encode bipartite and perfect-tensor type entanglement, but no other form of multipartite entanglement (see however \cite{Akers:2019gcv}). Hence, the class of microstates that we have access to would be a reduced subset of the most general class of CFT microstates.} however, a precise version of this statement is not settled yet. On the other hand, one can try to exploit this non-uniqueness to gain new insights on the gravity side. The utility of the non-uniqueness property stems from the observation that, if a version of this statement is true (even if we do not know it yet), then a particular solution to the max flow problem $v$ could potentially carry more information than the minimal surface $\gamma_A$ itself: it could encode in detail how the local correlations between the degrees of freedom in the region $A$ and in its complement are distributed for a particular choice of microstate. If so, then, one could imagine that specific questions related to bulk reconstruction and the emergence of spacetime could be answered in a more efficient way by properly selecting a class of configurations/states adapted to the specific problem at hand.

In this paper we will give some steps in this direction. Specifically, our main objective is to understand how the program of \emph{gravitation from entanglement} \cite{Lashkari:2013koa,Faulkner:2013ica,Swingle:2014uza,Caceres:2016xjz,Czech:2016tqr,Faulkner:2017tkh,Haehl:2017sot,Rosso:2020zkk} unfolds in the language of bit threads and to explore an alternative way of metric reconstruction based on this framework. The particular questions that we want to address are the following:
\begin{itemize}
  \item How are the metric and Einstein's equations encoded in generic thread configurations?
  \item Can bulk locality be manifest in particular constructions?
  \item Is it possible to reconstruct the bulk geometry from a max flow solution?
  \item If so, how does the method compare to the ones based on RT surfaces?
\end{itemize}
Following \cite{Lashkari:2013koa,Faulkner:2013ica,Swingle:2014uza,Caceres:2016xjz,Czech:2016tqr,Faulkner:2017tkh,Haehl:2017sot,Rosso:2020zkk}, we begin by considering these questions in a perturbative setting in which we study small deformations continuously connected to a reference state. An important motivation of such continuous construction comes from the study of the phase transition of RT surfaces that happens for disjoint regions as one varies their separation. It is known that, close to the phase transition, the RT surface can change from a connected to a disconnected configuration. Such jumps posit a puzzle to a potential quantum information interpretation of the RT surfaces from the bulk perspective, which is solved in the language of bit threads by imposing the additional property of being continuous across phase transitions \cite{Freedman:2016zud}. Continuity is, then, a desirable feature of bit threads under continuous deformations.

Before we study the above questions, let us review some of the features of the standard methods of metric reconstruction using RT surfaces \cite{Balasubramanian:2013rqa,Balasubramanian:2013lsa,Myers:2014jia,Headrick:2014eia,Czech:2014ppa,Czech:2015qta,Dong:2016eik,Czech:2016xec,Cao:2016mst,Espindola:2017jil,Espindola:2018ozt,Roy:2018ehv,Faulkner:2018faa,Czech:2019hdd,Balasubramanian:2018uus,Bao:2019bib,Cao:2020uvb}, and explain potential advantages of studying this problem with bit threads. While there are other methods for bulk reconstruction, e.g. \cite{Engelhardt:2016wgb,Trevino:2017mik,Hernandez-Cuenca:2020ppu}, our comments and comparisons refer only  to approaches that make explicit use of RT surfaces. Quite generally, if one hopes for a reconstruction of the metric everywhere in the bulk one must start with a sufficiently dense set of extremal surfaces that probe the full manifold $M$. This is in fact possible in some simple cases, at least for the subset of $M$ that can be foliated by boundary anchored extremal surfaces. For static $(2+1)-$dimensional bulk geometries this was achieved in \cite{Czech:2014ppa} starting from the full set of extremal surfaces associated with \emph{all} CFT intervals, and using ideas from hole-ography \cite{Balasubramanian:2013rqa,Balasubramanian:2013lsa,Myers:2014jia,Headrick:2014eia}. More recently, it was shown that the same ideas could be extended to the time-dependent case in \cite{Czech:2019hdd} and to higher dimensions \cite{Balasubramanian:2018uus}, here by focusing on the subset of extremal surfaces associated with spherical regions (or topologically equivalent, in the approach of \cite{Bao:2019bib}).

The problem of metric reconstruction using bit threads has a major advantage over the ones described above: it does not rely on the ability of the manifold $M$ of admitting foliations by boundary anchored extremal surfaces. In fact, threads can probe regions in the bulk that extremal surfaces cannot, such as entanglement shadows near the vicinity of (spherical) black hole horizons \cite{Agon:2018lwq}. It is important to point out that bulk shadows do not appear exclusively in cases where gravity is strong; one simple counterexample is the metric of a conical deficit geometry, which arises by the backreaction of a point particle in AdS \cite{Balasubramanian:2014sra}. Consequently, formulating the problem of metric reconstruction in the language of bit threads, even for the simpler case of perturbative states, is interesting on its own right. In particular, it will shed new light on the issue of emergence of spacetime from entanglement entropy \cite{VanRaamsdonk:2009ar,VanRaamsdonk:2010pw}, without resorting to other measures of entanglement such as \emph{entwinement} \cite{Balasubramanian:2014sra}.

Another important difference with respect to the problem of metric reconstruction using extremal surfaces is that the latter requires as a starting point the knowledge of a dense set of surfaces that probe the bulk geometry. While we can do the same in the language of bit threads, i.e., start from a \emph{dense set} of thread configurations, the fact that one single solution to the max flow problem already probes the full bulk geometry presents us with an interesting possibility: we can start from a \emph{finite set} of thread configurations, containing one, or possibly only a few solutions of the max flow problem. We will consider both approaches in this paper, and show that the explicit reconstruction is possible in both cases. In the remaining part of the introduction we will provide a quick guide to help navigate our paper and enumerate the most important findings of each section.

\subsection{Road map and summary}

We begin in section \ref{sec2} with a short discussion of various topics that we constantly refer to in our paper. Most of this material is
a review of previous works, covering known results about perturbations around AdS and the calculation of entanglement entropy,
both in the language of extremal surfaces and bit threads. We also include a short analysis of bit threads in perturbative excited states in
subsection \ref{pertstatesBT} which is new. The main message of this analysis is that, to leading order in the perturbation,
it is consistent to use the prescription (\ref{BitThreadReform}) on a constant-$t$ slice, even if the perturbation includes time dependence.

In section \ref{PBT} we study simple explicit realizations of max flows for bulk geometries that are perturbatively close to pure AdS. We begin in section \ref{generalities} by discussing some general properties about these max flows: the boundary condition at the minimal surface and how this condition is sufficient to encode the first law of entanglement entropy. We then proceed to study two particular constructions in subsections \ref{F1} and \ref{lsc}, respectively. The first method that we consider is a generalization of the geodesic method developed in \cite{Agon:2018lwq}. This method assumes a particular set of integral curves as a starting point, which we take to be the family of space-like geodesics that intersect normally the minimal surface. Given this assumption, one then determines the norm by imposing the divergenceless condition, implemented through the Gauss's law. We show that this construction works both for geodesics of the unperturbed and perturbed geometries under some mild assumptions. In subsection \ref{lsc} we study a slightly more general method. Here, our starting point is to propose a family of level set functions for the flow and then determine its norm based on the divergenceless condition, now implemented directly by solving a differential equation. The flows constructed via this method are a generalization of the maximally packed flows presented in \cite{Hubeny:2018bri,Agon:2018lwq}, where the level set functions are now arbitrary (not necessarily a nested set of minimal surfaces).  This method is therefore fully non-perturbative and easily adapted to any boundary entangling region.

Importantly, both constructions presented in section \ref{PBT} assume as an input a solution to the Einstein's equations in the bulk. Given an explicit metric one can determine the norm of the vector field from the divergenceless condition, which requires an integration from the minimal surface (where the norm is known) to the points of interest. Such integration generically introduces a nonlocal dependence on the background metric which renders these methods non-suitable for addressing questions of bulk reconstruction. However, this also suggests a way around it. More specifically, since the nonlocality is introduced in both cases through the implementation of the divergenceless constrain, it suggests that a construction that implements this condition in a background independent way would be absent of such nonlocalities, which is possible if pose the question in the language of differential forms.

Motivated by the above observation we start subsection \ref{ssec:diff} by rewriting the bit thread framework in the language of differential forms. We study in detail the case of perturbative states and show, in subsections \ref{sec:IWgeneral} and \ref{5.2}, that the Iyer-Wald formalism provides a candidate for a thread perturbation which is explicitly local in the metric and furthermore connects the closedness condition with the linearized Einstein's equations.
Further, we explore the problem of metric reconstruction in subsection \ref{metric-reconstruction} and show that it can be cast in terms of the inversion of a particular differential operator. We provide explicit inversion formulas for the case of spherical entangling regions in two distinct scenarios: $i$) assuming knowledge of a dense set of forms parametrized by their radii and centers and $ii$) assuming knowledge of a finite set with a minimal number of forms. The second approach turns out to be very powerful; for instance, it suffices to have a single form to provide a full solution for the bulk metric in asymptotically AdS$_3$ and AdS$_4$ spaces, which we construct explicitly. We also show that the problem is well-posed in higher dimensions, starting with a carefully selected finite set. We end the section with a detailed analysis of how to recover the time components of the metric via boosts and translations of the space-like hypersurface on which the threads are defined, and a thorough discussion on how to generalize the reconstruction problem to higher orders in the perturbation.

\section{Preliminaries\label{sec2}}

In this section, we will start with a brief discussion of a number of topics that we will be essential throughout the paper. We include this discussion for completeness. However, since most of this material is a review, it can be safely skipped by the cognoscenti.

\subsection{Linear perturbations around AdS}

Let us start by reviewing basic properties of linear perturbations around empty AdS. In Fefferman-Graham coordinates, any asymptotically AdS metric can be written as
\be\label{FG-PT}
ds^2=\frac{1}{z^2}\(\eta_{\mu\nu}dx^\mu dx^\nu+dz^2\)+\delta g_{\mu\nu}(x^\sigma,z)dx^\mu dx^\nu\,,\qquad \delta g_{\mu\nu}(x^\sigma,z)\equiv z^{d-2} H_{\mu\nu}(x^\sigma,z)\,,
\ee
where $x^\sigma$ are boundary coordinates and $z$ is the holographic coordinate. For concreteness we have assumed a Minkowski boundary geometry. With this parametrization, one can extract the expectation value of the stress-energy tensor
from the asymptotic form of the perturbation,
\be\label{T=H}
\langle T_{\mu\nu}(x^\sigma) \rangle = \f{d}{16\pi G_N}H_{\mu\nu}(x^\sigma,z=0)\,.
\ee
Plugging the above ansatz into the vacuum Einstein equations,
\be
R_{\mu\nu}-\frac{1}{2}g_{\mu\nu}R-\frac{d(d-1)}{2}g_{\mu\nu}=0\,,
\ee
we obtain the following expressions for the $zz$, $z\mu$ and $\mu\nu$ components \cite{Lashkari:2013koa}:
\be\label{perturbedeqns}
H^\mu\,\!_\mu=0\,,\qquad \p_\mu H^{\mu\nu}=0\,,\qquad \frac{1}{z^{d+1}}\p_z\(z^{d+1}\p_z H_{\mu\nu} \)+\Box H_{\mu\nu}=0\,,
\ee
respectively, where the box operator is the standard Laplace operator in Minkowski space, i.e., $\Box\equiv \partial_\mu\partial^\mu$. Alternatively, one can write down the perturbation as follows:
\be\label{deltaGprop}
\delta g_{\mu\nu}(x,z)=\int d^dy\, G(y-x,z)T_{\mu\nu}(y)\,,
\ee
where $G(x,z)$ is the Green's function of the graviton in empty AdS,
\be
G(x,z)=\frac{16\pi G_N}{d}2^{d/2}\Gamma[d/2+1]\int \frac{d^dp}{(2\pi)^d}\theta(-p^2)\frac{z^{d/2}}{p^{d/2}}J_{d/2}(|p|z)e^{-ip\cdot x},\quad\,\,\, |p|\equiv\sqrt{|p_\mu p^\mu|}\,.
\ee
A somewhat useful expression can be obtained by expanding $\delta g_{\mu\nu}$ in powers of $z$ \cite{Blanco:2013joa}:
\be\label{pertg}
\delta g_{\mu\nu}=\frac{16\pi G_N}{d}z^{d-2}\sum_{n=0}^\infty z^{2n}T^{(n)}_{\mu\nu}\,.
\ee
The strategy is to use the linear Einstein equations order by order in $z$ to determine $T^{(n)}_{\mu\nu}$ for $n >0$  in  terms  of  the expectation value of the stress-energy tensor $T^{(0)}_{\mu\nu}$. A simple calculation shows that the $zz$ and $z\mu$ equations imply
\be
T^{(n)}\,\!^\mu\,\!_\mu=0\,,\qquad \p_\mu T^{(n)}\,\!^{\mu\nu}=0\,,
\ee
so $T^{(n)}_{\mu\nu}$ is traceless and conserved for all $n$. Finally, the $\mu\nu$ equations imply that
\be\label{Tnexp}
T^{(n)}_{\mu\nu}=-\frac{\Box T^{(n-1)}_{\mu\nu}}{2n(d+2n)}=\frac{(-1)^n \Gamma[d/2+1]}{2^{2n}n!\Gamma[d/2+n+1]}\Box^n T^{(0)}_{\mu\nu}\,.
\ee
It is convenient to go to momentum space,
\be
T^{(0)}_{\mu\nu}(x)=\int \frac{d^dp}{(2\pi)^d}\, e^{-ip\cdot x} T_{\mu\nu}(p)\,,
\ee
where we can perform the sum
\be
\sum_{n=0}^\infty\[\frac{1}{n!\Gamma[d/2+n+1]}\(\frac{|p|z}{2}\)^{2n+d/2}\]=I_{d/2}(|p|z)\,,
\ee
if $p$ is space-like. For time-like momenta $p$ it gives instead $J_{d/2}(|p|z)$, recovering (\ref{deltaGprop}).

\subsubsection*{Perturbations in three-dimensional geometries\label{sec:2dpert}}

For $d=2$ there is a crucial simplification: the last term in the last equation of (\ref{perturbedeqns}) is absent! The reason is that for $d=2$, the first two equations (vanishing of the trace and conservation equations) imply that $\Box H_{\mu\nu}=0$. With this simplification, the last equation implies that
\be
\partial_z H_{\mu\nu}=\frac{C_{\mu\nu}}{z^{d+1}}\,,
\ee
where $\partial_z C_{\mu\nu}=0$. Moreover, since $H_{\mu\nu}$ must be finite at $z=0$, the only possibility is that $C_{\mu\nu}=0$. This means that only the $n=0$ term in (\ref{pertg}) survives, while all other higher order terms vanish. This can also be seen from the recursive formula (\ref{Tnexp}): for $d=2$ we have that $\Box T^{(0)}_{\mu\nu}=0$, therefore all $n\geq1$ terms vanish!

The above analysis implies that, to linear order in the perturbation, the general solution for the metric in $d=2$ is given by:
\be\label{deltag2}
\delta g_{\mu\nu}(x^\sigma,z)=\frac{16\pi G_N}{d}T_{\mu\nu}(x^\sigma)\,.
\ee
Since the stress tensor should be traceless and conserved the general form it can take is the sum of right-moving and left-moving waves,
\be
T_{\mu\nu}(t,x)=f(t-x)\left(
                   \begin{array}{cc}
                     1 & -1 \\
                     -1 & 1 \\
                   \end{array}
                 \right)+g(t+x)\left(
                   \begin{array}{cc}
                     1 & 1 \\
                     1 & 1 \\
                   \end{array}
                 \right)\,.
\ee
Specific examples can be obtained by specifying the profiles of $f(t-x)$ and $g(t+x)$. In Appendix \ref{app:examples} we will explore in detail the case corresponding to a local quench state.

\subsection{Linear corrections to entanglement entropy}

Entanglement entropy can be computed via the RT formula  \cite{Ryu:2006bv},
\be\label{hrt}
S_A = \frac{1}{4 G_N}\underset{\gamma_A \sim A}{\min} \left[ \text{area} \left(\gamma_A \right)\right] \,,
\ee
or its covariant HRT version \cite{Hubeny:2007xt}, where the minimality condition is replaced by extremality,
\be
S_A = \frac{1}{4 G_N}\underset{\gamma_A \sim A}{\text{ext}} \left[ \text{area} \left(\gamma_A \right)\right] \,.
\ee
We are interested in computing the leading correction to entanglement entropy, assuming that the geometry is a small perturbation over AdS.

At linear order in the expansion parameter, $\lambda$, entanglement entropy can in principle receive two types of contributions. To see this we can expand the area functional $\mathcal{L}$ and embedding functions $\phi(\xi^i)$ parametrizing the codimension-two surface $\gamma_A$ as follows,
\begin{equation}
\begin{split}
\mathcal{L}[\phi(\xi^i)]&=\mathcal{L}^{(0)}[\phi(\xi^i)]+\lambda\mathcal{L}^{(1)}[\phi(\xi^i)]+\mathcal{O}(\lambda^2)\,,\\
\phi(\xi^i)&=\phi^{(0)}(\xi^i)+\lambda \phi^{(1)}(\xi^i)+\mathcal{O}(\lambda^2)\,,\label{pertexps}
\end{split}
\end{equation}
where $\xi^i$ are coordinates describing the surface. Thus, on one hand, we have corrections due to the change in the geometry, while in the other hand, we have corrections to the surface itself.
However after evaluating $\mathcal{L}[\phi(\xi^i)]$ on-shell, only one term survives at linear order in $\lambda$,
\begin{equation}\label{arealambdaonshell}
\begin{split}
&\delta S_A=\lambda\int d^{d-1}\xi\,\mathcal{L}^{(1)}[\phi^{(0)}(\xi^i)]+\lambda\int d^{d-1}\xi\, \phi^{(1)}(\xi^i)\left[\cancel{\frac{d}{d\xi^i}\frac{\partial\mathcal{L}^{(0)}}{\partial \phi'(\xi^i)}-\frac{\partial\mathcal{L}^{(0)}}{\partial \phi(\xi^i)}}\right]_{\phi^{(0)}}\!\!\!+\mathcal{O}(\lambda^2)\,.
\end{split}
\end{equation}
This means that at this order, the embedding $\phi(\xi^i)$ can be taken to be the (unperturbed) embedding in pure AdS. This is a useful property, because there are many exact solutions for the embending functions of various regions in empty AdS. For our purposes, it will suffice to recall the explicit embedding for spheres in empty AdS in Poincar\'e coordinates,
\be
r^2+z^2=R^2\,.
\ee
We will make use of this expression in later sections when we discuss concrete realizations of perturbative bit threads.

\subsection{Bit threads in dynamical scenarios}

The original formulation of bit threads \cite{Freedman:2016zud} is equivalent to the (non-covariant) RT formula \cite{Ryu:2006bv}, equation (\ref{hrt}), so it only applies to situations with time reflection symmetry (e.g. spatial regions in static spacetimes). In this section we will explain one way to extend this prescription to fully dynamical cases and show that the formulation of \cite{Freedman:2016zud} extends straightforwardly to the case of perturbative excited states.

One way to include time dependence is by using the maximin reformulation of HRT \cite{Wall:2012uf}. To do so, we pick a particular Cauchy surface $\Sigma$ that contains the boundary of the region, $\partial A$, perform the area minimization on it, and then maximize over all possible $\Sigma$. We can then use the standard bit thread prescription for each Cauchy surface $\Sigma$ by maximizing the flux through the boundary region $\Sigma \cap \mathcal{D}[A]$\footnote{$\mathcal{D}[A]$ is the boundary domain of dependence of region $A$.} and then maximizing over all Cauchy surfaces:
\be\label{HRTBitThreadReform}
S_A={\frac1{4G_N}}\max_{\Sigma  \supset \partial A}\underset{\underset{\gamma_A\subset\Sigma}{\gamma_A\sim A}}{\min}\left[\text{area}(\gamma_A)\right]\quad  \iff \quad S_A=\frac1{4G_N}\max_{\Sigma  \supset \partial A} \max_{v \in {\cal F}_\Sigma}\int_{\Sigma \cap \mathcal{D}[A]}\!\!\!\!\!\!\!\!\!\!\!\!\! \sqrt{h} \, n_\mu v^\mu\,,
\ee
where
\be
{\cal F}_\Sigma\equiv\{v\in \mathfrak{X}(\Sigma) \, \vert\, \nabla_\mu v^\mu=0,\, \abs{v}\leq 1\}\,.
\ee
Here, $\mathfrak{X}(\Sigma)$ is the space of vector fields on $\Sigma$. We note that this formula was recently studied in the context of the membrane theory \cite{Agon:2019qgh}. There also exists a fully covariant bit thread version of the correspondence \cite{Headrick:toappear}, but we will not use it in this paper.

A solution to the maximin prescription given by the left-hand side of (\ref{HRTBitThreadReform}) consists of a codimension-two surface $\gamma_A$ that solves the two optimization steps. Such a solution would naturally be accompanied  by a specific choice of a codimension-one hypersurface $\Sigma$ on which $\gamma_A$ is a minimal surface. However, in \cite{Wall:2012uf} it was shown that such $\Sigma$ is highly non-unique away from the maximin surface $\gamma_A$. This fact was used in \cite{Wall:2012uf} to argue that one could pick a particular $\Sigma$ that simultaneously contains the maximin surfaces of various disjoint boundary regions required to prove the strong subadditivity property of holographic entanglement entropy. Below, we will use this freedom to argue that to first order in a general time-dependent perturbation of a static metric, one can always choose $\Sigma$ to be the constant-$t$ hypersurface associated with the unperturbed metric $\Sigma_0$, or more in general, any space-like surface that is perturbatively closed to it and passes through $\gamma_A$, $\Sigma_\lambda$.

\subsubsection{The case of perturbative excited states\label{pertstatesBT}}

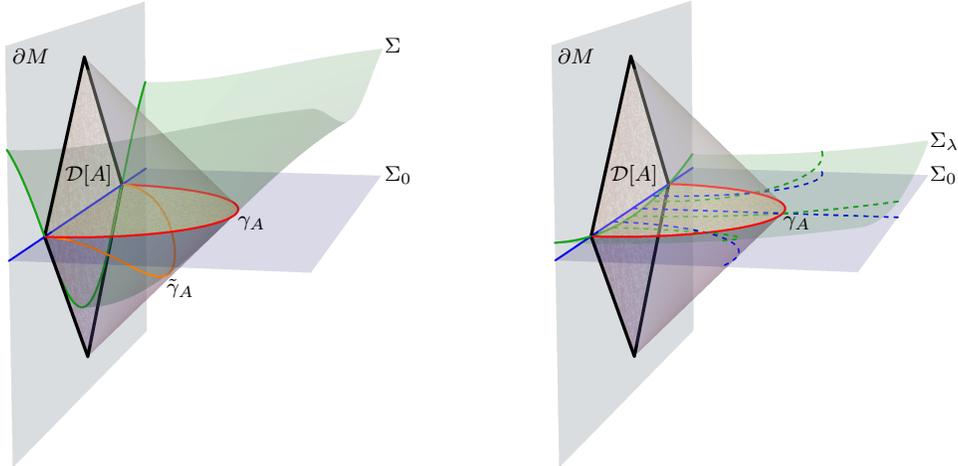
\begin{figure}[t!]
\begin{center}
  \includegraphics[width=5cm]{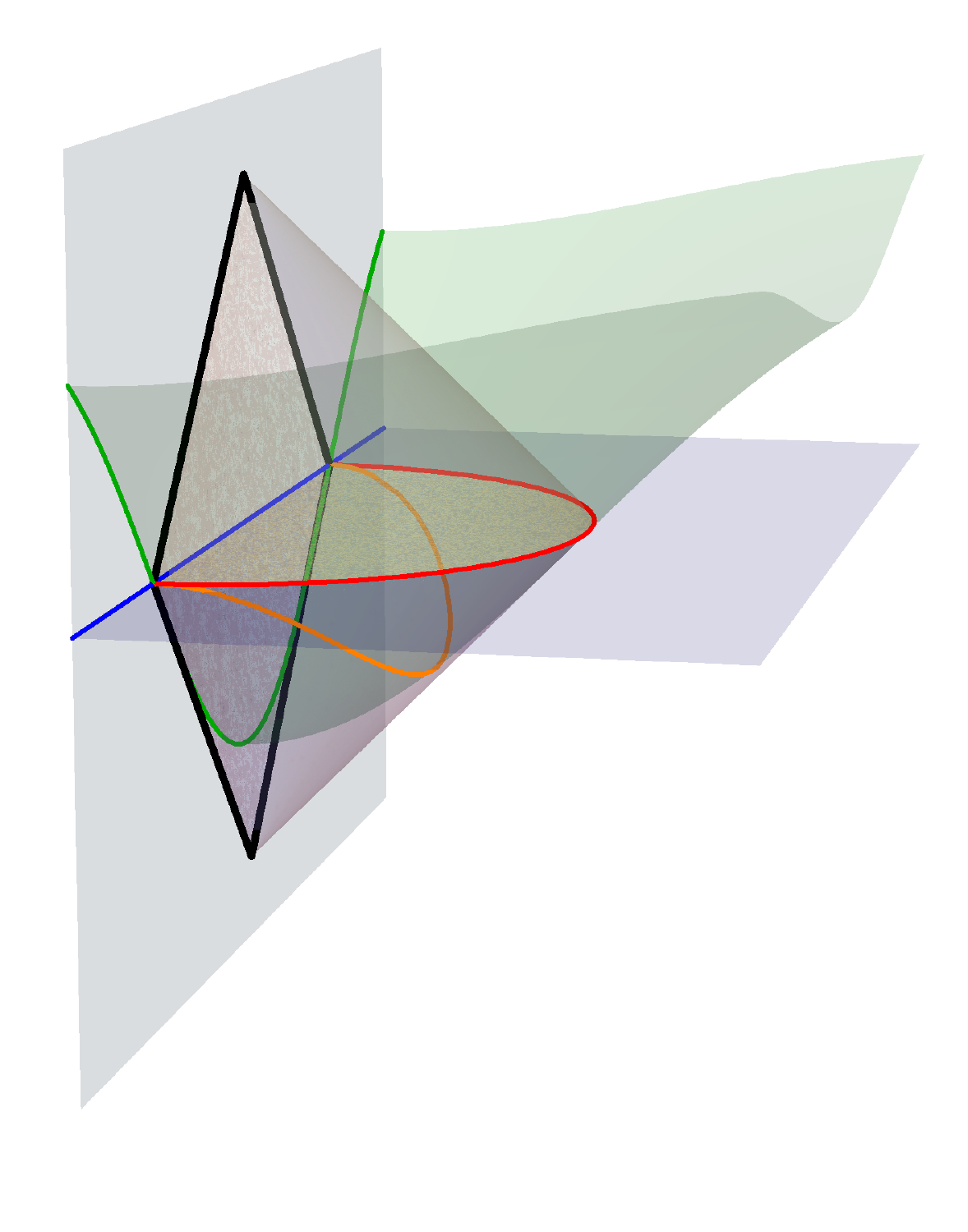}
  \hspace*{2cm}
  \includegraphics[width=5cm]{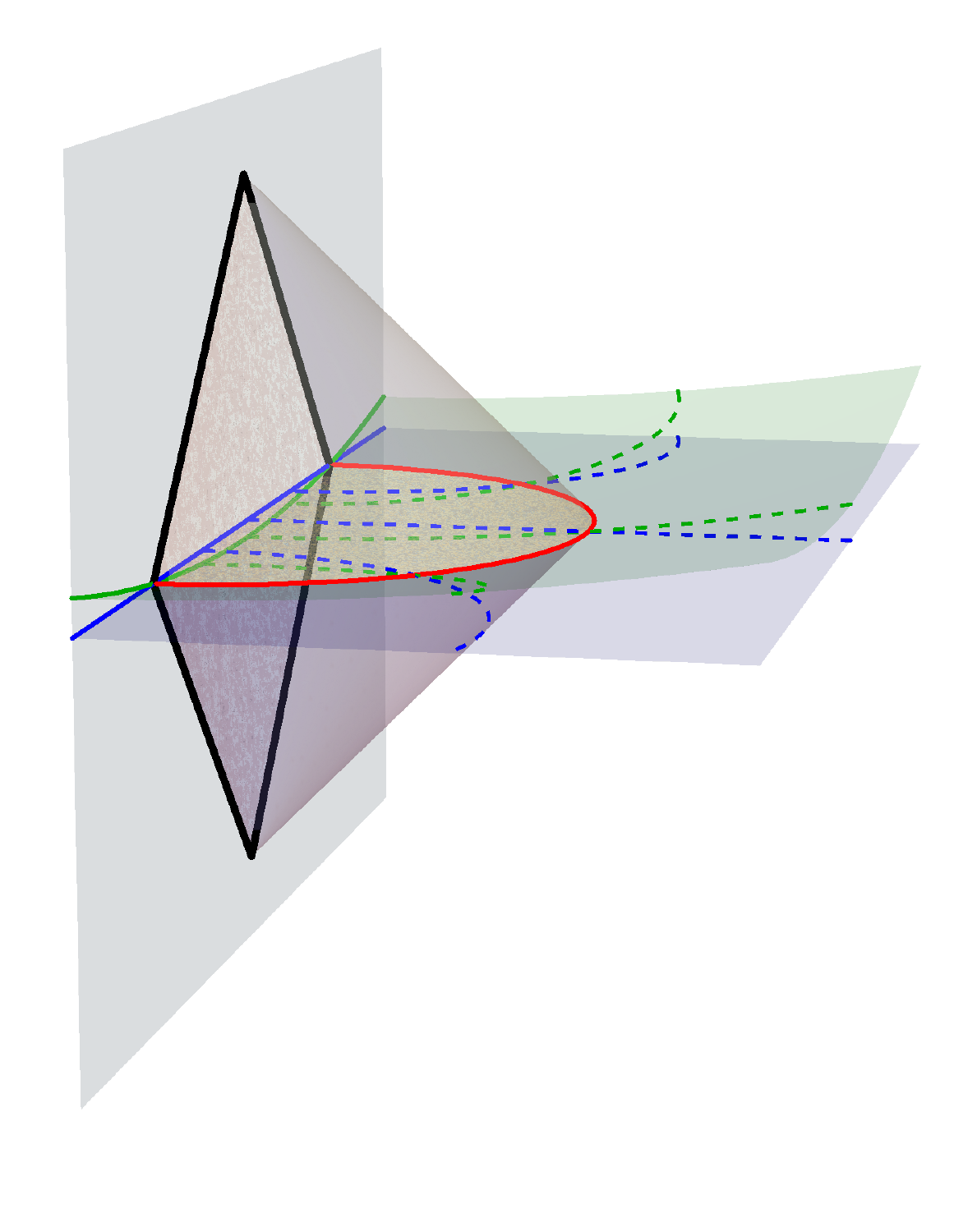}
  \setlength{\unitlength}{1cm}
  \setlength{\unitlength}{1cm}
\begin{picture}(0,0)
\put(-12.55,5.35){\scriptsize$\partial M$}
\put(-11.85,3.8){\scriptsize$\mathcal{D}[A]$}
\put(-9.55,3.2){\scriptsize$\gamma_A$}
\put(-10.5,2.3){\scriptsize$\tilde{\gamma}_A$}
\put(-7.6,5.5){\scriptsize$\Sigma$}
\put(-7.6,3.8){\scriptsize$\Sigma_0$}
\put(-5.3,5.35){\scriptsize$\partial M$}
\put(-4.6,3.8){\scriptsize$\mathcal{D}[A]$}
\put(-2.3,3.2){\scriptsize$\gamma_A$}
\put(-0.35,4.25){\scriptsize$\Sigma_\lambda$}
\put(-0.35,3.8){\scriptsize$\Sigma_0$}
\end{picture}
\end{center}
\vspace{-0.5cm}
\caption{\small A solution to the maximin problem $\gamma_A$ is naturally accompanied  by a specific choice of a codimension-one slice $\Sigma$ on which $\gamma_A$ is a minimal area surface. Such a slice is highly non-unique, however, not all slices that pass through $\gamma_A$ are allowed. Left: in this example $\Sigma$ is perturbatively close to the null congruence shot out from $\gamma_A$. In this case the minimal area surface $\tilde{\gamma}_A$ (orange curve) does not coincide with $\gamma_A$ (red curve). Right: for perturbative excited states, it can be shown that $\gamma_A$ is a minimal area surface on any slice $\Sigma_\lambda$ that is perturbatively close to $\Sigma_0$. This means that we can pick any of these surfaces, and in particular $\Sigma_0$, to construct relevant bit thread configurations.}
\label{fig:btslices}
\end{figure}

Even though the choice of $\Sigma$ is highly non-unique, it can be shown that not any slice that passes through $\gamma_A$ is a good one. The reason is that $\gamma_A$ is not necesarily minimal on any of such slices $\Sigma$. To see this, consider a null congruence shot out from $\gamma_A$. The surface $\gamma_A$ is extremal, hence, its expansion vanishes: $\theta=0$. However, the Raychaudhuri equation implies that $d\theta/d\lambda<0$ \cite{Wall:2012uf}. This means that in this case $\gamma_A$ is a local maximum of area rather than a minimum and, by continuity, the same should hold for space-like surfaces $\Sigma$ that are close enough to the null congruence. In the left panel of Figure \ref{fig:btslices} we give an example to illustarate this fact. Notice that in one of these slices $\Sigma$ the minimal area surface $\tilde{\gamma}_A$ is not the same as extremal surface $\gamma_A$. Therefore, finding a max flow in $\Sigma$ is not equivalent to computing the entanglement entropy of region $A$.

For the case of perturbative excited states, a natural candidate for a good Cauchy slice would be a slice $\Sigma_\lambda$ that is perturbatively close to the $t=t_0$ hypersurface associated with the unperturbed metric, $\Sigma_0$. We can parametrize such a slice as $t=t_0+\lambda\, \delta t(z,\vec{x})$, with the constraint that $\delta t(z,\vec{x})$ must vanish at $\gamma_A$. The question here is if we can find surfaces on $\Sigma_\lambda$ that are homologous to $A$ but have smaller area than $\gamma_A$ at order $\lambda$. Supposing there are such surfaces, we denote $\tilde{\gamma}^\lambda_A$ as the one with the minimal area. However, we know that $\gamma_A$ is a minimal area surface in the unperturbed background, therefore, by continuity we know that $\tilde{\gamma}_A^\lambda\to \gamma_A$ as $\lambda\to0$.  Without loss of generality we can then parametrize such a surface with embedding functions as in (\ref{pertexps}). On the other hand, the calculation in (\ref{arealambdaonshell}) shows that corrections to the embedding do not affect the area at linear order. This means that $\text{area}(\tilde{\gamma}^\lambda_A)=\text{area}(\gamma_A)+\mathcal{O}(\lambda^2)$, so we can conclude that $\gamma_A$ \emph{is} a minimal area surface on any $\Sigma_\lambda$ perturbatively close to $\Sigma_0$. We illustrate this result in the right panel of Figure \ref{fig:btslices}. This also implies that on any of these surfaces, and in particular on $\Sigma_0$, the solution to the max flow ploblem computes the entanglement entropy of region $A$, and hence all of them are equally good for the construction of bit thread configurations.

\section{Simple realizations of perturbative bit threads \label{PBT}}

Given the enormous simplification that happens at $\mathcal{O}(\lambda)$ from the point of view of the HRT prescription,
we would like to study and understand the general properties of perturbative thread configurations based on the constructions developed in \cite{Agon:2018lwq}.
We will start by stating simple constraints that the $\mathcal{O}(\lambda)$ HRT surfaces induce on general bit threads, and then proceed with the specific constructions. We will show that these methods  lead to thread configurations that successfully encode general properties of the CFT state and the bulk geometry, such as the first law of entanglement entropy and its relation to the (linearized) Einstein's equations, albeit in a highly nonlocal form.
Along the way, we will state the precise problem of metric reconstruction that we look to solve and enumerate the challenges that these simple constructions face, leading to a quest for a new method that exploits
bulk locality in a more explicit way.

\subsection{Generalities\label{generalities}}

Let us begin by considering empty AdS$_{d+1}$ in spherical coordinates. The geometry of a constant-$t$ slice $\Sigma$ is given by
\be
ds_{\Sigma}^2=\frac{1}{z^2}\(dr^2+r^2d \Omega^2_{d-2}+dz^2\)\,.
\ee
The minimal surface $\gamma_A$ for a ball of radius $R$ is given implicitly by
\be\label{emb2d}
r^2+z^2=R^2\,,
\ee
and its outward-pointing unit normal vector $\hat{n}$ at a point $(r,z)$ on the minimal surface is:
\be
\hat{n}^{a}=\frac{z}{R}\(r,z\)\,.
\ee
For simplicity we have omitted the angular coordinates, since both the minimal surface and the state are invariant under rotations.

A simple realization of a vector field/thread configuration, $v=|v|\hat{\tau}$, based on geodesics is given by (see \cite{Agon:2018lwq} for details)
\be\label{Vec1-d}
v^a=\(\frac{2Rz}{\sqrt{(R^2+r^2+z^2)^2-4R^2r^2}}\)^{d}\(\frac{r z}{R}\, , \frac{R^2-r^2+z^2}{2R} \)\,,
\ee
with
\bea
&&|v|=\(\frac{2Rz}{\sqrt{(R^2+r^2+z^2)^2-4R^2r^2}}\)^{d-1}\!\!\!\!\!\!\,,\label{Vec23-d}\\
&&\hat{\tau}^a =\frac{2Rz}{\sqrt{(R^2+r^2+z^2)^2-4R^2r^2}}\( \frac{r z}{R}\, ,\, \frac{R^2-r^2+z^2}{2R}\)\,.
\eea
As a check, notice that this vector field $(i)$ satisfies the divergenceless condition $\nabla \cdot v=0$ and $(ii)$ is equal to $\hat{n}$ at the location of the minimal surface $v|_{\gamma_A}=\hat{n}$. Combining these two, it immediately follows that the flux along any bulk surface $\Gamma_A$ homologous to $A$ (not necessarily the minimal surface $\gamma_A$) yields the entanglement entropy of the ball (in units of $4G_N$),
\be
S_A=\frac{1}{4G_N}\int v \cdot dS_{\Gamma_A}= \frac{1}{4G_N}\int \hat{n} \cdot dS_{\gamma_A} = \frac{1}{4 G_N}{\rm min} \left[ \text{area} \left(\gamma_A \right)\right]\,.
\ee
We emphasize that while the minimal surface $\gamma_A$ is in most cases \emph{unique}, the choice of vector field $v$ is highly \emph{non-unique}; it is  uniquely determined only at the bottle-neck $\gamma_A$.

Next, we would like to find the perturbed vector field in a perturbatively excited state, i.e., a state with bulk metric $g^{\lambda}_{\mu\nu}=g_{\mu\nu}+\lambda \delta g_{\mu\nu}+\mathcal{O}(\lambda^2)$ (satisfying Einstein's equations):
\be\label{eq:vlambda}
v_\lambda=v+\lambda \delta v+\mathcal{O}(\lambda^2)\,,
\ee
at linear order in $\lambda$. While the perturbation in the vector field $\delta v$ is on its own highly non-unique, any consistent realization must satisfy some nontrivial properties, including the first law of entanglement entropy in the CFT and the linearized Einstein's equations in the bulk. The problem that we want to address is the following:

\vspace{2mm}
\noindent\fbox{
    \parbox{0.975\textwidth}{
    \centering
    \emph{Given a consistent thread configuration for an excited state $v_\lambda$, is it possible to reconstruct locally the bulk geometry at the same order in the perturbation?}
    }%
}

\vspace{2mm}
A couple of comments are in order. First, note that we are focusing on excited states. While it is true that the same question would make sense even in the vacuum state, we recall that the bulk metric in this case is fixed by symmetries, rendering the problem exceptionally simple. Second, the non-uniqueness of $v_\lambda$ for a given metric indicates that the correspondence is not one-to-one. Even if we isolate a family of thread configurations that follow from the same bulk metric, the way they encode this information may be non-unique and, generically, highly nonlocal. In the following, we will identify basic constraints that generic realizations of $v_\lambda$ must satisfy and then, study how the particular constructions of \cite{Agon:2018lwq} encode the information about the bulk metric.

\subsubsection{Boundary conditions for the perturbed threads}

In order to find a solution $v=|v|\hat{\tau}$ for a thread configuration, we need to solve for the divergenceless condition $\nabla\cdot v=0$ subject to the norm bound $|v|\leq1$. One way to proceed is to use the fact that the norm bound is saturated $|v|=1$ at the bottle-neck $\gamma_A$. In other words, we need to impose that at the minimal surface $\gamma_A$, $v$ is equal to its unit normal,
\be\label{bcRT}
v^a_\lambda|_{\gamma_A}=\hat{n}^a\,.
\ee
Notice that this does not uniquely determine the vector field everywhere in the bulk; intuitively, the ambiguity of the thread configuration away from $\gamma_A$ corresponds to a choice of microstate in the dual CFT, such that all the macroscopic properties of the system are satisfied, including the entanglement entropy $S_A$.

Let us now determine how (\ref{bcRT}) looks like in the perturbed geometry. Fortunately, at the linear order in the perturbation the RT surfaces are unchanged and we can use this to our advantage. This implies that at this order, the change in the normal vector is only induced by the change in the geometry. To see this, consider
the metric on a constant-$t$ slice $\Sigma$ of the perturbed geometry\footnote{The indices $(a,b)$ here run over the space coordinates $x^a=\{x^i,z\}$. }
\be
ds^{2}_{\Sigma}=g^{\lambda}_{ab}dx^adx^b=(g_{ab}+\lambda\delta g_{ab})dx^adx^b\,,
\ee
where
\be
g_{ab}=\frac{1}{z^2}\left(
                            \begin{array}{cc}
                              \delta_{ij} & 0 \\
                              0 & 1 \\
                            \end{array}
                          \right)\,,\qquad  \delta g_{ab}=z^{d-2}\left(
                            \begin{array}{cc}
                             H_{ij} & 0 \\
                              0 & 0 \\
                            \end{array}
                          \right)\,.
\ee
We will keep the $\lambda$'s explicitly throughout our calculations as a bookkeeping device (to count the order of the perturbations), but at the end we will set it to unity. Also, for future reference, we give an explicit expression for the inverse metric at linear order in $\lambda$,
\be
g_\lambda^{ab}=g^{ab}+\lambda\delta g^{ab}\,,
\ee
where:
\be
g^{ab}=z^2\left(
                            \begin{array}{cc}
                              \delta_{ij} & 0 \\
                              0 & 1 \\
                            \end{array}
                          \right)\,,\qquad  \delta g^{ab}=-z^{d+2}\left(
                            \begin{array}{cc}
                             H_{ij} & 0 \\
                              0 & 0 \\
                            \end{array}
                          \right)\,.
\ee
As explained in the previous section, the embedding function (\ref{emb2d}) is not corrected at this order. Therefore its normal covector $\hat{n}_a$ remains the same, up to an overall constant $N$,
\be
\hat{n}_a=\frac{N}{Rz}x_a\,.
\ee
Ensuring that $\hat{n}$ is properly normalized to one, we find that at linear order in $\lambda$:
\be
N=1+\frac{\lambda z^2}{2R^2}\delta g_{ab}x^ax^b\,.
\ee
Finally, raising the index with the inverse metric we find that
\bea
\hat{n}^a&=&\frac{1}{Rz}\left(N g^{ab}x_b+\lambda \delta g^{ab} x_b\right)\,,\nonumber \\
&=&\frac{z}{R}x^a+\lambda\left(\frac{z\delta g_{cd}x^cx^d g^{ab}}{2R^3}+\frac{\delta g^{ab}}{Rz}\right)x_b\,.
\eea
For example, in $d=2$, we find that:
\be
\hat{n}^a=\frac{z}{R}\(x,z\)-\frac{\lambda x z^3 H(t,x)}{2 R^3}\left(x^2+2 z^2,-x z\right)\,, \quad H(t,x)\equiv H_{xx}(t,x)\,.
\ee
For $d\geq3$ we can obtain similar but more longwinded expressions but for the sake of simplicity we will not transcribe them here.
Finally, from (\ref{bcRT}) we find that at linear order in $\lambda$, our boundary condition at the bottle-neck $\gamma_A$  is:
\be\label{bcVector}\boxed{
v^a_\lambda|_{\gamma_A}=\frac{z}{R}x^a+\lambda\left(\frac{z\delta g_{cd}x^cx^d g^{ab}}{2R^3}+\frac{\delta g^{ab}}{Rz}\right)x_b\,.}
\ee
We emphasize that this condition does not uniquely determine $v_\lambda$ in the bulk, specially in regions far away from $\gamma_A$ where $v_\lambda$ is highly non-unique.

\subsubsection{First law of entanglement entropy}
Since $v_\lambda$ is divergenceless, the flux across any bulk surface homologous to $A$ is constant. Hence, the boundary condition (\ref{bcVector}) should be enough to
demonstrate the first law of entanglement, provided we pick $\gamma_A$ itself as our homologous region.

To illustrate this, we can perform a simple analysis in $d=2$ dimensions. The area element $dS_{\gamma_A}$ in this case is given by
\bea
dS_{\gamma_A}=\hat{n} ds_{\gamma_A}&=&\hat{n} \frac{dx}{z(x)}\sqrt{1+\lambda z^2 H_{xx}(t,x,z(x))+\mathcal{O}(\lambda^2)+z'(x)^2}\,,\\
&=&\hat{n} dx \[\frac{R}{R^2-x^2}+\lambda\frac{ (R^2-x^2) }{2 R}H_{xx}\(t,x,z(x)\)+\mathcal{O}(\lambda^2)\]\,.
\eea
The order $\mathcal{O}(\lambda)$ term gives the change in entanglement entropy,
\be
\delta S_A=\frac{1}{4G_N}\int \hat{n} \cdot dS_{\gamma_A}=\frac{1}{4G_N}\int dx\,\frac{ (R^2-x^2) }{2 R}H_{xx}\(t,x,z(x)\)\,.
\ee
Finally, according to (\ref{pertg}) we can expand $H_{xx}(t,x,z(x))$ as
\be
H_{xx}(t,x,z(x))=8\pi G_N \sum_{n=0}^\infty z(x)^{2n}T^{(n)}_{xx}(t,x)\,.
\ee
However, as emphasized in the previous section, for $d=2$ only the $n=0$ survives. By the traceless condition we know that $T^{(0)}_{xx}(t,x)=T^{(0)}_{00}(t,x)=T_{00}(t,x)$, so we arrive to the first law of entanglement entropy with the right modular Hamiltonian in 2d \cite{Faulkner:2013ica}
\be\label{1stlawEE}
\delta S_A=2\pi \int_{-R}^R dx\,\frac{ (R^2-x^2) }{2 R}T_{00}(t,x)\,.
\ee
For $d>2$ the proof is slightly more complicated, but it can be shown by working out the above expansions in momentum space, and resuming the resulting series. We refer the reader to \cite{Blanco:2013joa} for a detailed analysis in these higher dimensional cases.

The crucial insight here is that \emph{any} divergenceless vector field satisfying (\ref{bcVector}) will automatically encode the first law of entanglement entropy,
which for arbitrary dimensions takes the form
\be
\delta S_A=2\pi \int_{-R}^R d^{d-1}x\,\frac{ (R^2-r^2) }{2 R}T_{00}(x^\sigma)\,,\qquad r^2\equiv\sum_{i=1}^{d-1} x_i^2\,.
\ee
Since the first law of entanglement entropy has been shown to be equivalent to the bulk Einstein's equations at the linear level \cite{Faulkner:2013ica}, then \emph{all} consistent thread configurations should also encode them in some form. It remains to be seen how are the Einstein's equations encoded in the specific thread configurations, and how easy would be to recover the metric from particular constructions.

\subsection{Method 1: Geodesic bit threads \label{F1}}

Following \cite{Agon:2018lwq}, we will now present simple methods to construct explicit thread configurations satisfying the boundary condition (\ref{bcVector}) for perturbative excited states. The first method consists on picking a family of integral curves with good properties, and then fixing the norm by ensuring that Gauss's law is satisfied everywhere. In the following, we will describe this construction in some detail and study how the information of the bulk metric is encoded in the resulting thread configuration.

\subsubsection{Integral curves\label{sec:form1-intc}}

A good family of integral curves must satisfy the following properties:
  \begin{enumerate}
  \item They must be orthogonal to the minimal surface $\gamma_A$.
  \item They must be continuous and not self-intersecting.
  \item They must start and end at the boundary, or possibly at a bulk horizon.
\end{enumerate}
Given a family with these properties, it is then straightforward to construct a divergenceless vector field with the desired boundary condition. There is a small caveat here, however: one can only check if the norm bound  is satisfied $|v|\leq 1$ \emph{a posteriori}.

One crucial result of \cite{Agon:2018lwq} is that a thread construction based on space-like geodesics automatically satisfy the norm bound, provided that the metric background satisfies some simple geometric properties. This conclusion followed from a systematic analysis of geodesic foliations of an arbitrary Riemannian geometry, so it must also hold true for the case in consideration, i.e., for geometries dual to perturbative excited states. Therefore, our first candidate for the family of integral curves will be the space-like geodesics of the perturbed background.

\paragraph{Corrected geodesics:}

Let us consider the $d=2$ and $d>2$ cases separately. In \cite{Agon:2018lwq} it was shown that space-like geodesics in an arbitrary $(2+1)$-dimensional $(d=2)$ background lead to a vector field satisfying the norm bound $|v|\leq1$, provided that the Ricci scalar
on a constant-$t$ slice (a Riemannian submanifold) is negative everywhere, i.e.
\be\label{critd2}
R<0\,.
\ee
We can check that this condition is indeed satisfied for the perturbative states that we are considering.
Working in coordinates adapted to the geodesics, and using the same notation of \cite{Agon:2018lwq}, we will write the bulk metric as follows:
\be
ds^2\equiv G_{\mu\nu}dx^{\mu}dx^{\nu}=-\psi(\lambda,x) dt^2+d\lambda^2+\gamma(\lambda,x) dx^2\,,
\ee
where $x$ labels different points along the minimal surface and $\lambda$ is an affine parameter that runs along geodesics orthogonal to it.\footnote{This coordinate system does not need to foliate the full manifold; points that are not covered by these coordinates have by definition a vanishing vector field $v=0$.} The above metric is a solution of Einstein's equations:\footnote{We have set $8\pi G_N=1$ for simplicity.}
\be\label{EEQ-CC}
\mathcal{R}_{\mu \nu}-\frac{1}{2} \mathcal{R} G_{\mu \nu}+\Lambda G_{\mu \nu}=\mathcal{T}_{\mu\nu}\,,
\ee
where $\mathcal{T}_{\mu\nu}$ is the bulk energy momentum tensor.
A quick calculation shows that the induced Ricci on a constant-$t$ slice is:
\be
R=\frac{2}{\psi(\lambda,x)}(\mathcal{T}_{00}(\lambda,x)+\Lambda \psi(\lambda,x))\,,
\ee
hence, for negative cosmological constant $\Lambda<0$, we have that $R<0$ if and only if the local energy density is bounded from above:
\be\label{boundE}
\varepsilon(\lambda,x)\equiv -\mathcal{T}^0_{\,\,\;\;0}(\lambda,x)<-\Lambda\,.
\ee
Since the kind of perturbations that we are considering are all vacuum solutions, i.e. we have $\mathcal{T}_{\mu\nu}=0$, then we conclude that the corrected geodesics can indeed be taken as a good family of integral curves.

For spheres in higher dimensional spaces ($d>2$) the situation is a bit more complicated. Assuming that the state is invariant under rotations, we can pick a plane that intersects the origin and find the geodesics within this plane. Then we foliate the full spacetime by surfaces of revolution generated by rotating such geodesics along all possible angles. With this construction, the bulk metric can be written as
\bea\label{metricsph}
ds^2=-\psi(\lambda,x) dt^2+d\lambda^2+\gamma(\lambda,r)dr^2+e^{2\tau(\lambda,r)}d\Omega_k^2\,,\qquad ds_2^2\equiv d\lambda^2+\gamma(\lambda,r)dr^2\,.
\eea
After some algebra, one finds that the criterion (\ref{critd2}) generalizes to \cite{Agon:2018lwq}
\bea\label{R2cond}
R_2<2k\left[\partial^2_\lambda\tau+\partial_\lambda\tau(k+\partial_\lambda\log\gamma)\right]\,,
\eea
where $R_2$ is the induced Ricci on the auxiliary 2-dimensional metric defined in (\ref{metricsph}). On a pure AdS background, one finds that $R_2=-2$, while the terms on the right hand side of (\ref{R2cond}) are strictly positive. This means that there is a finite gap, or in other words, that the bound is $\mathcal{O}(1)$ far from saturation. On the other hand, linear perturbations of the metric would lead to corrections on both sides of the equation but these corrections can only be of order $\mathcal{O}(\lambda)$. This means that for sufficiently small $\lambda$, the condition (\ref{R2cond}) will still hold true, regardless of the fluctuations. Similar arguments could be made for metrics that are perturbatively close to AdS but are not rotationally invariant, however, the analysis would be certainly more complicated. In these situations one would need to find corrected geodesics within infinitely many planes intersecting the origin and repeat the above steps. But, again, since the pure AdS case is far from saturating (\ref{R2cond}), the analysis at linear order would only lead to corrections of order $\mathcal{O}(\lambda)$, meaning that the bound would always be satisfied for sufficiently small $\lambda$.

The above arguments show that the $\mathcal{O}(\lambda)$ geodesics are good candidates for integral curves for any number of dimensions.
There is a slight technical problem, however: it is practically impossible to obtain closed expressions for the corrected geodesics in a \emph{generic} perturbed background. In practice, rather than working with the corrected geodesics, it is more convenient to propose an alternative family of integral curves. In the following we will explore this possibility in more detail.

\paragraph{Uncorrected geodesics:}

The corrected geodesics are far from saturating the bound (\ref{critd2}) in $d=2$ or, more generally, (\ref{R2cond}) in higher dimensions. Therefore, it is clear that a continuous family of curves that are perturbatively close to them will similarly do the job. The most natural and simplest candidate for this are the \emph{uncorrected} space-like geodesics.

To illustrate this point we will consider the $d=2$ case, where we can make a precise analytic statement. In this case, the minimal surface (\ref{emb2d}) is given implicitly by
\be\label{minimalsurf}
z_m^2+x_m^2=R^2\,.
\ee
We have added subindexes `$m$' to point out that these coordinate points are on $\gamma_A$. The geodesics in pure AdS are given by semicircles anchored at the boundary. These semicircles form a two-parameter family of curves and are defined implicitly by
\bea\label{geodesic}
(x-x_s)^2+z^2=R_s^2
\eea
where $x_s$ is the center of the circle and $R_s$ its radius. The tangent vector with unit norm at an arbitrary point is given by
\bea\label{tau}
\hat{\tau}^a=\left(\frac{z}{R_s}-\frac{\lambda z^5 H(t,x)}{2 R_s^3}\right)\(z , x_s-x\)\,,
\eea
where $H(t,x)\equiv H_{xx}(t,x)$. As expected, the tangent vector still points in the same direction but its normalization is corrected at leading order in the perturbation.
Since the integral curves must be orthogonal to the minimal surface, we must enforce that $\hat{\tau}|_{\gamma_A}=v_\lambda|_{\gamma_A}$, where the latter is given in (\ref{bcVector}). At order $\mathcal{O}(\lambda)$, this requirement leads to\footnote{With these definitions $R_s$ can take negative values. We can take an absolute value of $x_m$ in the denominator of (\ref{80}) to make $R_s$ positive. However, allowing $R_s$ to take any value will be useful below, in the definitions of $x_a$ and $x_{\bar{a}}$.}
\bea\label{80}
&&R_s(x_m)=\frac{R \sqrt{R^2-x_m^2}}{x_m} \left[1+\frac{\lambda (R^2-x_m^2)^2 H(t,x_m)}{R^2}\right]\,,\\
&&x_s(x_m)=\frac{R^2}{x_m}\left[1+\frac{\lambda (R^2-x_m^2)^2 H(t,x_m)}{R^2}\right]\,.\label{81}
\eea
In order to arrive to these expressions we have made use of the equation (\ref{minimalsurf}) to eliminate $z_m$. Finally, plugging (\ref{80})-(\ref{81}) into  (\ref{geodesic}) we obtain an implicit expression for the family of geodesics orthogonal to $\gamma_A$, parametrized by the point $x_m\in[-R,R]$ on the minimal surface.

Next, we need to check if the proposed integral curves are properly nested \cite{Agon:2018lwq}. In order to check this, we find the point $x_a$ at which they intersect $A$,\footnote{If we insist that $R_s\geq0$, these definitions for $x_a$ and $x_{\bar{a}}$ would only be valid for $x_s\geq0$, while for $x_s\leq0$ one should interchange the two.}
\bea\label{r0}
x_a=x_s-R_s=\frac{R}{x_m}\(R-\sqrt{R^2-x_m^2} \)\left[1+\frac{\lambda (R^2-x_m^2)^2 H(t,x_m)}{R^2}\right]\,,
\eea
and the dual point $x_{\bar{a}}$ at which the curves intersect $\bar{A}$,
\bea\label{barr0}
x_{\bar{a}}=x_s+R_s=\frac{R}{x_m}\(R+\sqrt{R^2-x_m^2} \)\left[1+\frac{\lambda (R^2-x_m^2)^2 H(t,x_m)}{R^2}\right]\,.
\eea
One can check that self-intersection is avoided if and only if $dx_a/dx_m>0$ and $dx_{\bar{a}}/dx_m<0$. A quick calculation leads to
\bea
\!\!\!\!\!\!\!\!\!\!\!\!\!\!\frac{dx_a}{dx_m}&=&\frac{R^2}{x_m^2} \frac{(R-\sqrt{R^2-x_m^2})}{\sqrt{R^2-x_m^2}}\times\nonumber\\
&&\quad\left[1+\frac{\lambda (R^2-x_m^2)^2 H(t,x_m)}{R^2}+\frac{\lambda x_m\sqrt{R^2-x_m^2}}{R^3}\frac{d}{dx_m}\left((R^2-x_m^2)^2 H(t,x_m)\right)\right],
\eea
\bea
\!\!\!\!\!\!\!\!\!\!\!\!\!\!\frac{dx_{\bar{a}}}{dx_m}&=&-\frac{R^2}{x_m^2} \frac{(R+\sqrt{R^2-x_m^2})}{\sqrt{R^2-x_m^2}}\times\nonumber\\
&&\quad\left[1+\frac{\lambda (R^2-x_m^2)^2 H(t,x_m)}{R^2}+\frac{\lambda x_m\sqrt{R^2-x_m^2}}{R^3}\frac{d}{dx_m}\left((R^2-x_m^2)^2 H(t,x_m)\right)\right].
\eea
One can check that at order $\mathcal{O}(1)$ both conditions are satisfied, i.e., $dx_a/dx_m>0$ and $dx_{\bar{a}}/dx_m<0$. At linear order in the perturbation, we get a term that does not have a definite sign (the last term in the square brackets), but one can always choose a small enough $\lambda$ such that these inequalities are still satisfied. As an example, let us consider a plane wave, $H(t,x)=\epsilon \sin[\omega(t-x)]$.\footnote{In this example even the second term in the square brackets can have a negative sign, but this problem goes away when one impose energy conditions. The last term, however, will still be indefinite after imposing energy conditions.} The last term in the square brackets can become order $\mathcal{O}(1)$ if the frequency $\omega$ is large enough. To prevent this to happen, one must take $
\lambda\ll1/(\epsilon\, \omega R^3)$. If the background is decomposed in Fourier modes, then the maximum frequency will be the relevant one, and the above condition is replaced by
\be
\lambda\ll\frac{1}{\epsilon\, \omega_{\text{max}} R^3}\,.
\ee
This means that for smooth functions we can always find a small $\lambda$ that satisfies the conditions. For sharply peaked functions this might not be the case, since the Fourier spectrum could contain arbitrarily high frequency modes. We will therefore restrict our attention to states with smooth stress energy tensor. Notice that this is not an important restriction. In CFT language, a state with a sharply peaked stress energy tensor will not be perturbatively close to the vacuum, and hence, the gravity dual would have important higher order contributions that we have ignored in the approximation of linearized gravity.

\subsubsection{Magnitude}

Given a set of integral curves the next step is to find the appropriate norm of the vector field $|v_\lambda|$. We will denote $X(x_{m},\xi)$ the proposed family of curves; $x_m$ labels points on the minimal surface and $\xi$ is a parameter that runs along the curve. As explained above, the curves $X(x_{m},\xi)$ can be the uncorrected geodesics. The parameter $\xi$ can be taken as the proper length from the given point to the minimal surface.

Following \cite{Agon:2018lwq}, we now fix the norm by implementing a version of Gauss's law for an infinitesimal cylinder enclosing each curve.\footnote{Alternatively, we could fix the norm by solving the first order differential equation for $|v_\lambda|$ resulting from the divergenceless condition, subject to the appropriate boundary condition at $\gamma_A$. This would be completely equivalent to the Gauss's law method described here, since the latter condition is the differential form of Gauss's law. However, since we have the explicit form of the integral lines, the Gauss's law turns out to be more convenient in this case, providing a final answer in closed form, as shown below in equation (\ref{magnitudeV}).} More specifically, we impose that the flux through an infinitesimal area element $\delta A$ transverse to one of the threads is constant,
\bea\label{divcond}
\int_{\delta A} |v_\lambda| \sqrt{h |_{\lambda}} d^{d-1}x={\rm constant}\,,
\eea
where $h_{ab}=g_{ab}-\hat{\tau}_a\hat{\tau}_b$ is the projection of the metric on a plane orthogonal to the integral curve. Using the fact that at the minimal surface $|v_\lambda(x_m,\xi_m)|=1$, and letting $\delta A\to0$, we arrive to the following expression for the norm
\bea\label{magnitudeV}
 |v_\lambda(x_{m},\xi)|  =\frac{\sqrt{h_\lambda(x_{m},\xi_{m}) }}{\sqrt{h_\lambda(x_{m},\xi)}}\,,
\eea
where $\xi_{m}$ is the parameter at which the curve intersects the minimal surface $\gamma_A$. Notice that we do not need to verify whether the norm bound $|v_\lambda|\leq1$ is satisfied everywhere. This is already guaranteed given our choice of integral curves and the argument based on the negativity of the scalar curvature presented in Section \ref{sec:form1-intc}. Reference \cite{Agon:2018lwq} provides various explicit examples of geodesic flows constructed with this method, including the case of spherical regions in empty AdS, given in equation (\ref{Vec1-d}). In Appendix \ref{app:examples} we complement this study by constructing a new explicit example, now for the case of the specific perturbative excited state corresponding to a local quench.

We can now inquire about how the bulk metric and the Einstein's equations are encoded in this particular construction. Unfortunately, at this level we can already see that such information is encoded in the vector field $v_\lambda$ in a highly nonlocal fashion. On one hand, one needs to solve for the geodesics in the unperturbed background subject to a boundary condition that depends on a particular metric perturbation. And, on the other hand, the magnitude of the vector is found by transporting the boundary condition along the geodesic, ensuring that the vector field is divergenceless. This process is inherently nonlocal; in particular, the final result for $|v_\lambda|$ exhibits an explicit bilocal dependence on the metric perturbation, since it must be evaluated at the points labeled by $\xi_m$ and $\xi$. The latter parameter, in particular, encodes the proper distance between the point in consideration and the minimal surface $\gamma_A$, which is nonlocal information on its own. These observations imply that it would be rather difficult to invert the problem and recover the metric from the resulting thread configuration. Similarly, the same remarks apply for the Einstein's equations: even though they are assumed as a starting point for this construction (the perturbations we consider are on-shell), they are ultimately encoded nonlocally in the resulting thread configuration.

\subsection{Method 2: Level set construction \label{lsc}}

The second method of constructing thread configurations consists on starting with a specific family of level set hypersurfaces and then building up a vector field
that is orthogonal to them and, of course, divergenceless. This is a slight generalization of a method initially proposed in \cite{Agon:2018lwq}, as we will see below. In the following, we will spell out the details of the general construction for arbitrary metrics, and then specialize to the case of perturbative excited states, where the construction simplifies drastically.

\subsubsection{General metrics}

We begin by proposing a family of level set surfaces with the following properties:
  \begin{enumerate}
  \item They must contain the minimal surface $\gamma_A$ as one of its members.
  \item They must be continuous and not self-intersecting.
  \item They must not include closed bulk surfaces.
\end{enumerate}
Given a family with these properties, it is then straightforward to construct a divergenceless vector field with the desired boundary condition. We can understand this as follows: given a family of level set hypersurfaces, one can first generate the corresponding integral lines by imposing that they must be orthogonal to each member of the family. Having the integral lines, then, the problem reduces to that of section (\ref{F1}) so we could follow the steps outlined there. This means that, in general, we can only check if norm bound is satisfied \emph{a posteriori}. There is however, one clever exception to the rule. We can ensure that $|v|\leq 1$ is satisfied everywhere by construction if we impose the following extra condition on the level set surfaces:
\begin{enumerate}
\setcounter{enumi}{3}
\item They must be homologous to $A$.
\end{enumerate}
If this condition is satisfied, then, the max flow-min cut theorem guarantees that $|v|$ will be maximal at the bottle neck $\gamma_A$. Since $|v|_{\gamma_A}=1$ then, this implies that $|v|\leq 1$ at any other member of the family. Notice that condition 4 is not a strict requirement, but a useful one. In fact, simple examples of vector fields generated by level sets that are \emph{not} homologous to $A$ are the \emph{maximally packed flows} constructed in \cite{Agon:2018lwq}. In that construction
the level set surfaces were picked as a family of nested minimal surfaces, containing $\gamma_A$ as one of its members. The motivation there was to find a flow with maximal norm $|v|=1$ in a given codimension-one region of the bulk, which was possible due to the nesting property of bit threads \cite{Freedman:2016zud,Headrick:2017ucz}.\footnote{Maximally packed flows also satisfy the norm bound by construction. If one picks level sets that are not homologous to $A$ and are not minimal surfaces, then indeed, the norm bound should be checked \emph{a posteriori}.} For the purposes of this paper, however, we are not interested in the above requierement, so we can explore other possibilities. In the remaining part of this section we will in fact assume that the condition 4 is satisfied, so we do not have to deal with the norm bound.

Let us now describe in detail the construction from level sets. To begin with, we need an efficient way to specify our level set hypersurfaces. In practice, we can do so by picking an appropriate scalar function $\varphi(x^i)$ such that the $\varphi=$ constant surfaces give us our desired level sets. We can then write the following equation for $v_\lambda$:
\be\label{v:levelsets}
v=\Upsilon(x^i)\nabla \varphi(x^i)\,.
\ee
At first glance, \eqref{v:levelsets} seems more general than a gradient flow, but in fact it is not. In principle one could always redefine the scalar function $\varphi \to \tilde{\varphi}=\int^{\varphi}\Upsilon(\psi)d \psi $ and therefore simply write $v=\nabla \tilde{\varphi}$. However, the function $\tilde{\varphi}$ would not only encode information about the level sets, but also about the norm, so it would be extremely difficult to guess a good function that gives us our desired level sets \emph{and} that also satisfies the divergenceless condition $\nabla^2\tilde{\varphi}=0$. In practice, then, it is much easier to start with (\ref{v:levelsets}) and determine $\Upsilon(x^i)$ through the divergenceless condition. We emphasize that, in this scenario, the specific values of $\varphi$ do not have a particular meaning and are in particular \emph{not} related to the norm of $v$. The field $\varphi$ here only determines the unit vector in $\vec{\tau}=v/|v|$, through
\be
\vec{\tau}=\frac{\nabla{\varphi}}{|\nabla{\varphi}|}\,.
\ee
One crucial observation that follows from the definition (\ref{v:levelsets}) is that the covector $v_a$ (i.e. $v$ with lower index) only depends on the metric $g_{ab}$ through $\Upsilon(x^i)$. To make this point self-evident, and in a form which is partially ``independent'' of the metric, we can write
\be\label{cov:levelsets}
v_a=\Upsilon(\varphi,g)\partial_a \varphi\,.
\ee
We will exploit this observation below, for the case of perturbative states. For now, let us notice that the boundary condition at the minimal surface implies:
\bea
\Upsilon^2(\varphi, g) g^{a b} \partial_a \varphi \partial_b \varphi \Big|_{\gamma_A}=1\,,
\eea
or, equivalently,
\bea\label{eq:bcvsets}
\Upsilon(\varphi, g)\Big|_{\gamma_A}=\frac{1}{| \partial \varphi|_g}\bigg|_{\gamma_A}, \qquad\qquad| \partial \varphi|_g\equiv \sqrt{g^{ab}  \partial_a \varphi \partial_b \varphi}\,.
\eea
All we have left is to determine $\Upsilon$ away from the minimal surface, which can be done by imposing the divergenceless condition. Here we have two options: $i)$ we can use Gauss's law as we did in Section \ref{F1} or $ii)$ we can directly attempt to solve $\nabla\cdot v=0$, which should give us a first order differential equation for $\Upsilon$. As mentioned in Section \ref{F1}, the two methods are completely equivalent, since Gauss's law is the integral form of the divergenceless condition. However, since we do not have explicit expressions for the integral lines, then, the first option turns out to be more complicated in this case.\footnote{We could get the integral lines $X(x_m,\xi)$ in terms of the field $\varphi$ and its derivatives, but in order to do so we would need to solve a first order differential equation, which would by itself have the same level of complexity as solving directly the divergenceless condition.} We therefore proceed by deriving a differential equation for $\Upsilon$, which can be derived from $\nabla\cdot v=0$. Plugging (\ref{cov:levelsets}) into this condition and massaging the equation leads to:
\be\label{eq:PsiPDE}
(\nabla\varphi)\cdot(\nabla\Upsilon)+(\nabla^2\varphi) \Upsilon=0\,.
\ee
As advertised, this is a first order differential equation for $\Upsilon$ in terms of the scalar field $\varphi$ and the background metric $g$. Solving this equation subject to the boundary condition (\ref{eq:bcvsets}) would then give a unique solution for the vector field $v$.

\subsubsection{Perturbative excited states}

The above construction simplifies drastically for the case of perturbative excited states. In the following we will specialize to this situation and study in detail how the information about the bulk perturbation is encoded in the resulting thread configuration.

For a metric of the form $g^\lambda_{ab}=g_{ab}+\lambda \delta g_{ab}$ we are only interested in obtaining the vector field $v_\lambda$ (\ref{eq:vlambda}) to linear order in the perturbation around the zeroth order solution. Since the minimal surface $\gamma_A$ does not change at linear order in $\lambda$, a simple choice for the level sets consistent with all requirements would be to pick the same surfaces as for the unperturbed geometry. In this case we have that:
\be
v^\lambda_a=v_a+\lambda\delta v_a=\Upsilon_\lambda(\varphi,g_\lambda)\partial_a \varphi\,,\qquad \Upsilon_\lambda(\varphi,g_\lambda)=\Upsilon(\varphi,g)+\lambda \delta\Upsilon(\varphi,g_\lambda)\,.
\ee
In other words, with this choice of level sets, only the function $\Upsilon(x^i)$ gets corrected at linear order in $\lambda$, so the first correction of the vector field $\delta v_a$ turns out to be proportional to the zeroth order solution,
\be\label{eq:deltavaPsi}
\delta v_a= \delta\Upsilon(\varphi,g_\lambda)\partial_a \varphi= \Psi(\varphi,g_\lambda)v_a\,,\qquad \Psi(\varphi,g_\lambda)\equiv\frac{\delta\Upsilon(\varphi,g_\lambda)}{\Upsilon(\varphi,g)}\,.
\ee
The function $\Psi$ is determined at the minimal surface by the boundary condition $|v_\lambda|=1$. Expanding at linear order, we obtain
\bea\label{norm}
g^{ab}_\lambda v^\lambda_{a}v^\lambda_{b}\Big|_{\gamma_A}=g^{ab} v_{a}v_{b}+\lambda\left(2g^{a b}v_{a} \delta v_{b}+\delta g^{ab} v_{a }v_{b}\right)\Big|_{\gamma_A}=1\,.
\eea
Since the zeroth order term is already normalized to one, the terms inside the parenthesis must vanish.  Using (\ref{eq:deltavaPsi}) we arrive to:
\bea\label{bc-a}
\Psi(\varphi,g_\lambda)|_{\gamma_A}=-\frac{1}{2}\delta g^{ab} v_a v_b=\frac{1}{2}\delta g_{ab} v^a v^b\,.
\eea
In the last equality we have used the fact that $\delta (\delta^a_{\,\,\, b})=\delta\(g^{ab}g_{bc}\)=\delta g^{ab}g_{bc}+g^{ab}\delta g_{bc} =0$. We did this, because it would be particularly convenient to have an expression for the boundary condition of $\Psi(\varphi,g_\lambda)$ in terms of the background $v$ with upper index.

Next we would like to determine the function $\Psi$ away from the minimal surface, which can be done by imposing the divergenceless condition. Again, we proceed by deriving a differential equation for $\Psi$ akin to (\ref{eq:PsiPDE}). In order to do so, first notice that
\bea
v^a_\lambda =v^a +\lambda \delta v^a= (g^{a b} +\lambda \delta g^{ab} )(v_b+\lambda \delta v_b)\,,
\eea
so\footnote{Notice that $\delta v_a$ is proportional to $v_a$ but $\delta v^a$ is \emph{not} proportional to $v^a$. This is why we have mostly worked with covectors in this section.}
\be\label{deltavaup}
\delta v^a = g^{ab}\delta v_b+\delta g^{ab}v_b=\Psi g^{ab}v_b+\delta g^{ab}v_b=\Psi v^a-g^{ ab}\delta g_{bc} v^c\,.
\ee
Taking the divergence of $v_\lambda$ and using the fact that $\nabla\cdot v=0$ (at zeroth order), we obtain:
\be\label{eq:DivCOnPer}
\nabla_\lambda\cdot v_\lambda=\frac{1}{\sqrt{g_\lambda}}\partial_a\left(\sqrt{g_\lambda}\, v_\lambda^a\right)=\frac{\sqrt{g}}{\sqrt{g_\lambda}}(\cancel{\nabla\cdot v})+\frac{\lambda}{\sqrt{g_\lambda}}\partial_a\left[\delta\({\sqrt{g_\lambda}}\)v^a +\sqrt{g}\,\delta v^a\right]=0\,.
\ee
Taking the explicit variation of $\sqrt{g_\lambda}$ and using (\ref{deltavaup}) we obtain:
\bea
\partial_a\( \tfrac 12 \sqrt{g} \, g^{bc} \delta g_{bc} v^a + \Psi  \sqrt{g}\,v^a- \sqrt{g}\,g^{a b}\delta g_{b c} v^c\)=0\,,
\eea
or, equivalently,
\bea\label{eq:DEPsi}
v \cdot \nabla \Psi+\nabla_a (\delta g^{a b} v_b)+\tfrac{1}{2} v\cdot \nabla ( \delta g ) =0\,,
\eea
where $\delta g\equiv g^{ab} \delta g_{ab}$. In summary, given a background metric $g_{ab}$ and a solution to the max flow problem $v^a$, one can always solve the problem of maximizing the flux in a state where the metric $g_{ab}^\lambda$ is perturbatively closed to the original one. Assuming that the level set surfaces remain the same in the perturbed geometry, the solution for the perturbation of $v$ is given by equation (\ref{deltavaup}), which is determined in terms of a scalar function $\Psi$ and the metric perturbation $\delta g_{ab}$. This function can be obtained by solving the first order differential equation (\ref{eq:DEPsi}) subject to the boundary condition (\ref{bc-a}).

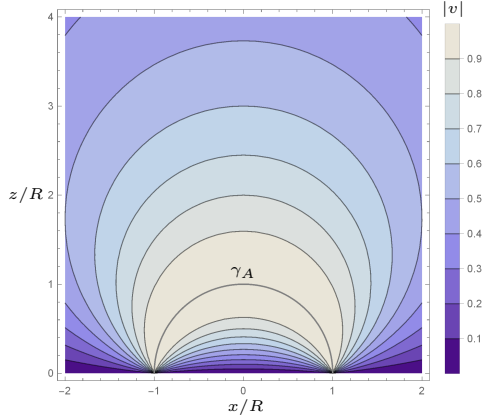
\begin{figure}[t!]
  \centering
  \includegraphics[scale=0.4]{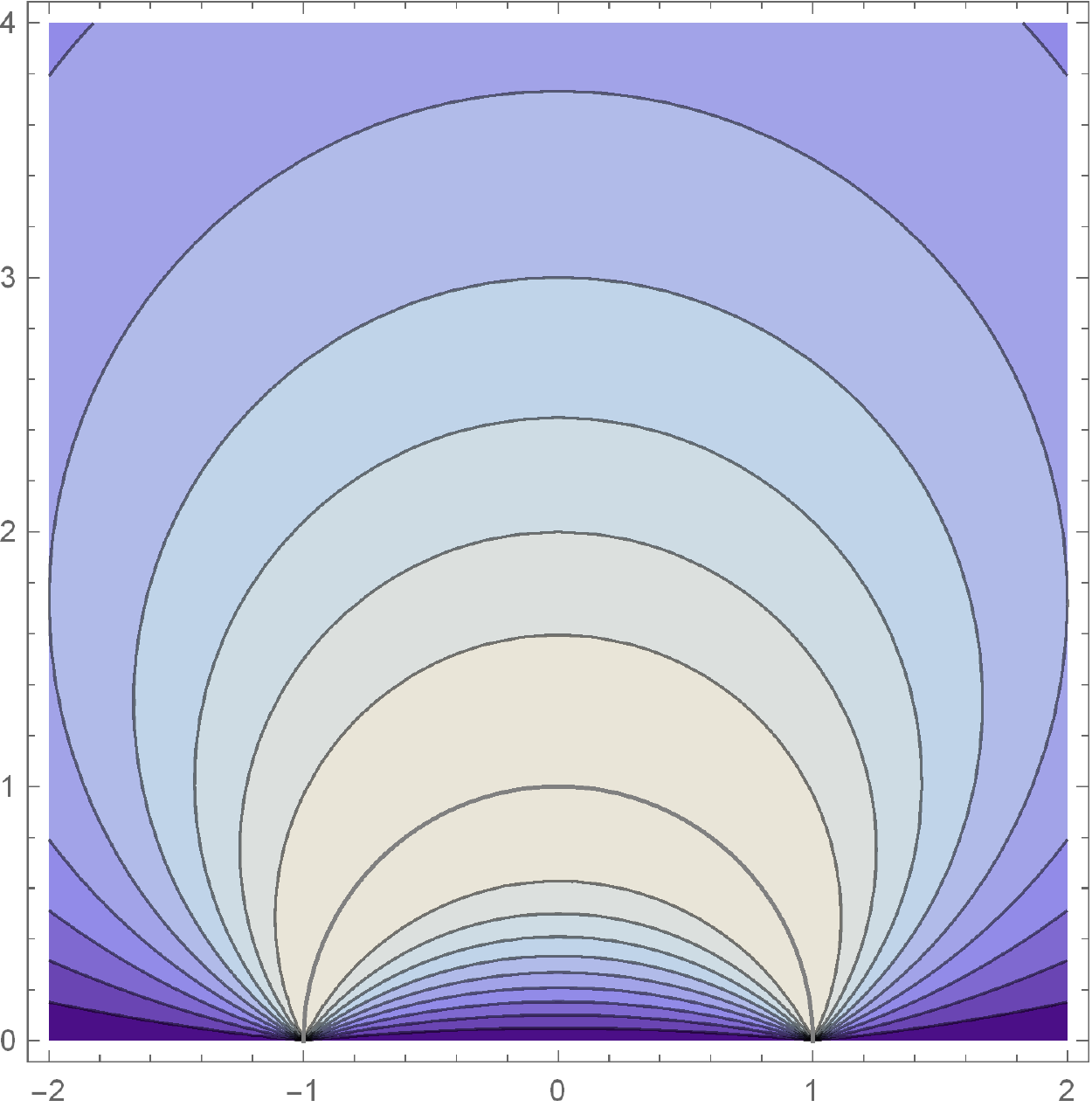}$\,\,$\includegraphics[scale=0.4]{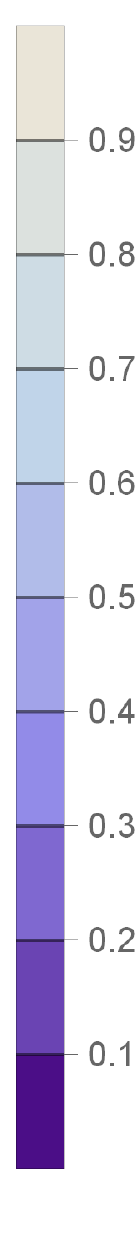}
  \begin{picture}(0,0)
\put(-100,-5){{\tiny $x/R$}}
\put(-183,75.5){{\tiny $z/R$}}
\put(-99,47){{\tiny $\gamma_A$}}
\put(-19,146){{\tiny $|v|$}}
\end{picture}
  \vspace{0mm}
  \caption{Contour plot for the magnitude $|v|$ of the geodesic flow given in (\ref{Vec1-d}), in $d=2$ dimensions (i.e., empty AdS$_3$). The contours correspond to the level set surfaces of $v$, which are all homologous to $A$ and, in particular, include $\gamma_A$ as one of its members. This implies that this vector field $v$ can indeed be used as a seed to generate a good solution $v_\lambda$ in a perturbative excited state.}\label{fig:contours}
\end{figure}

In retrospective, the only non-trivial input required for this kind of construction is the choice of background vector field $v$, which is in turn used as a seed for the perturbed solution $v_\lambda$. Specializing to spherical regions, one simple choice would be to pick $v$ as a geodesic flow, which is known in closed form if the background metric is empty AdS. This background $v$ is given explicitly in equation (\ref{Vec1-d}). It is easy to check that the level sets of this vector field are all homologous to $A$, as is shown in Figure \ref{fig:contours}. Since this construction assumes that the level sets are kept fixed, this implies that any perturbative solution $v_\lambda$ build up from this background field $v$ will automatically respect the norm bound $|v_\lambda|\leq1$. In Appendix \ref{app:examples} we present an explicit example of such perturbative solutions, for the case of a local quench.

Finally, we can comment on how the metric perturbation and the Einstein's equations are encoded in this particular construction. Although the explicit use of metric is reduced in comparison to the construction via integral curves, the last step in the level sets method introduces the same level of nonlocality. In particular, the way we fixed the scalar field $\Psi$ was by solving the divergenceless condition (\ref{eq:DivCOnPer}). Even though this equation is local, the nontrivial boundary condition (\ref{bc-a}) introduces nonlocalities in the solutions, because the equation effectively transports information from $\gamma_A$ to other regions in the bulk. From the Gauss's law perspective the situation is perhaps easier to understand. In that case, the final answer for $v_\lambda$ exhibits an explicit bilocal dependence with respect to the metric perturbation, through its magnitude (\ref{magnitudeV}). The way we solve for $\Psi$ in this formalism is completely equivalent to that case, because Gauss's law is nothing but the integral form of the divergenceless condition. Hence, even though this construction seems particularly efficient for building up perturbative solutions $v_\lambda$, it ultimately contains the same kind of nonlocalities than the construction via integral curves. Hence, the inversion problem to recover the bulk metric and the Einstein's equations is equally difficult in both constructions.

\section{Bit threads and bulk locality\label{SecForms}}

The simple perturbative realizations of bit threads of the previous section highlight the need of a bit thread construction that does not make explicit use of the metric.
Fortunately, we know how to reformulate this formalism in a framework that makes background independence explicit: using the language of differential forms. The equivalence between divergenceless vector fields $v$ and closed $(d-1)-$forms $\bm w$ was already emphasized in \cite{Freedman:2016zud} and was used in \cite{Headrick:2017ucz} to efficiently deal with some subtleties of the max flow problem for null intervals.\footnote{We also point out that a reformulation of the Ryu-Takayanagi prescription in terms of calibrations (closed forms) was worked out independently in \cite{Bakhmatov:2017ihw}.} In this section we will first break down this equivalence in detail, giving explicit formulas that translate various relevant expressions between the two languages. We then argue that the Iyer-Wald formalism provides us with a particular realization of the perturbed thread configuration $\delta\bm w$ that makes explicit use of bulk locality. In particular, we show that the linearized Einstein's equations are explicitly encoded in this construction through the closedness condition, i.e., $d \delta\bm w=0$. We exploit this unique property of the Iyer-Wald construction to tackle the question of metric reconstruction and show that this problem can be phrased in terms of the inversion of a particular differential operator. Finally, we carry out the explicit inversion at linear order and discuss how to generalize our results to higher orders in the perturbation.

\subsection{Bit threads in the language of differential forms \label{ssec:diff}}

In the presence of a metric $g_{ab}$, the explicit map between \emph{flows}, i.e., divergenceless vector fields $v$ and closed $(d-1)-$forms $\bm w$, is given by
\bea\label{hodgestar}
v^a =g^{a b}(\star \bm w)_{b}\,,
\eea
where $\star \bm w$ represents the Hodge star dual  of $\bm w$, defined via
\bea
(\star \bm w)_{b}\equiv\frac{1}{(d-1)!} \sqrt{g}\,  w^{a_1\ldots a_{d-1}}\varepsilon_{a_1\ldots a_{d-1} b}\,.
\eea
In the above formula $\varepsilon_{a_1\ldots a_{d}}$ represents the totally antisymmetric Levi-Civita symbol, with sign convention $ \varepsilon_{i_1 \ldots i_{d-1}z}=1$. Furthermore, the indices of $\star\bm w$ are raised with the Riemannian metric $g_{ab}$, and its determinant is denoted by $g$. At this point we can already notice an important difference between the two objects, namely that, while the notion of a flow requires a background metric, $\bm w$ can be defined independently of $g_{ab}$. This will play a crucial role below, specifically, when we address the problem of metric reconstruction.

Let us carry on with our analysis. The inverse of the map (\ref{hodgestar}) can be stated in terms of the natural volume form $\bm \epsilon$, given by
\bea
\bm \epsilon=\frac{1}{d!}\,  \epsilon_{a_1 \ldots a_{d}} dx^{a_1}\wedge \cdots \wedge dx^{a_{d}}\,,
\eea
where $ \epsilon_{a_1 \ldots a_{d}}$ is  proportional to $\varepsilon_{a_1\ldots a_d}$ and normalized such that $ \epsilon_{i_1 \ldots i_{d-1}z}=\sqrt{g}$. In terms of $\bm \epsilon$, the $(d-1)-$form $\bm w$ is given by
\bea\label{flow-forms-1}
{\bm w}= \frac{1}{(d-1)!}\epsilon_{a_1 \ldots a_{d-1}b}  \, v^b \,dx^{a_1}\wedge \cdots \wedge dx^{a_{d-1}}\,,
\eea
or in components,
\bea \label{flow-forms-2}
{w}_{a_1 \ldots a_{d-1}}=  \epsilon_{a_1 \ldots a_{d-1}b} v^b\,.
\eea
Following standard manipulations one can relate the divergence of $v^a$ with the exterior derivative of $\bm w$.\footnote{See e.g. Appendix B.2 of \cite{Wald:1984rg} for an explicit derivation of various identities that we use in this section.} Explicitly, taking the exterior derivative of equation (\ref{flow-forms-1}) leads to
\bea \label{flows-forms}
d {\bm w}=\(\nabla_a v^a  \) {\bm \epsilon}\,.
\eea
This formula shows explicitly the anticipated fact that divergenceless vector fields, or ``\emph{flows}'', are mapped to closed $(d-1)-$forms. The precise relation between the two is given by (\ref{flow-forms-1}).

Now, it is well known that $k-$forms have well defined integrals over $k-$dimensional hypersurfaces. Therefore it is convenient to write down an explicit formula for the restriction of $\bm w$ on a codimension-one surface $\Gamma$ in terms of intrinsic geometric quantities of that surface. Such a formula can be derived using the fact that the
volume $d-$form $\bm \epsilon$ induces a volume $(d-1)-$form $\tilde{\bm \epsilon}$ on $\Gamma$ via
\bea\label{inducedeps}
\epsilon_{a_1\ldots a_{d-1}b}=d\,  \tilde{\epsilon}_{[a_1\ldots a_{d-1}}n_{b]}\,,
\eea
where $n$ is the unit normal to the surface. Contracting the last index of (\ref{inducedeps}) with $v$ and using (\ref{flow-forms-2}) leads to an explicit expression for the form $\bm w$ evaluated at an arbitrary codimension-one surface $\Gamma$, with local unit normal $n$, in terms of the $(d-1)-$form $\tilde{\bm \epsilon}$
\bea \label{boundaryw}
\bm w|_{\,\!_\Gamma}=(n_a v^a) \tilde{\bm \epsilon}\,.
\eea

Next, consider Gauss's theorem applied to the divergenceless vector field $v^a$, in a bulk region $N$ with $\partial N= A\cup \(-m\)$, where $m$ is a surface homologous to $A$ ($m\sim A$):
\bea
\int_{N}\nabla_a v^a \bm{\epsilon}=\int_{\partial N} \(n_a v^a\) \tilde{\bm \epsilon}=\int_{A} \(n_a v^a\) \tilde{\bm \epsilon}-\int_{m} \(n_a v^a\) \tilde{\bm \epsilon}=0\,.
\eea
This leads to the homology condition
\bea\label{homo-cond}
\int_{A}\(n_a v^a\) \tilde{\bm \epsilon}=\int_{m\sim A}\!\!\!\!\!\!\(n_a v^a\) \tilde{\bm \epsilon}\,.
\eea
This result is equivalently derived in the language of forms, using Stoke's theorem:
\bea
\int_{N} d {\bm w}=\int_{\partial N} {\bm w}=\int_{ A} {\bm w}-\int_{m} {\bm w}=0\,.
\eea
This leads to
\bea
\int_{A} {\bm w}=\int_{m\sim A}\!\!\!\!\!\! {\bm w}\,,
\eea
which is equivalent to (\ref{homo-cond}), given (\ref{boundaryw}).

With the ingredients described above, we are now in a position to translate the max flow-min cut theorem to the language of differential forms. First, we have
\bea\label{eq:boundw}
\int_{m} \bm w = \int_m \(n_a v^a\) \tilde{\bm \epsilon}\leq  \int_m \tilde{\bm \epsilon}\,.
\eea
The inequality here comes from the standard norm bound, $|v|\leq1$, which in terms of forms can be rewritten as
\bea\label{w-norm1}
\frac{1}{(d-1)!}g^{a_1 b_1}\cdots g^{a_{d-1} b_{d-1}} w_{a_1 \ldots a_{d-1}}w_{b_1 \ldots b_{d-1}} \leq 1\,.
\eea
In short, equation (\ref{eq:boundw}) implies that, locally, the form $\bm w$ evaluated on any codimension-one hypersurface is bounded by the natural volume form defined on it $ \tilde{\bm \epsilon}$. The max flow-min cut theorem then implies that
\bea\label{eq:MFMC}
\underset{\bm w \in \bm W}{\rm max} \int_{A} \bm w =\underset{m \sim A}{\rm min } \int_m \tilde{\bm \epsilon} \,,
\eea
where $\bm W$ is the set of closed forms obeying the bound (\ref{w-norm1}). This means that at the bottle-neck $\gamma_A$, an optimal bit thread form $\bm w^*$ should be equal to the volume form $\tilde{\bm \epsilon}$, i.e.,
\bea\label{formmA}
{\bm w^*}|_{\gamma_A}=\tilde{\bm \epsilon}|_{\gamma_A}\,.
\eea
Finally, combining with the RT formula for entanglement entropy, (\ref{eq:MFMC}) becomes
\bea
S_A=\frac{1}{4G_N}\,\underset{\bm w \in \bm W}{\rm max} \,\, \int_{A} \bm w\,.
\eea
which is the differential form version of the max-flow formula (\ref{BitThreadReform}).

There are many situations in which one might want to define the threads in terms of forms $\bm w$ instead of vector fields $v$. In particular, this reformulation will prove extremely useful for the problem at hand, namely, for the study of perturbations around a given background and the corresponding solutions to the flow maximization problem.

\subsubsection{The case of linear perturbations}

Having understood how the bit threads formalism translate to in the language of differential forms, it is now time to go back to our original problem.
We will assume that the following data is given: a background metric $g_{ab}$ on a manifold $M$ with boundary $\partial M$, and an optimal flow $v$ that maximizes the flux through a boundary region $A$. Using (\ref{flow-forms-2}), then, this would also imply the knowledge of an optimal closed form $\bm w$. In the following,  we will consider the max flux problem in geometries that are perturbatively close to $g_{ab}$, i.e., $g^{\lambda}_{ab}=g_{ab}+\lambda\delta g_{ab}$. We will denote a solution to the problem as $\bm w_\lambda$, where $\bm w_\lambda=\bm w+\lambda\delta \bm w$.

First, notice that the closedness condition implies
\bea\label{ddelta}
d\( \bm w+\lambda\delta \bm w \)=0 \qquad \to \qquad d\( \delta \bm w \)=0\,.
\eea
We can also use the fact that the the minimal surface $\gamma_A$ does not change at first order in the perturbation, so $\gamma_A^\lambda=\gamma_A$. Since this is
a bottle-neck for the flow, both $v$ and $\bm w$ are fixed at its location. In particular, from (\ref{formmA}) it follows that
\bea\label{deltaboundary}
\(\bm w+\lambda \delta \bm w\)|_{\gamma_A}=\(\tilde{\bm \epsilon }+\lambda \delta \tilde{\bm \epsilon }\)\qquad \to\qquad  \delta \bm w |_{\gamma_A}=\delta \tilde{\bm \epsilon }\,.
\eea
Then, given a max flow $\bm w$  for the unperturbed geometry, we are set to find a closed $(d-1)-$form $\delta \bm w$ that satisfies the boundary condition
(\ref{deltaboundary}) and it is such that the norm bound constraint (\ref{w-norm1}) holds everywhere in the bulk for the sum $\bm w_\lambda=\bm w +\lambda\delta \bm w$.
For simplicity, let us introduce the following notation for the inner product
\bea
\langle \bm w , \tilde{\bm w}\rangle_g=\frac{1}{(d-1)!}g^{a_1 b_1}\cdots g^{a_{d-1} b_{d-1}} w_{a_1 \ldots a_{d-1}}\tilde{w}_{b_1 \ldots b_{d-1}}\,,
\eea
and for its first order variation with respect to the metric
\bea
\langle \bm w , \tilde{\bm w}\rangle_{\delta g}=\frac{1}{(d-1)!}\delta(g^{a_1 b_1}\ldots g^{a_{d-1} b_{d-1}}) w_{a_1 \ldots a_{d-1}}\tilde{w}_{b_1 \ldots b_{d-1}}\,.
\eea
With these notations, the norm bound (\ref{w-norm1}) at first order in $\lambda$ is given by
\bea\label{P-norm-bound}
\langle \bm w , \bm w\rangle_g+\lambda\[2\langle \bm w , \delta \bm w\rangle_g +\langle \bm w , \bm w\rangle_{\delta g}\] \leq 1\,,
\eea
which looks more difficult to implement than its vector field counterpart. From (\ref{P-norm-bound}) it is clear that the norm bound will typically depend on $\bm w$ so a priori it seems unlikely that a generic $\delta \bm w$ obeying  (\ref{ddelta}) and (\ref{deltaboundary}) could satisfy (\ref{P-norm-bound}) independent of $\bm w$. The task becomes even more untractable if one requires $\delta \bm w$ to be given in terms of a linear local functional of $\delta g_{ab}$ and its covariant derivatives $\nabla_{(a_1}\cdots \nabla_{a_n)}\delta g_{ab}$ (see however \cite{Wald:2005nz}).
In the remaining part of this section we will show that, despite the above remarks, the Iyer-Wald formalism provides a concrete realization of such perturbed form.

\subsection{Iyer-Wald formalism and Einstein's equations\label{sec:IWgeneral}}

One of the crucial breakthroughs in the joint program of holography and quantum information is that the first law of the entanglement entropy, together with the Ryu-Takayanagi formula, imply the linearized Einstein's equations in the bulk. This was originally proven using  Hamiltonian perturbation theory \cite{Lashkari:2013koa}.
In a beautiful paper \cite{Faulkner:2013ica}, it was further shown that it is possible to make this connection more explicit by the proper implementation of the Noether's charge formalism in the bulk, also known as the Iyer-Wald formalism. In this new language, the problem of linearized perturbations is cast in terms of differential forms, a more natural and elegant approach that bridge the CFT and bulk quantities in an efficient way. In this section we will briefly review the basic ingredients of \cite{Faulkner:2013ica}, making the connection between entanglement entropy and Einstein's equations manifest. Later in Section \ref{5.2} we will show that the Iyer-Wald formalism provides
us with a canonical choice for the differential form $\delta\bm w$ that solves the max flux problem in a perturbed geometry. As a byproduct, we will show that such a canonical form will automatically encode (locally) the linearized Einstein's equations in the bulk which, in turn, will prove useful for the problem of metric reconstruction.

Let us first state the problem that \cite{Lashkari:2013koa} sought to solve and then discuss the approach of \cite{Faulkner:2013ica}. In general quantum field theories (holographic or not),
for small perturbations over a reference state, $\rho=\rho^{(0)}+ \lambda \delta\rho$, entanglement entropy satisfies the first law
\be\label{eq:1stlaw}
\delta S_A=\delta\langle H_A\rangle\,,
\ee
where $\langle \bullet \rangle$ represents the expectation value of the operator in the respective quantum state and $H_A$ is the so-called modular Hamiltonian. By definition, this operator is related to the reduced density matrix $\rho_A=\text{tr}_{A^c}[\rho]$ through
\be\label{defmodularH}
\rho_A=\frac{e^{-H_A}}{\text{tr}[e^{-H_A}]}\,.
\ee
However, there are very few cases for which (\ref{defmodularH}) can be explicitly inverted to obtain $H_A$ in closed form. The most famous example is the case where $A$ is half-space, say $x_1>0$, and $\rho$ corresponds to the vacuum state of the QFT. In this case \cite{Bisognano:1975ih,Unruh:1976db}
\begin{equation}\label{modular1}
H_A=2\pi\int_A x_1 \, T_{00}(t,\vec{x}) \, d^{d-1}x\,.
\end{equation}
For generic CFTs, this setup can be conformally mapped to the case where
$A$ is a ball of radius $R$, centered at an arbitrary point $\vec{x}=\vec{x}_c$, in which case \cite{Hislop:1981uh,Casini:2011kv}
\begin{equation}\label{modular2}
  H_A=2\pi\int_A  \frac{R^2-(\vec{x}-\vec{x}_c)^2}{2R} T_{00}(t,\vec{x})\,d^{d-1}x\,.
\end{equation}
On the other hand, the left-hand side of (\ref{eq:1stlaw}) is computed via the Ryu-Takayanagi formula in the bulk. For ball shaped regions in pure AdS, or small perturbations around it, the RT surface $\gamma_A$ is given by a half hemisphere of radius $R$ extended on the extra dimension, centered at $\vec{x}=\vec{x}_c$ and $z=0$. The Ryu-Takayanagi formula then adopts the form
\bea\label{RT-Ball}
\delta S_A=\frac{1}{4G_N}\int_{\gamma_A} \delta \tilde{\bm \epsilon}
\eea
where $\delta\tilde{\bm \epsilon}$ is the variation of the natural volume form on the surface $\gamma_A$. A further ingredient is the relation between the
expectation value of the boundary stress tensor, appearing in the right-hand side of (\ref{eq:1stlaw}), and the fluctuations of the bulk metric $\delta g_{\mu\nu}$. In the Fefferman-Graham gauge, where the latter is given by (\ref{FG-PT}), the former can be identified as the first subleading (normalizable) mode in a near boundary expansion (\ref{T=H}). Taking into account that the boundary field theory is conformal and that the stress tensor conserved, then, this identification imposes non-trivial boundary condition for the metric fluctuations $H_{\mu\nu}$,
\bea
\begin{cases}
  \displaystyle \langle T^\mu_{\,\,\, \mu}(x)\rangle =0  & \displaystyle \qquad \to \qquad H^\mu_{\,\,\, \mu}(x,z=0)=0\,,\\[3ex]
  \displaystyle \partial_\mu \langle T^{\mu \nu} (x) \rangle =0 & \displaystyle \qquad \to \qquad \partial_\mu H^{\mu \nu}(x,z=0)=0\,.
 \end{cases}
\eea
Using the above,  the right-hand side of (\ref{eq:1stlaw}) becomes
\bea\label{H-mod}
\delta \langle H_A \rangle &=&2\pi \int_{A} d^{d-1}x \frac{R^2-|\vec{x}-\vec{x_0}|^2}{2R}\delta \langle T_{tt}(t_0,\vec{x})\rangle \,, \nonumber  \\
& =&\frac{d}{16 G_N R} \int_{A} d^{d-1}x \(R^2-|\vec{x}-\vec{x_0}|^2\) H^i_{\,\,\, i}(t_0,\vec{x},z=0)\,.
\eea
Similarly, evaluating the left-hand side of (\ref{eq:1stlaw}) using (\ref{RT-Ball}) yields
\bea\label{Ent-H}
\delta S_A&=&\frac{1}{4G_N}\int_{\gamma_A} \delta \tilde{\bm \epsilon} \nonumber \\
          &=&\frac{1}{8 G_N R} \int_{\gamma_A} d^{d-1}x\( R^2 \delta^{ij}  - (x^i-x^i_0)(x^j-x^j_0)\)H_{i j} (t_0, \vec{x}, z)\,.
\eea
The first law (\ref{eq:1stlaw}), together with (\ref{H-mod}) and (\ref{Ent-H}), then establishes a relation between integral functionals of $H_{\mu \nu}$ on $A$ and on $\gamma_A$. It turns out that this functional dependence in turn implies the Einstein's equations for $H_{\mu\nu}$, linearized around pure AdS. This was shown originally in \cite{Lashkari:2013koa} by a direct comparison between the two sides of (\ref{eq:1stlaw}).

From the gravitational perspective, (\ref{eq:1stlaw}) was then proven to be equivalent to the generalized first law of black hole thermodynamics applied to the bifurcate killing horizon of Rindler AdS \cite{Faulkner:2013ica}. This was made explicit by a clever implementation of the Noether's theorem in the bulk, using a formalism developed a couple of decades back by Iyer and Wald in \cite{Iyer:1994ys}. In order to apply this formalism to the problem at hand, the key observation was that the RT surface for ball-shaped regions in pure AdS, or perturbations around it, coincides with the bifurcate horizon of the time-like killing vector
\bea
\xi=-\frac{2\pi}{R}\(t-t_0\)[z\partial_z+(x^i-x^i_0)\partial_i]+\frac{\pi}{R}[R^2-z^2-(t-t_0)^2-(\vec{x}-\vec{x}_0)^2]\partial_t\,,
\eea
with respect to which a notion of energy and entropy are possible. In fact, a specific conformal transformation (known as the CHM map \cite{Casini:2011kv}) maps the interior of the Rindler wedge associated with $A$ to the exterior of an hyperbolic black hole in AdS, where the killing vector $\xi$ coincides with the generator of time translations.
Following Iyer and Wald \cite{Iyer:1994ys,Iyer:1995kg,Wald:2005nz} one then investigates the Noether's theorem for the Killing symmetry generated by $\xi$. This leads to the definition of a $(d-1)-$form
\bea\label{chi}
{ \bm \chi}=-\frac{1}{16 \pi G_N} \left[ \delta (\nabla^A \xi^B {\bm \epsilon}_{AB} )+\xi^B {\bm \epsilon}_{AB}(\nabla_c h^{AC}+\nabla^A h^C_{\,\, C})\right]\,,
\eea
where $h_{AB}=z^{d-2}H_{A B}$ and ${\bm \epsilon}_{AB}$ is the volume $(d-1)-$form
\bea
{\bm \epsilon}_{AB}=\frac{1}{(d-1)!}\epsilon_{A B C_3 \cdots C_{d+1}}dx^{C_3} \wedge\cdots \wedge dx^{C_{d+1}}\,,
\eea
with $\epsilon_{z t i_1 \cdots i_{d-1}}=\sqrt{-G}$. As noted in \cite{Faulkner:2013ica}, the form ${\bm \chi}$ satisfies the following properties
\bea\label{chi-eqs}
\int_{\gamma_A} {\bm \chi} =\delta S_A\,, \qquad \int_{A} {\bm \chi}=\delta \langle H_A\rangle\,,
\eea
which can be more easily verified by evaluating (\ref{chi}) on a Cauchy hypersurface $\Sigma$, containing both $\gamma_A$ and $A$. For instance, taking $\Sigma$ to be the $t=t_0$ slice, one obtains the $(d-1)-$form
\bea\label{chiSigma}
{ \bm \chi}|_{\,\!_\Sigma}\equiv\tilde{\bm {\chi}} &=&\frac{z^d}{16 \pi G_N} \Bigg\{{\bm \epsilon}^t_{\,\,z}\left[ \(\frac{2\pi z}{R}+\frac{d}{z} \xi^t+\xi^t\partial^z \) H^i_{\,\,i} \right] +\nonumber \\ \label{chi-Sigma}
 && + {\bm \epsilon}^t_{\,\, i} \left[ \(\frac{2\pi (x^i-x^i_0)}{R}+\xi^t\partial^i \) H^j_{\,\,j} -\(\frac{2\pi (x^j-x^j_0)}{R}+\xi^t\partial^j\) H^i_{\,\,j} \right]  \Bigg\}\,,
\eea
from which both equations in (\ref{chi-eqs}) follow trivially. The key point here is that ${ \bm \chi}$ is closed provided that the bulk equations of motion are satisfied. For instance, in the constant-$t$ Cauchy slice used above one finds
\bea\label{closeness}
d \tilde{\bm {\chi}}=-2\xi^t \delta E^g_{tt} \, {\bm \epsilon}^t\,,
\eea
where $\delta E^g_{tt}$ is the $tt$ component of the linearized Einstein's equations, and  ${\bm \epsilon}^t$ is the induced volume form on $\Sigma$. Similarly, other components of the Einstein's equations are obtained by specializing to different Cauchy slices. Thus, provided that the metric perturbations satisfy these equations, the form $\bm \chi$ is closed and the Stokes theorem implies the equality between the left and right equations (\ref{chi-eqs}). This equivalence also applies in the converse way, given the nonlocal form of (\ref{chi-eqs}) and the arbitrariness of $R$ and $\vec{x}_0$. This concludes the proof of the statement that they were after, namely, that \emph{for theories where the Ryu-Takayanagi formula computes entanglement entropy, the first law of entanglement entropy in the CFT is equivalent to the Einstein's equations in the bulk, linearized over empty AdS.}

\subsection{Method 3: Canonical bit threads from Iyer-Wald  \label{5.2}}

Taking into account the nice properties of $\bm \chi$ defined via the Iyer-Wald formalism, here, we propose that specializing this form to a spacelike hypersurface $\Sigma$ containing both $\gamma_A$ and $A$ can be taken as a \emph{canonical} candidate for the perturbed thread $(d-1)-$form\footnote{See \cite{Oh:2017pkr} for a previous attempt at deriving a bit thread configuration from Iyer-Wald.}
\bea \label{w-chi}
\tilde{\bm{\chi}}=\frac{1}{4G_N}{\delta \bm w}\,.
\eea
Given the integral properties of this form, it is straightforward  to check that the flux through any surface homologous to $A$ yields the change of the entanglement entropy in the perturbed state, $\delta S_A$, as expected. Furthermore, this construction fully exploits the property of bulk locality, in particular, connecting the required closedness of ${\delta \bm w}$ with the linearized Einstein's equations via (\ref{closeness}). We will see below that this property will play a very important role for the problem of bulk reconstruction.

It only remains to be checked whether the norm bound constraint (\ref{w-norm1}) is satisfied at the desired order (\ref{P-norm-bound}) everywhere in the bulk. This condition will depend on the background form $\bm w$ and then might not hold in general. However, for our purposes it will suffice to find \emph{one} $\bm w$ such that the combination $\bm w_\lambda=\bm w + \lambda \delta \bm w$ respects the bound for any perturbation. We will devote the remaining part of this section to check that this is indeed possible.

To begin with, we note that the norm constraint in the form (\ref{P-norm-bound}) is slightly more complicated than its equivalent in terms of vectors. Hence, we will first translate the form (\ref{w-chi}) into the language of flows and then check the condition in terms of the latter. For this purpose, we will need an explicit expression relating $\delta v$ and $\delta \bm w$ in the presence of a perturbed metric $g^\lambda_{ab}=g_{ab}+\lambda\delta g_{ab}$. In terms of the Levi-Civita symbol $\varepsilon$, the variation of (\ref{flow-forms-2}) reads
\bea\label{chi-v}
\lambda \delta {w}_{a_1 \ldots a_{d-1}}&=&  \varepsilon_{b a_1 \ldots a_{d-1}} \delta( \sqrt{g_\lambda}\, v_\lambda^b)=\lambda \varepsilon_{b a_1 \ldots a_{d-1}} \sqrt{g} \left(\delta   v^{b}+\tfrac{1}{2} g^{cd}\delta g_{cd} v^{b}\right)\,.
\eea
It is convenient to define a new vector field $\delta v_{\Phi}^a$, given by
\bea\label{deltaVPhi}
\delta v_{\Phi}^a\equiv \frac{\delta\( \sqrt{g_\lambda} \, v_\lambda^a\)}{\sqrt{g}}\bigg|_{\lambda\to1}=\delta v^{a}+\tfrac{1}{2} g^{b c}\delta g_{b c} v^{a}\,,
\eea
which is divergenceless with respect to the unperturbed metric $g_{ab}$, i.e.,
\be
\nabla\cdot \delta v_\Phi=\frac{1}{\sqrt{g}}\partial_a(\sqrt{g}\,\delta v_\Phi^a)=0\,,
\ee
and is related to $\delta {\bm w}$ via its Hodge dual (again, with metric $g_{ab}$),
\bea\label{hodgestar-pert}
\delta v_{\Phi}^a =g^{a b}(\star \delta \bm w)_{b}\,.
\eea
The subindex $\Phi$ here highlights the fact that the flux of this vector field across any bulk surface homologous to $A$, computed with the original metric, equals the change in the entanglement entropy $\delta S_A$. We emphasize that this vector field should \emph{not} be thought as the variation of the flow $v$, but just as an auxiliary object. However, given a $\delta v_{\Phi}^a$ obtained from (\ref{hodgestar-pert}), we can easily recover the \emph{true} variation of the flow $\delta v^a$ from (\ref{deltaVPhi}). In the Fefferman-Graham gauge (\ref{FG-PT}), the metric perturbation takes the form $\delta g_{ij}=z^{d-2} H_{ij}$ (with $\delta g_{zz}=\delta g_{zi}=0$) and $\delta v^a$ reads
\bea\label{separation}
\delta v^a=\delta   v_{\Phi}^{a}-\tfrac{1}{2}z^d H^i_{\,\,\, i} v^{a}\,,
\eea
where $H^{i}_{\,\,\, i}=\delta^{ij}H_{ij}$. Thus, $\delta v^a$ depends not only on $\delta \bm w$ but also on the background flow $v^a$. In fact, the extra piece in (\ref{separation}) is precisely what is needed such that the divergence of $v^a$ taken with the full metric $g^\lambda_{ab}$ vanishes at the desired order,
\bea
\nabla_\lambda\cdot v_\lambda=\frac{1}{\sqrt{g_\lambda}}\partial_a(\sqrt{g_\lambda} v_\lambda^a)&=&\frac{\sqrt{g}}{\sqrt{g_\lambda}}(\cancel{\nabla\cdot v})+\frac{\lambda}{\sqrt{g_\lambda}}\partial_a[\delta(\sqrt{g_\lambda})v^a+\sqrt{g}\delta v^a]\,,\nonumber\\
&=&\frac{\lambda}{\sqrt{g_\lambda}}\partial_a[\cancel{\tfrac{1}{2}z^d H^i_{\,\,\, i} v^{a}}+\sqrt{g}(\delta   v_{\Phi}^{a}-\cancel{\tfrac{1}{2}z^d H^i_{\,\,\, i} v^{a}})]\,,\nonumber\\
&=&\lambda\frac{\sqrt{g}}{\sqrt{g_\lambda}}(\nabla\cdot \delta v_{\Phi})=0\,.
\eea

Next, we need to make a choice for the background flow $v$ in order to get an explicit $\delta v^a$ and test the norm bound $|v_\lambda|\leq1$. Since the background $v$ should already respect the bound $|v|\leq1$, it is clear that $v_\lambda$ can only exceed this bound by an amount of order $\mathcal{O}(\lambda)$. This can indeed be the case for bulk points that saturate the bound at leading order $|v|=1$ (e.g., at the bottle-neck $\gamma_A$), or in their vicinity. On the other hand, points that are parametrically far from saturating the bound at leading order are safe, in the sense that we can always take $\lambda$ to be arbitrarily small such that $|v_\lambda|=|v|+\mathcal{O}(\lambda)\leq1$.

Given the above discussion, then, we should ideally pick a background flow $v$ such that its magnitude decays rapidly away from the minimal surface $\gamma_A$. Fortunately, we already know good examples of flows respecting this property, e.g., the so-called ``geodesic flows'' \cite{Agon:2018lwq}, which for spheres in empty AdS take the form (\ref{Vec1-d}). In the following we will take these background solutions and verify that the norm bound is satisfied at the desired order in the perturbation. First, notice that from (\ref{Vec1-d}), and using (\ref{hodgestar-pert})-(\ref{separation}), it immediately follows that
\bea
\delta v_{\Phi}&=&\frac{z^{d+1}}{4\pi} \Big\{ \left[ \(\frac{2\pi z}{R}+\frac{d}{z} \xi^t+\xi^t\partial^z \) H^i_{\,\,i} \right] \partial_z +\nonumber \\
 && + \left[\(\frac{2\pi (x^i-x^i_0)}{R}+\xi^t\partial^i \) H^j_{\,\,j} -\(\frac{2\pi (x^j-x^j_0)}{R}+\xi^t\partial^j\) H^i_{\,\,j} \right] \partial_i\Big\}\,.
\eea
and
\bea\label{eq:deltav_IW}
\delta v&=&\frac{z^{d+1}}{4\pi} \Big\{ \left[ \(\frac{2\pi z}{R}+\frac{d}{z} \xi^t+\xi^t\partial^z -\frac{2\pi}{z} v^z\) H^i_{\,\,i} \right] \partial_z -\(\frac{2\pi}{z} v^i \)H^j_{\,\,j} \partial_i \nonumber \\
 && + \left[\(\frac{2\pi (x^i-x^i_0)}{R}+\xi^t\partial^i \) H^j_{\,\,j} -\(\frac{2\pi (x^j-x^j_0)}{R}+\xi^t\partial^j\) H^i_{\,\,j} \right] \partial_i\Big\}\,.
\eea
By construction, $v_\lambda=v+\lambda\delta v$ saturates the norm bound on $\gamma_A$ since the form $\bm w_\lambda=\bm w+\lambda\delta \bm w$ from which it is derived obeys the appropriate boundary condition for a max flow (\ref{formmA}).
We can check this explicitly: at $\gamma_A$ we have that $\xi^t=0$, so\footnote{A brief comment is in order. The expression for $\delta v_a|_{\gamma_A}$ in (\ref{deltav}) does not agree with the expected boundary condition at the bulk bottle-neck (\ref{bcVector}).
The explanation of this mismatch is simple: the difference between the two vector fields is proportional to a vector that is tangential to the minimal surface $\delta v^a_{T} =\(\delta^a_{\,\,\, b} - v^a v_b \) \delta v^b$ so its first order contribution to the norm constraint vanishes, $g_{ab}v^a \delta v^b_{T}=0$, because $\delta  v_{T}$ is orthogonal to $v$. Therefore, $\delta v_a|_{\gamma_A}$ in (\ref{deltav}) is equally good as (\ref{bcVector}) to our order of approximation.}
\bea\label{deltav}
v|_{\gamma_A}=\frac{z}{R}\[(x^i-x_0^i)\partial_i + z\partial_z \]\,,\qquad \delta v|_{\gamma_A}=-\frac{z^{d+1}}{2R}\( x^j-x_0^j \) H^i_{\,\,\, j} \,\, \partial_i\,.
\eea
This leads to the expected saturation at first order in $\lambda$,
\bea\label{norm-pert}
g_{ab}^{\lambda}v^a_\lambda v^b_\lambda\Big|_{\gamma_A}=g_{ab}v^av^b+\lambda(\cancel{\delta g_{ab}v^av^b+2g_{ab}v^a\delta v^b})\Big|_{\gamma_A}=1\,.
\eea
Away from $\gamma_A$ the norm bound is not guaranteed to hold, but since $|v|$ decays as a power law, it would suffice to study $|v_\lambda|$ in a neighborhood of $\gamma_A$. In order to see this in detail, we note that the level sets of the background flow $v$ (depicted in Figure \ref{fig:contours}) have the form
\bea\label{surface-ls}
(z+\Delta)^2+|\vec{x}-\vec{x}_0|^2=R^2+\Delta^2\,,\qquad z\geq0\,,
\eea
where $\Delta\in\mathbb{R}$, i.e., spheres with radius $\sqrt{R^2+\Delta^2}$, centered at $(\vec{x}_c=\vec{x}_0,z_c=-\Delta)$.
It can be checked that, in the vicinity of the minimal surface ($\Delta^2 \ll R^2$)
\bea
|v|\approx 1-\frac{ (d-1)\Delta^2}{2R^2}\,.
\eea
Since $d>1$, then $|v|<1$ for any $\Delta\neq0$ as expected. Now, we want to check whether the norm bound is still satisfied for $v_\lambda$ at the leading order in the perturbation. More precisely, what we really want is to check that for a fixed $\lambda$, $|v_\lambda|\leq1$ at linear order in $\lambda$ for an arbitrary $\Delta$. A short calculation shows that
\be\label{norm-const-A}
|v_\lambda|=1-\frac{ (d-1) \Delta^2}{2R^2}+\lambda\left(\tfrac{1}{2}\delta g_{ab} v^av^b +g_{ab}v^a\delta v^b\right)   \overset{?}{\leq} 1\,,
\ee
which after some algebra can be put in the following form:
\bea\label{norm-constraint-B}
 \lambda\frac{\Delta z^{d+1}}{R^2}\((x^i-x_0^i)\partial_i H^j_{\,\, j} -(x^i-x_0^i)\partial^j H_{ij}+ z\partial_z H^j_{\,\, j}+(d-1) H^j_{\,\, j}\)  \overset{?}{\leq} \frac{ (d-1) \Delta^2}{R^2 }\,.
\eea
The parenthesis in the left-hand side of (\ref{norm-constraint-B}), or (\ref{norm-const-A}), can in fact be non-negative, which implies that the above inequality will not hold for arbitrarily small $\Delta$. Nevertheless, it is interesting to estimate the order of magnitude of the potential violation of the norm bound constraint as it could still be consistent with our order of approximation. From (\ref{norm-constraint-B}) it follows that the norm bound can be violated provided that
\be\label{eq:order}
\frac{\Delta}{R}\lesssim\mathcal{O}(\lambda)\,.
\ee
Plugging (\ref{eq:order}) back into (\ref{norm-const-A}), we observe that this would only lead to a violation of order $\mathcal{O}(\lambda^2)$. Since all of our analysis is at linear order in $\lambda$, we can safely ignore this issue. In other words, up to our order of approximation the norm bound is \emph{not} violated and this means that the ``canonical'' thread configuration constructed from the Iyer-Wald formalism satisfies all the defining properties for a max flow. We relegate to Appendix \ref{app:examples} the study of a explicit example of these canonical thread configurations.

Finally, we can comment on how the information of the metric perturbation and the Einstein's equations are encoded in this particular thread construction. First, notice that the variation of the flow $\delta v$ constructed from Iyer-Wald fully exploits the property of bulk locality. This is evident, since for this particular construction the divergenceless condition, $\nabla_\lambda \cdot v_\lambda$, maps directly to the Einstein's equations, which are defined locally in the bulk. Moreover, the fact that the $\delta v$ constructed here (\ref{eq:deltav_IW}) can be written in terms of a linear local functional of $\delta g_{ab}$ and its derivatives, present us with an interesting possibility: we can use the information of this canonical solution to invert the problem and recover the bulk metric from it! We will explore this problem in more detail in the next subsection, and comment on the implications and possible generalizations to the full non-linear regime.

\subsection{Metric reconstruction\label{metric-reconstruction}}

\subsubsection{Explicit reconstruction at linear order\label{explicit-reconstruction}}

Our bit thread construction based on differential forms makes explicit use of the property of bulk locality, hence, it should be possible to invert the problem and recover the metric for a generic linear excitation of the boundary quantum state. In this section we will study this problem in detail. More specifically, we will consider a manifold $M$ with boundary $\partial M$ and a set of forms $\delta \bm w$ that encode the local pattern of entanglement of boundary regions. We will assume the knowledge of the zeroth order ---pure AdS--- metric $g_{\mu\nu}$, which is otherwise fixed by conformal symmetry (i.e. kinematics), and set up the problem of how to reconstruct the metric perturbations $\delta g_{\mu\nu}$ from the above data.

Our starting point is the knowledge of the change in the entanglement structure of the CFT, which in this case is encoded in the set of $(d-1)-$forms $\delta \bm w$. We emphasize that these canonical forms can be \emph{uniquely} specified solely from CFT data. Given a perturbative excited state in the CFT, one can first evaluate the expectation value of the stress-energy tensor $T_{\mu\nu}$ and thus the modular Hamiltonian $H_A$ associated with a spherical region $A$. This information can then be used as a boundary condition for $\delta \bm w$ on $A\subset\partial M$. For instance, specializing to a constant-$t$ slice $\Sigma$, this yields\footnote{Notice that a choice of boundary condition on $A$ is equivalent to picking a specific \emph{entanglement contour} in the dual CFT. We emphasize that this corresponds to focusing on a particular class of microstates with a given local entanglement pattern. Although this boundary condition is in general non-unique, (\ref{bcIyerWald}) is \emph{the} boundary condition singled out by the Iyer-Wald construction.}
\be
\delta \bm w|_{A}=\frac{4\pi G_N}{R}\(R^2-|\vec{x}-\vec{x}_0|^2\)\langle T_{00}\rangle\,  \bar{{ \bm \epsilon}}\,,
\ee
where $\bar{{ \bm \epsilon}}$ is the natural volume form in the boundary CFT. In fact, we can analytically continue this form to the whole boundary $\partial M$, so that\footnote{More covariantly, on a general Cauchy slice $\Sigma'$ containing $\partial A$, the boundary condition would be
$$
\delta \bm w|_{\partial M}=4G_N \,N^\mu\zeta_A^\nu \langle T_{\mu\nu}\rangle\,  \bar{{ \bm \epsilon}}\,,
$$
where $N^\mu$ is a future pointing unit normal vector, and $\zeta_A$ is the conformal killing vector that generates $\mathcal{D}[A]$,
$$
\zeta_A=\frac{\pi}{R}\[(R^2-(t-t_0)^2-|\vec{x}-\vec{x}_0|^2)\partial_t+2(t-t_0)(x^i-x_0^i)\partial_i\]\,.
$$
Upon conformally mapping the causal development of the region $\mathcal{D}[A]$ to the hyperbolic cylinder $H^{d-1}\times R_\tau$, it can be shown that $\zeta_A$ coincides with the time-like Killing vector $2\pi R\, \partial_\tau$.}
\be\label{bcIyerWald}
\delta \bm w|_{\partial M}=\frac{4\pi G_N}{R}\(R^2-|\vec{x}-\vec{x}_0|^2\)\langle T_{00}\rangle\,  \bar{{ \bm \epsilon}}\,.
\ee
One way to see that this is consistent would be to conformally map the interior of the sphere to the exterior. Upon implementing this transformation one finds the same functional form for the modular Hamiltonian but integrated along $\vec{x}\in A^c$, hence, providing a boundary condition also at $A^c=\partial M\setminus A$.
With the above boundary condition, the full $(d-1)-$form in the interior of the manifold $M$ is then uniquely determined if we assume \emph{bulk locality} \cite{Wald:2005nz}. To see this, notice that the Iyer-Wald construction yields a form $\delta \bm w$ such that $d\delta \bm w=0$ on-shell, which is a local condition. If we want to maintain this condition, then, the only ambiguity in $\delta \bm w$ would be the addition of a term $\delta \bm w \to \delta \bm w +d \bm C$ where $\bm C$ is a $(d-2)-$form such that $d \bm C$ vanishes on $\partial M$. This is of course a gauge redundancy, which we fix by working in Fefferman-Graham coordinates. Therefore, the boundary condition (\ref{bcIyerWald}) together with the condition of bulk locality are enough to uniquely specify the full $(d-1)-$form on $M$.

Before proceeding with the specifics of this analysis, let us first quickly review how the problem of metric reconstruction is normally addressed. In the usual HRT story, given background metric $g_{\mu\nu}$, the change in the entanglement entropy of a region $A$ at first order in the perturbation $\delta g_{\mu\nu}$ is given by\footnote{The analysis in terms of extremal surfaces can be done for general states, not necessarily perturbative. Here we are discussing only this simpler case to highlight an important difference with our approach.}
\bea
\delta S_A=\int_{\gamma_A} \delta \sqrt{h} \,d^{d-2}x\,,\qquad \delta \sqrt{h} =\frac{1}{2}\sqrt{h} h^{ij} \delta h_{ij}\,.
\eea
This means that $\delta S_A$ encodes information about the first order change in the trace of the induced metric $h_{ij}$ over the extremal surface $\gamma_A$. On the other hand, the induced metric on $\gamma_A$ depends on the bulk metric as well as the explicit embedding of $\gamma_A$ on the geometry. Therefore, by cleverly considering different boundary regions with extremal surfaces intersecting at a bulk point, one could access to the various components of the bulk metric at the given bulk point and hence derive an inversion formula for the metric perturbation $\delta g_{\mu\nu}$.

As is evident from the previous paragraph, the problem of metric reconstruction by extremal surfaces heavily relies on the possibility of foliating the full manifold $M$ with boundary anchored extremal surfaces. In particular, we would necessarily need to start from a dense family of surfaces that pass through all (reachable) bulk points multiple times. While we can do this in the language of bit threads, i.e., start from a \emph{dense set} of thread configurations, the fact that one single solution to to the max flow problem already probes the full bulk geometry presents us with an interesting possibility: we can start from a \emph{finite set} of thread configurations, containing one, or possibly only a few solutions of the max flow problem. The minimal number of thread configurations needed in such a set will generally depend on symmetry considerations as well as the number of dimensions, as will be discussed below. For the time being, let us summarize the two approaches to metric reconstruction that we can explore. For simplicity, we will frame the discussion by focusing on a constant-$t$ slice $\Sigma$, so that we will aim to recover the spatial components of the metric $\delta g_{ij}$. The $\delta g_{tt}$ and $\delta g_{ti}$ components can be recovered in a similar way, by choosing appropriate boosted slices $\Sigma'$, as we will explain at the end of the section.
The two methods that we will explore are:
\vspace{-3mm}
\begin{itemize}
  \item \emph{Reconstruction from a dense set of thread configurations.} Here we will assume knowledge of $\delta \bm w$ for \emph{all} spheres in the CFT, with arbitrary radius $R$ and center point $\vec{x}_0$.\vspace{-3mm}
  \item \emph{Reconstruction from a minimal set of thread configurations.} Here we will assume knowledge of $\delta \bm w$ for \emph{a few} spheres with radius $R^{(i)}$, and center point $\vec{x}_0^{(i)}$, with $i=1,\ldots,n$. The precise value of $n$ will be fixed so that the inversion problem is well defined.
\end{itemize}
\vspace{-3mm}
We will now discuss these two methods in detail.

\subsubsection*{Reconstruction from a dense set of thread configurations}

Given the infinite set of $(d-1)-$forms $\delta \bm w$ encoding the \emph{canonical} entanglement pattern of spheres of arbitrary radius $R$ and center point at $\vec{x}_0$, on a constant-$t$ slice $\Sigma$, our goal is to extract the components of the perturbed metric $\delta g_{ij}$. We recall that, in the presence of a metric, the set of $(d-1)-$forms $\delta \bm w$
define a set of covector fields $\delta w_a(R,\vec{x}_0,z,\vec{x})$, instead of the numbers $\delta S(R,x_0)$, so it is clear that in this framework we have infinitely more information in comparison to the standard setup using extremal surfaces. Hence, we can expect to be able to reconstruct the metric in a more straightforward way.

If the full metric is given in the Fefferman-Graham gauge (\ref{FG-PT}),
the components of the metric perturbation take the form $\delta g_{ij}=z^{d-2} H_{ij}$ (with $\delta g_{zz}=\delta g_{zi}=0$).
In this gauge, the components of the covectors $\delta w_a$ can be related to the metric perturbations as follows:
\bea\label{wz}
\delta w_z(R,\vec{x}_0,z,\vec{x})& =&\frac{z}{4\pi}\(\frac{2\pi z}{R}+\frac{d}{z} \xi^t+\xi^t\partial^z \) H^i_{\,\,i}\,, \\ \label{wi}
\delta w_i(R,\vec{x}_0,z,\vec{x}) &= &\frac{z}{4\pi}\left[ \(\frac{2\pi (x^i-x^i_0)}{R}+\xi^t\partial^i \) H^j_{\,\,j} -\(\frac{2\pi (x^j-x^j_0)}{R}+\xi^t\partial^j\) H^i_{\,\,j} \right],
\eea
where
\be
\xi^t=\frac{\pi}{R}\(R^2-z^2-|\vec{x}-\vec{x}_0|^2\)\,.
\ee
These equations can be easily inverted using the dependence on $R$ and $\vec{x}_0$ of (\ref{wz}) and (\ref{wi}). In fact, there are infinitely many ways to invert these equations.
The simplest way is to get rid of the derivative terms, so that we obtain a system of algebraic equations. However, we have several ways to accomplish this. Below we will discuss two different options.

The first option is by evaluating both sides of (\ref{wz}) and (\ref{wi}) on the set of parameters $(R,\vec{x}_0)$ that satisfy $\xi^t(R,\vec{x}_0)=0$, i.e.,
\bea\label{xit0}
R^2=z^2+|\vec{x}-\vec{x}_0|^2\,.
\eea
Notice that the requirement given by (\ref{xit0}) means that our reconstruction is limited to the points that are accessible via extremal surfaces. This means that this option is, in a sense, analogous to the metric reconstruction via the HRT prescription and does not exploit the full reach of the bit threads. We will continue for now and then explain an alternative that does not impose this limitation. Let us first analyze (\ref{wz}). From this equation, we can immediately find an algebraic expression that gives the perturbed trace $H^i_{\,\,i}(z,\vec{x})$,
\bea\label{inv-Hii}
H^i_{\,\,i}(z,\vec{x})=\frac{2 R}{z^2}\delta w_z\(R,\vec{x}_0; z,\vec{x}\)\Bigg|_{\xi^t(R,\vec{x}_0)=0}\,.
\eea
In fact, we can extract the trace $H^i_{\,\,i}$ at a point $(z,\vec{x})$ from (\ref{inv-Hii}) using any single covector with parameters $(R,\vec{x}_0)$ such that (\ref{xit0}) is satisfied. This is an example of the non-uniquess of the inversion formulas. Notice that equation (\ref{inv-Hii}) provides the solution to the full inversion problem for $d=2$, in which case (\ref{wi}) is identically zero. In fact, for $d=2$ the only component of the metric perturbation that we need to solve for corresponds to $H_{xx}(z,x)$ which equals the trace (\ref{inv-Hii}). In higher dimensions, we can use (\ref{wi}) in addition to (\ref{wz}), and proceed in a similar way to extract the information of the individual components of the perturbed metric $H^i_{\,\,j}(z,\vec{x})$. In order to do that, first replace the solution for the trace (\ref{inv-Hii}) in equation (\ref{wi}), so that the latter equation becomes:
\bea\label{wixit0}
\delta w_i \Big|_{\xi^t=0}&= &\frac{(x^i-x^i_0)}{z} \delta w_z \Big|_{\xi^t=0} -\frac{z}{2 R}\(x^j-x^j_0\)H^i_{\,\,j} \,.
\eea
Further, for a given $j$ and within the set of allowed parameters, we can take $x_0^j\neq x^j$ and $x_0^k=x^k$ for $k\neq j$. This leads to the following solutions for the diagonal and non-diagonal components of the perturbation:
\bea
\(\textrm{no sum over}\,\, j\)\quad H^j_{\,\, j}(z,\vec{x})&=&-\frac{2R}{z(x^j-x_0^j)}\, \delta w_j \Big|_{\xi^t=0}+\frac{2R}{z^2}\delta w_z \Big|_{\xi^t=0}\,, \\
H^i_{\,\, j}(z,\vec{x})&=&-\frac{2R}{z(x^j-x_0^j)}\, \delta w_i \Big|_{\xi^t=0}\,,
\eea
which provides the solution to the full inversion problem for $d> 2$.

As mentioned above, the method outlined in the previous paragraph is limited to the points that are accessible via extremal surfaces.
An alternative way of inverting the system of equations in a less restrictive setting is the following. First notice the relations
\bea
\frac{\partial \xi^t }{ \partial x_0^k}=\frac{2\pi (x^k-x_0^k)}{R}\,, \qquad \frac{\partial \xi^t }{ \partial x_0^k}\bigg|_{x_0^k=x^k}\!\!\!=0\,,
\eea
Using the above, one finds that
\bea\label{wiz-derivatives}
\frac{\partial \delta w_z }{ \partial x_0^k}\bigg|_{x_0^k=x^k}\!\!\!\!\!\!&=& 0\,,\nonumber \\
\frac{\partial \delta w_i }{ \partial x_0^k}\bigg|_{x_0^k=x^k}\!\!\!\!\!\!&=&-\frac{z}{2R}\delta^i_k\, H^{j}_{\,\, j}+\frac{z}{2R}\delta^j_k\, H^{i}_{\,\, j}\,,
\eea
from which one can easily invert the diagonal and off diagonal components of the perturbation for $d>2$.\footnote{Notice that for $d=2$ equation (\ref{wi}) is identically equal to zero and then equations (\ref{wiz-derivatives}) are trivial.}  More explicitly, we obtain that
\bea \label{rep-indices}
\(\textrm{no sum over}\,\, j\)\quad H^{j}_{\,\, j}(z,x)&=&\frac{2R}{z}\(\frac{\partial \delta w_j }{ \partial x_0^j}\bigg|_{x_0^j=x^j}\!\!\!-\frac{1}{d-2}\sum_{i} \frac{\partial \delta w_i }{ \partial x_0^i}\bigg|_{x_0^i=x^i}\! \)\,, \\ \label{other-Hij}
H^{i}_{\,\, j}(z,x)&=&\frac{2R}{z}\frac{\partial \delta w_i }{ \partial x_0^j}\bigg|_{x_0^j=x^j}\!\!\! ,
\eea
which gives the solution to the full inversion problem for $d> 2$. 
On the other hand, for $d=2$ we can simply use equation (\ref{inv-Hii}) or, alternatively, the second reconstruction method that we explain below. Notice that summing over $j$ in equation (\ref{rep-indices}) leads to a different expression for the trace of the perturbation as (\ref{inv-Hii}). This represents another example of the non-uniqueness in the inversion equations.

\subsubsection*{Reconstruction from a minimal set of thread configurations}

One may wonder whether the extended nature of the entanglement pattern information present in a single form $\delta {\bm w}(R,\vec{x}_0;z,\vec{x})$ for fixed $(R,\vec{x}_0)$, or a finite number of forms $n$, could suffice to recover the components of the metric perturbations $\delta g_{ij}$. Naively, the number of unknown variables that we need to solve for  is $d(d-1)/2$, which is the number of symmetric components of $H_{ij}$ ($\{i,j\}$ run over $d-1$ values). On the other hand, the number of equations that we would have at our disposal is given by $nd$, i.e., $d$ components of a single covector $\delta w_a$, times the total number of forms $n\in\mathbb{N}$. Assuming they are all linearly independent of each other, we have that for a fixed $d$, the minimum number of forms $\bar{n}$ that we would need is
\be\label{minimaln}
\bar{n}=\left\lceil\frac{d-1}{2}\right\rceil\,,
\ee
where $\lceil\bullet\rceil$ represents the ceiling function (i.e., the smallest integer greater or equal to its argument). For $d=2$ and $d=3$ ($3d$ and $4d$ gravity respectively) we obtain $\bar{n}=1$, which means that we can hope to recover the full metric from a single form. In higher dimensions the problem would be underdetermined so we would need to increase the number of forms, although, to a finite number set by $d$. In the following, we will focus on the cases for which $\bar{n}=1$ but we will come back to the higher dimensional cases at the end of the section.

We start with equation (\ref{wz}) and notice that its right-hand side can be rewritten as
\bea\label{eq:inversion}
\delta w_z(R,\vec{x}_0,z,\vec{x})=\frac{(\xi^t)^2}{4\pi z^{d-1}}\partial_z\left(\frac{z^d H^i_{\,\, i}}{\xi^t}\right)\,.
\eea
This equation can be in principle easily inverted for the trace $H^i_{\,\, i}$. However, we must proceed with some care because $\xi^t$ can attain a zero value in multiple points in the bulk. We refer the reader to appendix \ref{app:inversion} for a detailed analysis of this equation, and we will simply state the final result here.
The analysis is naturally split for bulk points $\vec{x}\in A^c$ and $\vec{x}\in A$ ($\forall z$). In the former case, $\xi^t$ never vanishes in the bulk, so the analysis is particularly simple. In this case we obtain
 \bea \label{Hii-4p}
H^i_{\,\, i} (z,\vec{x})&=&4R(z_*^2-z^2) \, \int_{0}^1  d\lambda \, \frac{ \lambda^{d-1}\delta w_z(\lambda z,\vec{x})}{[z_*^2-(\lambda z)^2]^2}\,,\qquad z_*^2\equiv R^2-|\vec{x}-\vec{x}_0|^2\,.
\eea
This equation is also valid for $\vec{x}\in A$ provided that $z^2< z_{*}^2$. For $z\geq z_*$ the integrand in (\ref{Hii-4p}) has a double pole at $\lambda_*=z_*/z$ (with $\lambda_*\in[0,1]$). This divergence can be removed by an appropriate regularization, e.g., using the principle value prescription. However, given the simplicity of our problem we can find the answer directly from (\ref{eq:inversion}) by taking one of the end points of the integral to be arbitrarily close to the zero locus of $\xi^t$. After a series of careful manipulations explained in appendix \ref{app:inversion} we arrive at a formula that is valid in the region $\vec{x}\in A$ and for all $z$:
\bea\label{Hii-9p}
\!\!\!\!\!\!\!\!H^{i}_{\,\,i}(z,\vec{x})=\frac{2R z_{*}^{d-4}\delta w_z(z_{*},\vec{x})}{z^{d-2}}+4R(z_{*}^2-z^2)\! \int_{0}^1\! d\lambda\frac{\lambda [\lambda^{d-2} \delta w_z(\lambda z,\vec{x})-\lambda_{*}^{d-2} \delta w_z(z_{*},\vec{x})]}{[z_{*}^2-(\lambda z)^2]^2}.
\eea
This new integral seems to still have a single pole at $\lambda=\lambda_{*}$. However, a close inspection shows that this term is proportional to
\be
(d-2) \delta w_z(z_*,\vec{x})+z_* \partial_z\delta w_z(z_*,\vec{x})=0\,,
\ee
which can be checked from (\ref{wz}). Therefore, the integral is manifestly finite. We note that, indeed, the naive principle value regularization of (\ref{Hii-4p}) results in (\ref{Hii-9p}), and likewise there are many ways to derive (\ref{Hii-9p}) from (\ref{Hii-4p}). Perhaps the simplest way to do it is by slightly changing the integration contour:
 \bea \label{Hii-4b}
H^i_{\,\, i} (z,\vec{x})=4R(z_*^2-z^2) \int_{i\epsilon}^{1+i\epsilon}\!\!\!  d\lambda \, \frac{ \lambda^{d-1}\delta w_z(\lambda z,\vec{x})}{[z_*^2-(\lambda z)^2]^2}\,,
\eea
where $\epsilon\in\mathbb{R}$, and then letting $\epsilon\to0$. It can be shown that this prescription is consistent with both (\ref{Hii-4p}) and (\ref{Hii-9p}) so it is valid everywhere in the bulk.

Notice that in the above formulas we have not specified the number of dimensions $d$. This means that given the knowledge of a single $\delta \bm w(z,\vec{x})$ one can always find an inversion formula for the trace of the perturbation, given explicitly by (\ref{Hii-4p})-(\ref{Hii-9p}) or its equivalent (\ref{Hii-4b}). We will now recover the other components of the metric for the cases $d=2$ and $d=3$.

\begin{itemize}
  \item \emph{The $d=2$ case:}
\end{itemize}
The inversion problem in $d=2$ is exceptionally simple because in this case there is only one metric component to solve for. Therefore, the trace of the perturbation provides the full solution to the problem, i.e., $H^i_{\,\, i}(z,x)=H_{xx}(z,x)$. Nevertheless, we will analyze this case in some detail and check that our formulas for the trace ar consistent with the expected results.

First, notice that in this case there are further simplifications that considerably reduce our problem: equation (\ref{wi}) vanishes exactly so $\delta w_{x}=0$ everywhere in the bulk. The closedness relation  $d \delta  {\bm w}=0$ then implies $\partial_z \delta w_z=0$, so $\delta w_z(z,x)=\delta w_z(0,x)$. Using this fact, and applying the formula (\ref{Hii-4b}) which is valid everywhere in the bulk, we obtain
\bea\label{Hxx2d}
 \!\!\!\!\!\!\!\!\!\!\!H_{x x} (z,x)=4R \delta w_z(0,x)(z_*^2-z^2) \!\left.\int_{i \epsilon}^{1+i \epsilon}\!\!\!  d\lambda \frac{ \lambda}{[z_*^2-(\lambda z)^2]^2}\right|_{\epsilon\to0}\!\!\!\!=\frac{2R \delta w_z(0 ,x)}{z_*^2}=H_{x x} (0,x).
\eea
This is precisely what is expected from the analysis of Section \ref{sec:2dpert}, specifically from equation (\ref{deltag2}).
For $x\in A^c$, (\ref{Hii-4p}) coincides explicitly with (\ref{Hii-4b}) so the integral above is the same.
For $x\in A$ we can alternatively use (\ref{Hii-9p}). Notice that for $d=2$ the integrand in (\ref{Hii-9p}) is identically zero since $\delta w_z(z,\vec{x})$ is constant
so, in this case, the full result is given by the first term in (\ref{Hii-9p}). Indeed, this term coincides with (\ref{Hxx2d}), as expected.

Again, since for $d=2$ this is the only component of the perturbed metric that we need to solve for, then, equation (\ref{Hxx2d}) completes the inversion problem.

\begin{itemize}
  \item \emph{The $d=3$ case:}
\end{itemize}
To solve the inversion problem in $d=3$ we need equations (\ref{wi}), in addition to (\ref{wz}). These equations involve derivatives with respect to the spatial coordinates $\partial_{i}$ so, as a system of first order differential equations, we would need information about $H_{ij}$ at some fixed $x_i$ in order to have a well defined boundary value problem. We will deal with the choice of such boundary conditions below, but for now, let us explain how to setup the inversion problem.

First, since equation (\ref{Hii-4b}) already provides the solution for the trace part of the metric, we can already replace this back into (\ref{wi}) and solve for the remaining metric components. We find it convenient to write the fluctuations as
\be\label{Hij-3d}
H_{ij}=\left(
                            \begin{array}{cc}
                              \frac{h}2+\phi & \chi \\
                              \chi & \frac{h}2-\phi \\
                            \end{array}
                          \right)\,,
\ee
so that $h=H^i_{\,\,i}$ gives the trace while $\phi$ and $\chi$ are the two fields that we still need to solve for. Defining $x_1\equiv x$ and $x_2\equiv y$,
equations (\ref{wi}) then take the form:
\bea
\delta w_x& = &\frac{z}{4\pi}\left[ \(\frac{2\pi (x-x_0)}{R}+\xi^t\partial_x \) \(\frac{h}2-\phi\) -\(\frac{2\pi (y-y_0)}{R}+\xi^t\partial_y\) \chi \right], \label{Hij-3d-2} \\
\delta w_y&= &\frac{z}{4\pi}\left[ \(\frac{2\pi (y-y_0)}{R}+\xi^t\partial_y \) \(\frac{h}2+\phi\) -\(\frac{2\pi (x-x_0)}{R}+\xi^t\partial_x\) \chi \right].\label{Hij-3d-2b}
\eea
Thus, we have a system of two coupled partial differential equations of first order. It is possible to decouple these equations by taking one further derivative and combining appropriately the two equations. To do so, we first rewrite (\ref{Hij-3d-2})-(\ref{Hij-3d-2b}) as:
\bea\label{constrains-phi-xi}
\partial_x\( \frac{\phi}{\xi^t}\) + \partial_y\( \frac{\chi}{\xi^t}\)&=&\delta\Omega_x\,,\qquad \delta\Omega_x\equiv  \partial_x\(\frac{h}{2\xi^t}\)-\frac{4\pi}{z(\xi^t)^2}\delta w_x\,,\label{phichi}\\
 -\partial_y\( \frac{\phi}{\xi^t}\)+ \partial_x\( \frac{\chi}{\xi^t}\)&=&\delta\Omega_y\,,\qquad\delta\Omega_y \equiv \partial_y\(\frac{h}{2\xi^t}\)-\frac{4\pi}{z(\xi^t)^2}\delta w_y\,.\label{phichib}
\eea
Equations (\ref{phichi}) and (\ref{phichib}) can now be combined as a pair of Poisson's equations:
\bea
\(\partial_x^2+\partial_y^2\)\( \frac{\phi}{\xi^t}\)&=&\rho_\phi\,,\qquad \rho_\phi\equiv \partial_x\delta\Omega_x-\partial_y\delta\Omega_y\,,\\
\(\partial_x^2+\partial_y^2\)\( \frac{\chi}{\xi^t}\)&=&\rho_\chi\,,\qquad \rho_\chi\equiv\partial_x\delta\Omega_y+\partial_y\delta\Omega_x\,,
\eea
where $\rho_\phi$ and $\rho_\chi$ act as sources for $\phi$ and $\chi$, respectively. These equations can be solved using standard Green's function methods. Assuming knowledge of the solutions at a closed surface $\partial \mathcal{V}$, one can formally write the solution in the interior of the surface $\mathcal{V}$ as follows:
\bea\label{solution1}
\frac{\phi(\vec{x})}{\xi^t(\vec{x})}&=&\!\int_{\mathcal{V'}}\!\! G(\vec{x},\vec{x}')\rho_\phi(\vec{x}')dV'+\!\int_{\partial \mathcal{V'}}\[ \frac{\phi(x')}{\xi^t(x')}\nabla^{'} G(\vec{x},\vec{x}')-G(\vec{x},\vec{x}')\nabla^{'} \frac{\phi(x')}{\xi^t(x')}\]\cdot dS'\,, \\
\frac{\chi(\vec{x})}{\xi^t(\vec{x})}&=&\!\int_{\mathcal{V'}}\!\!G(\vec{x},\vec{x}')\rho_\chi(\vec{x}')dV'+\!\int_{\partial \mathcal{V'}}\[ \frac{\chi(x')}{\xi^t(x')}\nabla^{'} G(\vec{x},\vec{x}')-G(\vec{x},\vec{x}')\nabla^{'} \frac{\chi(x')}{\xi^t(x')}\]\cdot dS'\,,\label{solution2}
\eea
where $G(\vec{x},\vec{x}')$ is the Green's function for the Laplace operator in 2d, given by
\bea\label{Green2d}
G(\vec{x},\vec{x}')=\frac{1}{2\pi}\log|\vec{x}-\vec{x}'| +c_0\,.
\eea

Next, we need to impose appropriate boundary conditions. To do so, let us start discussing the region $z>R$ for which $\xi^t<0$ so that there will be no subtleties with poles in the integrals. We will consider a closed surface $\partial \mathcal{V'}$ at infinity, so that $dS'=\hat{r}\, r'd\theta'$, with $r'=|\vec{x}'|\to\infty$. We note that, normally, the integral over $\partial \mathcal{V'}$ in 2d would give a finite contribution, assuming that the fields that we solve for are finite at infinity. This is because $\nabla^{'} G(\vec{x},\vec{x}') \cdot dS'\to$ constant at large $r'$. However, in our particular problem, we need to solve for the combination $\phi/\xi^t$, or $\chi/\xi^t$, which decay (at least) as $1/r'^2$ at large $r'$.\footnote{To see this, notice that both $\phi$ and $\chi$ are components of the perturbation $H_{ij}$, which can generally be written as (\ref{pertg}), i.e., $H_{ij}\sim \sum z^{2n}T^{(n)}_{ij}$. The leading order term gives the CFT stress-energy tensor, which scales as $T^{(0)}_{ij}\sim1/{r^{d-2}}\sim1/r$ for perturbations of compact support, or $T^{(0)}_{ij}\sim$ constant otherwise (e.g. for plane waves). For $n>0$, equation (\ref{Tnexp}) tells us that $T^{(n)}_{ij}\sim\Box^n T^{(0)}_{ij}$, so these terms they decay faster at infinity. Hence, $H_{ij}$ scales like $T^{(0)}_{ij}$ and $\phi/\xi^t\sim1/r^2$ and $\chi/\xi^t\sim1/r^2$ at large $r$ (in the worst case scenario).} This means that the surface integrals in (\ref{solution1}) and (\ref{solution2}) vanish, regardless of the specific values of $\phi$ and $\chi$ at $r\to\infty$, and therefore,
\bea\label{solution1b}
\phi(\vec{x})&=&\frac{\xi^t(\vec{x})}{2\pi}\int_{\mathcal{V'}} \log|\vec{x}-\vec{x}'| \rho_\phi(\vec{x}')dV'\,,  \\
\chi(\vec{x})&=&\frac{\xi^t(\vec{x})}{2\pi}\int_{\mathcal{V'}}\log|\vec{x}-\vec{x}'| \rho_\chi(\vec{x}')dV'\,,\label{solution2b}
\eea
where the boundary conditions once again force $c_0=0$.\footnote{Notice that if $c_0\neq0$, both $\phi\sim r^2\to\infty$ and $\chi\sim r^2\to\infty$ at large $r$, which would be unphysical.} For the region $z<R$, we note that $\xi^t$ can become zero at multiple points in the bulk, so we would need to deal with potential divergences of the integrands. Following the calculation of the trace (explained in detail in Appendix \ref{app:inversion}), we can expect to get rid of these non-physical divergences by a simple regularization procedure. One easy way to do this is by adding a small imaginary part to $z$,
\bea\label{solution1c}
\phi(\vec{x})&=&\frac{\xi^t(\vec{x})}{2\pi}\int_{\mathcal{V'}} \log|\vec{x}-\vec{x}'| \rho_\phi(\vec{x}')dV'\bigg|_{z\to z+i\epsilon}\,,  \\
\chi(\vec{x})&=&\frac{\xi^t(\vec{x})}{2\pi}\int_{\mathcal{V'}}\log|\vec{x}-\vec{x}'| \rho_\chi(\vec{x}')dV'\bigg|_{z\to z+i\epsilon}\,,\label{solution2c}
\eea
and at the end of the calculation let $\epsilon\to0$. The integration region should now be free of singularities, so these formulas apply for all values of $z$. Hence, together with the trace formula (\ref{Hii-4b}), they provide a complete solution of the reconstruction problem in $d=3$.

\begin{itemize}
  \item \emph{Higher dimensional cases:}
\end{itemize}
As discussed above, by comparing the number of independent components of the perturbed metric $H_{ij}$ against the number of equations that we obtain from a single form $\delta {\bm w}$, one can conclude that the minimum number  of forms $\bar{n}$ that are in principle required to invert the problem is given by (\ref{minimaln}). However, this does \emph{not} imply that any set of $\bar{n}$ forms should lead to a well defined inversion problem since, depending on the choice, one could end up with a set of equations that are not completely linearly independent. In the following we will spell out the precise conditions that we must impose on a minimal set of forms and give a concrete example of how these conditions can be satisfied.

First, notice that for a spherical region, a single form $\delta {\bm w}$ is parametrized by $d$ real numbers $\mathcal{P}=\{R, x_0^1,\cdots ,x_0^{d-1}\}$. Moreover, one can easily check that for each choice of such numbers, the $d$ components of the form $\delta w_a$, with $a\in\{z,1, \ldots, d-1\}$, are all linearly independent since each component involves different sets of metric components $H_{ij}$. Therefore, the task at hand reduces to finding a convenient set of parameters $\mathcal{P}_k$, with $k\in \{1\,,\dots\,, \bar{n}\}$, such that the individual components of each form are linearly independent across the set. More concretely, if we label the $\bar{n}$ forms as $\delta \bm w^{k}$, then we would need to impose that for a fixed $a$ the set $\{\delta w_a^{1}\,,\dots\,, \delta w_a^{\bar{n}} \}$ must be linearly independent.

In order to visualize explicitly the dependence of each component $\delta w_a$ on the parameters $\{R, x_0^1,\cdots ,x_0^{d-1}\}$, we rewrite equations (\ref{wz}) and (\ref{wi}) as follows:
\bea\label{wz-i}
\delta w_z&=&\(\frac{R^2-|\vec{x}_0|^2}{4R}\)\(d\, H^i_{\,\,i}+z\partial_zH^i_{\,\,i} \)+\frac{x_0^l}{2R}\( d\, x_l H^i_{\,\,i} +x_l z\partial_zH^i_{\,\,i}\)\nonumber \\ &&\, \qquad-\frac{1}{4R}\Big[\((d-2)z^2+|\vec{x}|^2  \)H^i_{\,\,i}+ z\(z^2+|\vec{x}|^2\)\partial_zH^i_{\,\,i}\Big]\,, \\ \label{wi-i}
\delta w_i&= &\(\frac{R^2-|\vec{x}_0|^2}{4R}\)\[z\big(\partial^i\, H^j_{\,\,j}-\partial^j\,H^i_{\,\,j}\big)\]+\frac{x_0^l}{2R}\[ z\, x_l\big(\partial^i\, H^j_{\,\,j}-\partial^j\,H^i_{\,\,j}\big)-z\big(\delta^i_{\,\,l}H^j_{\,\,j}-\delta^j_{\,\,l}H^i_{\,\,j}\big)\] \nonumber \\
&&\, \qquad-\frac{1}{4R}\[\(z^2+|\vec{x}|^2\)\big(\partial^i\, H^j_{\,\,j}-\partial^j\,H^i_{\,\,j}\big)+2z\big(x^iH^j_{\,\,j}-x^jH^i_{\,\,j}\big)\]\,.
\eea
In this way, we can express each component of the form as a linear combination of $(d+1)$ linearly independent functions
\be
\delta w_a(R,\vec{x}_0;z,\vec{x})=\sum_{l=1}^{d+1} \alpha_l(R,\vec{x}_0)F^l_a(z,\vec{x})\,,
\ee
with coefficients $\alpha_l$ given by
\bea\label{coeffs-2}
\{\alpha_1\,, \alpha_2\,, \dots\,, \alpha_{d+1}\}=\left\{\(\frac{R^2-|\vec{x}_0|^2}{4R}\)\,, \frac{x_0^1}{2R}\,, \dots\,,\frac{x_0^{d-1}}{2R} \,, \frac{1}{4R}\right\}\,.
\eea
Note, however, that by choosing a set of parameters $\mathcal{P}$ we can only specify up to $d$ of the above coefficients, while one of them will necessarily be determined in terms of the others. This is in fact not a problem. If we consider the set of forms $\delta \bm w^{k}$ and repeat the above analysis, we find now that
\be\label{sumkl}
\delta w^{k}_a(R^{k},\vec{x}_0^{\,k};z,\vec{x})=\sum_{l=1}^{d+1} \alpha^k_l(R^{k},\vec{x}_0^{\,k})F^l_a(z,\vec{x})\,,
\ee
where $k\in\{1,\ldots,\bar{n}\}$. We note that the number of coefficients $\alpha^k_l$ that we can fix for each $k$ is larger than the total number forms that we have at our disposal, i.e., $d>\left\lceil\frac{d-1}{2}\right\rceil$. Therefore, we still have a lot of freedom on the choice of parameters $\mathcal{P}_k$ to be able to make the set $\{\delta w_a^{1}\,,\dots\,, \delta w_a^{\bar{n}} \}$ linearly independent. One way to achieve this is by focusing only on a subspace of forms
obtained by varying $\bar{n}$ out of the $d$ free coefficients of each $\delta w_a^{k}$, which we denote as $\beta^k_l$, while keeping the rest fixed. More explicitly, we can split the sum in (\ref{sumkl}) as
\be
\delta w^{(k)}_a=\sum_{l}\beta^k_l\,F^l_a\,+\sum_{l}
\tilde{\beta}^k_l\,F^l_a\,,
\ee
where $\beta^k_l$ is now a $\bar{n}\times \bar{n}$ matrix, with the coefficients that we vary, and $\tilde{\beta}^k_l$ denote the ones that we keep fixed. The condition for the above linear independence is then given by the non-vanishing of the determinant of the matrix $\beta^{k}_{l}$. For instance, we could take
\bea\label{cut-alpha}
\beta^{k}_{l}= \left\{\frac{(x_0^{1})^k}{2R^{k}}, \ldots,\frac{(x^{\bar{n}-1}_0)^k}{2R^{k}} , \frac{1}{4R^{k}}\right\}\,,
\eea
and vary the parameters $\{R^k,(x_0^{i})^k\}$ for $i\in\{1,\ldots,\bar{n}-1\}$ such that $\text{det}(\beta^{k}_{l})\neq0$ while keeping
$(x_0^{i})^k$ fixed for $i\in\{\bar{n},\ldots,d-1\}$. More generally, notice that there can be many other possible choices. One one hand, the choice of the subset $\beta^{k}_{l}\subset\alpha^{k}_{l}$ can be arbitrary and, on the other hand, the remaining free parameters can be random; they must not necesarily be kept fixed.

\subsubsection*{General time slices: recovering the full bulk metric}

Let us now go back to the problem of recovering the full bulk metric $\delta g_{\mu\nu}$. First, notice that when we picked a constant-$t$ slice $\Sigma$,
we were able to recover the components of the metric tangent to it, namely, $\delta g_{ij}$. This means that we still need
to find the time components, $\delta g_{tt}$ and $\delta g_{ti}$, in order to complete the reconstruction problem. Below, we present
a simple algorithm to recover these extra metric components.

From the bulk point of view it is easy to see that $\delta g_{tt}$ and $\delta g_{ti}$ are, in fact, constrained from the equations of motion (\ref{perturbedeqns}). Specifically,
they can be determined from the $zz$ and $z\mu$ components of Einstein's equations,
\be
\delta g^{\mu}_{\,\,\,\mu}=0\,, \qquad \partial_\mu \delta g^{\mu\nu}=0\,.
\ee
These equations imply that $\delta g_{tt}$ must equal the spatial trace $\delta g_{tt}=\delta g^i_{\,\,i}$, while $\delta g_{ti}$ can be determined from a first order equation $\partial_t \delta g_{ti} = \partial_j \delta g_{ij}$. This can be easily implemented in practice. However, the problem is that these particular bulk equations of motion are not known from the CFT perspective, so their use cannot be justified. Indeed, once the surface $\Sigma$ is chosen to be a constant-$t$ slice, the closedness condition $d \delta\bm w=0$ \emph{only} encodes the $tt$ component of the Einstein's equations, as explained at the end of Section \ref{sec:IWgeneral}.

One simple solution to this problem, is to pick a more general time slice $\Sigma'$ and repeat the reconstruction analysis outlined above. For simplicity, we will pick here a boosted slice parametrized by coordinates $(t',x_i',z')$, but a similar analysis can be implemented from more general choices of $\Sigma'$. We will denote the boosted coordinate $x_i=x$ and label the coordinates orthogonal to it as $x_j=y_j$ for $j\neq i$. With this notation, a boost with rapidity $\beta$ (which we can take to be arbitray) is given by the bulk transformations
\bea
t&=&t' \cosh\beta +x' \sinh \beta\,,\label{boost1}\\
x&=&x' \cosh\beta +t' \sinh \beta\,,\\
y_i&=&y_j'\,,\\
z&=&z'\,.\label{boost4}
\eea
We can now perform the reconstruction analysis in this new slice, and recover the components $\delta g_{i'\!j'}$ on $\Sigma'$. Indeed, a quick calculation shows that the components that we can recover are
\bea
\delta g_{x'x'}|_{\beta}&=& \sinh^2\beta \,\delta g_{tt}+ 2 \sinh \beta \cosh \beta\, \delta g_{tx} + \cosh^2\beta \, \delta g_{xx}\,,\\
\delta g_{x'y_i'}|_{\beta}&=& \sinh\beta \,\delta g_{ty_i}+ \cosh \beta\, \delta g_{xy_i}\,,\\
\delta g_{y_i'y_j'}|_{\beta}&=& \delta g_{y_iy_j}\,.
\eea
If this analysis is done for at least a few values of the rapidity $\beta_1\neq\beta_2\neq0$, then it is clear that we will have enough information to recover the extra components $\delta g_{tt}$ and $\delta g_{tx}$ from linear combinations of $\delta g_{i'\!j'}|_{\beta_i}$.

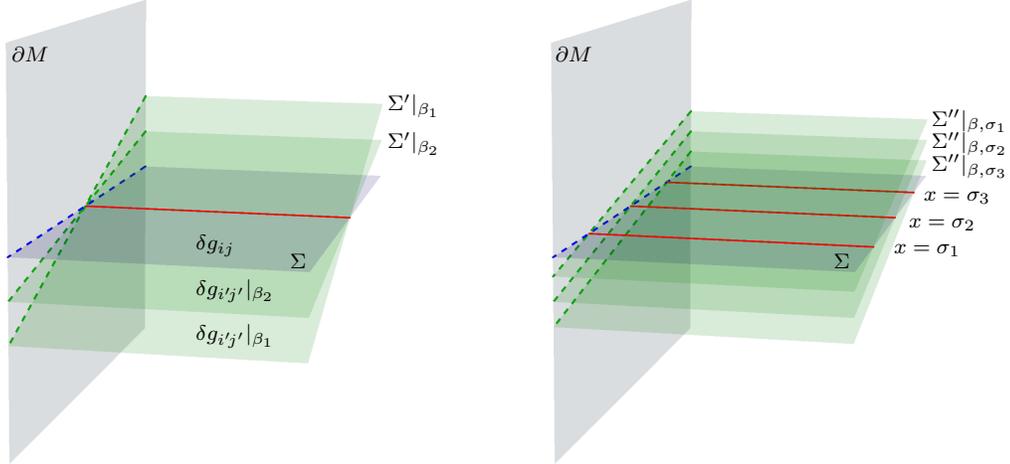
\begin{figure}[t!]
\begin{center}
  \includegraphics[width=5cm]{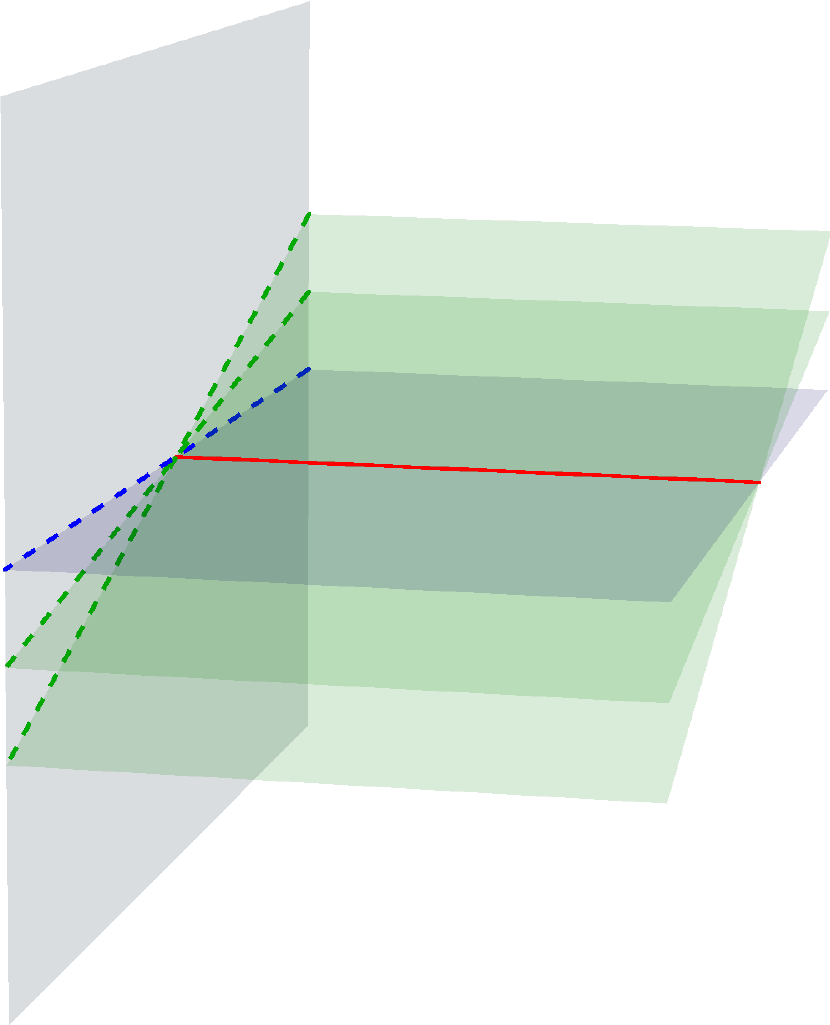}
  \hspace*{2cm}
  \includegraphics[width=5cm]{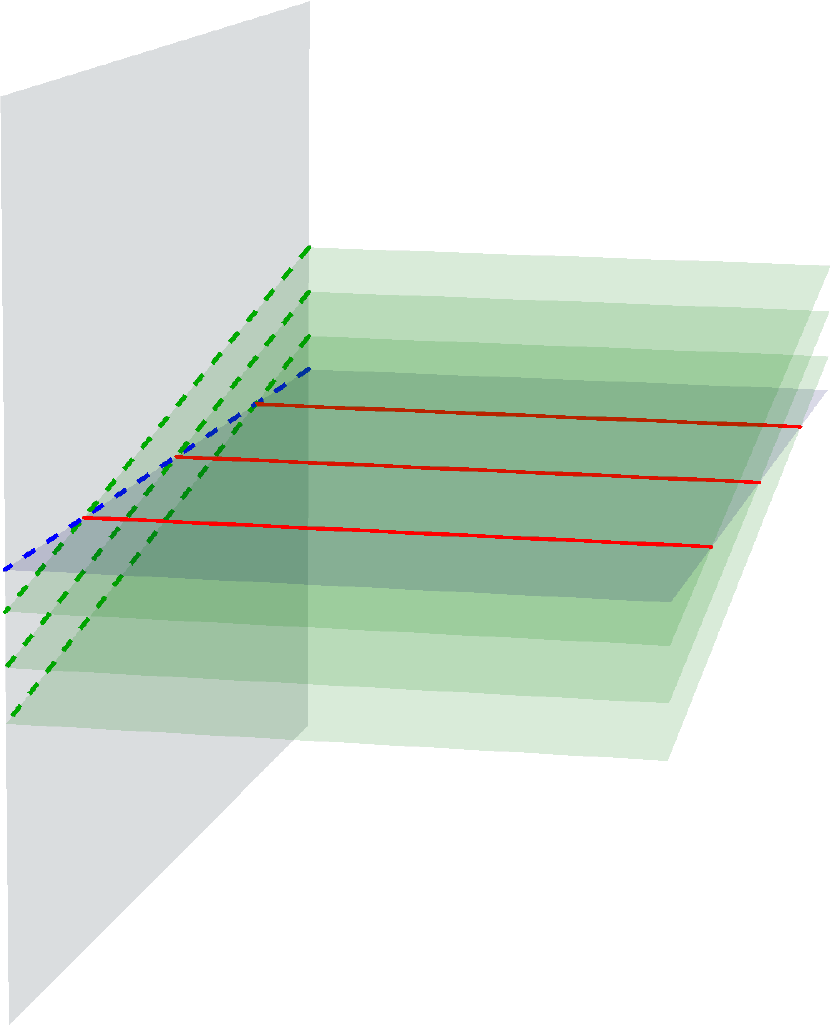}
  \setlength{\unitlength}{1cm}
\begin{picture}(0,0)
\put(-12.43,5.35){\scriptsize$\partial M$}
\put(-8.73,2.6){\scriptsize$\Sigma$}
\put(-9.98,2.85){\scriptsize$\delta g_{ij}$}
\put(-9.98,2.25){\scriptsize$\delta g_{i'\!j'}|_{\beta_2}$}
\put(-9.98,1.65){\scriptsize$\delta g_{i'\!j'}|_{\beta_1}$}
\put(-7.43,4.7){\scriptsize$\Sigma'|_{\beta_1}$}
\put(-7.43,4.2){\scriptsize$\Sigma'|_{\beta_2}$}
\put(-5.2,5.35){\scriptsize$\partial M$}
\put(-1.5,2.6){\scriptsize$\Sigma$}
\put(-0.2,4.5){\scriptsize$\Sigma''|_{\beta,\sigma_1}$}
\put(-0.2,4.2){\scriptsize$\Sigma''|_{\beta,\sigma_2}$}
\put(-0.2,3.9){\scriptsize$\Sigma''|_{\beta,\sigma_3}$}
\put(-0.3,3.5){\scriptsize$x=\sigma_3$}
\put(-0.5,3.15){\scriptsize$x=\sigma_2$}
\put(-0.7,2.8){\scriptsize$x=\sigma_1$}
\end{picture}
\end{center}
\vspace{-0.2cm}
\caption{\small Algorithm to reconstruct the full bulk metric, including its time components $\delta g_{tt}$ and $\delta g_{ti}$. In the left panel we show the components of the metric that can be recovered on a generic boosted slice $\Sigma'|_{\beta}$. Combining the data obtained from multiple boosted slices it is possible to recover the full metric on the locus of points in the intersection between the original slice $\Sigma$ and the new slices $\Sigma'$, depicted in red. In the right panel we show new slices $\Sigma''|_{\beta,\sigma}$ that are obtained by a simple translation of the original boosted slices $\Sigma'$. The extra data obtained from the inversion problem in these new slices is sufficient to recover all components of the metric in the full slice $\Sigma$.}
\label{fig:timemetric}
\end{figure}

This completes the reconstruction problem for the locus of all points in the intersection between the original slice $\Sigma$ and the new slices $\Sigma'$, as shown in left panel of Figure \ref{fig:timemetric}. In particular, from the transformations (\ref{boost1})-(\ref{boost4}), it follows that the constant-$t'$
slices parametrized by $\beta$ intersect the surface $\Sigma$ at the line
\be
x=0\,,
\ee
depicted in red. If we want to reconstruct other points of $\Sigma$, then, we can generalize the boost transformations (\ref{boost1})-(\ref{boost4}) to include a translation in $x$, so that
\bea
t&=&t'' \cosh\beta +x'' \sinh \beta\,,\label{boost1b}\\
x&=&x'' \cosh\beta +t'' \sinh \beta+\sigma\,,\\
y_i&=&y_j''\,,\\
z&=&z''\,.\label{boost4b}
\eea
The new slices $\Sigma''$, depicted in the right panel of Figure \ref{fig:timemetric}, intersect the original slice $\Sigma$ at
\be
x=\sigma\,,
\ee
so it is clear that, if we perform the reconstruction problem for a general slice $\Sigma''$ with arbitrary $\beta$ and $\sigma$ we would have enough
information to reconstruct the full metric in $\Sigma$.

Finally, note that the above algorithm has a trivial extension to generic choices of $\Sigma$. Thus, picking a family of slices $\Sigma$ that foliates the full manifold $M$, and repeating the same analysis for each $\Sigma$ gives a complete solution to the reconstruction problem in $M$.

\subsubsection{Going beyond linear order\label{beyond}}

In the past section we have shown that the problem of metric reconstruction can be carried out explicitly at linear order for the case of perturbative excited states. This was accomplished using the canonical bit thread construction based on differential forms. But, can this methodology be generalized to the non-linear regime?

To answer this question, we will start with a brief review of our findings and then discuss how the different aspects of our proposal can be generalized. To begin with
the reconstruction problem, we assumed knowledge of a set of canonical forms that codify the entanglement pattern of subregions in the dual CFT. We recall that these canonical forms $\delta \bm w$ can be uniquely specified from CFT data. Specifically, they are completely fixed by the boundary condition on $\partial M$ (\ref{bcIyerWald}), which is given in terms of the one-point function of the CFT stress-energy tensor $\langle T_{\mu\nu}\rangle$, and the requirement of bulk locality. In the presence of a metric, we found that one of these forms specifies a covector field that can be schematically written as
\be\label{metricreconst}
(\star \delta \bm w)_{a}=\mathcal{F}_a^{bc}\,\delta g_{bc}\,,
\ee
where $\mathcal{F}_a^{bc}$ is a specific linear differential operator. In low-dimensional cases, the equations resulting from a single form provide enough data to invert the problem, so that
\be
\delta g_{ab}=[\mathcal{F}^{-1}]^c_{ab}\,(\star \delta \bm w)_{c}\,.
\ee
This kind of inversion works for $d=2$ and $d=3$, i.e., in AdS$_{3}$ and AdS$_{4}$, respectively. For higher dimensional cases the problem becomes underdetermined, but it can be easily generalized by starting from a set of differential forms $\delta \bm W=\{\delta \bm w^1,\ldots,\delta \bm w^n\}$, that encode the entanglement pattern of a \emph{family} of subregions in the CFT. In this case, we find that
\be
(\star \delta \bm w^i)_{a}=(\mathcal{F}^i)_a^{bc}\,\delta g_{bc}\,,
\ee
so, for large enough $n$ one can always invert the system as
\be
\delta g_{ab}=[\mathcal{F}^{-1}_i]^c_{ab}\,(\star \delta \bm w^i)_{c}\,.
\ee
The optimal value of $n$ depends on the number of dimensions, and is given by (\ref{minimaln}). Of course, the larger
the value of $n$, the more information that we have at our disposal, and the easier the inversion problem becomes. In fact, in the limit of $n\to\infty$, or
when the set of differential forms is dense enough, there is even enough maneuver to turn the inversion problem into a simple algebraic system of equations as shown explicitly in section (\ref{explicit-reconstruction}).

Let us now comment on what to expect in the non-linear regime. To start with, we can treat the problem perturbatively but extending the above results to higher orders in the perturbation.
In this case, the perturbation of the metric can be given as an expansion\footnote{As shown in \cite{Lashkari:2015hha}, when going to higher orders in $\lambda$ it is convenient to work in the so-called Hollands-Wald gauge \cite{Hollands:2012sf} so that the coordinate location of $\gamma_A$ is fixed and $\mathcal{L}_{\xi_A}(g^\lambda_{\mu\nu})|_{\gamma_A}=0$. In this gauge the argument presented in Section \ref{pertstatesBT} about the choice of $\Sigma$ can be extended to higher orders in $\lambda$.}
\be\label{eq:metrichigher}
g^\lambda_{\mu\nu}= g_{\mu\nu}+\lambda\delta g_{\mu\nu}^\lambda\,,\qquad \delta g_{\mu\nu}^\lambda\equiv\delta^{(1)} g_{\mu\nu}+\lambda\delta^{(2)} g_{\mu\nu}+\lambda^2\delta^{(3)} g_{\mu\nu}+\cdots\,,
\ee
and, similarly, any solution to the max-flow problem
\be\label{eq:formhigher}
\bm w_\lambda=\bm w+\lambda\delta \bm w_\lambda\,,\qquad \delta \bm w_\lambda\equiv \delta^{(1)} \bm w+ \lambda \delta^{(2)} \bm w+ \lambda^2 \delta^{(3)} \bm w+\cdots\,.
\ee
On general grounds we expect that, in the presence of the metric (\ref{eq:metrichigher}), the change in the form $\delta \bm w_\lambda$ should follow an equation similar to (\ref{metricreconst}), i.e.,
\be
(\star \delta \bm w_\lambda)_{a}=\tilde{\mathcal{F}}_a^{bc}\,\delta g_{bc}^\lambda\,,
\ee
but now with $\tilde{\mathcal{F}}_a^{bc}$ being a higher order differential operator. For example, at second order in $\lambda$ we expect a second order differential operator, and similarly for higher orders. This introduces the standard non-uniqueness problem for the inversion of a non-linear operator. However, this issue can be circumvented by solving the reconstruction problem recursively in $\lambda$. To see this, notice that the different terms in (\ref{eq:formhigher}) should depend on
the different metric perturbations and their derivatives as follows:
\be\label{eq:formcomponents}
\delta^{(i)}\bm w=\delta^{(i)}\bm w(\delta^{(j)}g_{\mu\nu},\nabla^{k} \delta^{(j)} g_{\mu\nu})\,,
\ee
for $j\in\{1,\ldots,i\}$ and $k\in\{1,\ldots,1+i-j\}$. In other words, for a given value of $i$, in the right-hand side of (\ref{eq:formcomponents}) we expect up to $i^{\text{th}}$ derivatives of $\delta^{(1)} g_{\mu\nu}$ but only $1^{\text{st}}$ derivatives of $\delta^{(i)} g_{\mu\nu}$. Thus, if we solve for the metric at first order in $\lambda$, then we can reformulate the problem of bulk reconstruction at second order as a linear problem above the $g_{\mu\nu}+\lambda\delta^{(1)} g_{\mu\nu}$ solution. This also generalizes to higher orders in $\lambda$, so that the inversion problem at a given order can also be cast as a linear problem above one lower order.

We should also comment on how the boundary condition (\ref{bcIyerWald}) generalizes to higher orders in $\lambda$. At linear level, we saw that it is fully specified by the one-point function of the CFT stress-energy tensor $\langle T_{\mu \nu}\rangle$. However, at higher orders we would need to specify further data, e.g. \cite{Beach:2016ocq}. A useful case study would be to consider the reconstruction problem at second order in $\lambda$, which was already worked out in \cite{Faulkner:2017tkh} in the framework of extremal surfaces. This paper generalized the Iyer-Wald construction, and thus \cite{Faulkner:2013ica}, to include second order variations in the metric, hence their construction can be used to obtain canonical thread configurations at second order in $\lambda$. More specifically, \cite{Faulkner:2017tkh} focused on a class of CFT states which are expected to have a classical gravity description, and are defined by adding sources for primary operators to the Euclidean path integral defining the vacuum state \cite{Botta-Cantcheff:2015sav,Christodoulou:2016nej,Marolf:2017kvq},
\be\label{cftState}
|\psi_\lambda\rangle=T e^{-\int_{-\infty}^0dt_Ed^{d-1}x\phi^{(0)}_\alpha\mathcal{O}_\alpha}|0\rangle
\ee
Among other things, their calculation led to a closed expression for the entanglement entropy in these CFT states at second order in the sources:\footnote{In  their  calculation,  they  expressed  the  second  order  terms  in  terms  of  the  first order  sources. After that, they related the sources at first order to the one-point functions at first order via a map that depends on the CFT two-point functions (which are fixed by conformal symmetry).}
\bea\label{bc:secondorder}
\!\!\!\!\!\!\!\!\!\!\!\!S_A(\langle T\rangle,\langle\mathcal{O}\rangle)&=&a^{*}S_A^{(0)}+\lambda\int K^{(1)}_A(x)\langle T(x)\rangle+\lambda^2\int\!\!\!\int K^{(2)}_A(x_1,x_2)\langle\mathcal{O}_\alpha(x_1)\rangle\langle\mathcal{O}_\alpha(x_2)\rangle\nonumber\\
&&+\frac{\lambda^2}{C_T}\int\!\!\!\int \tilde{K}^{(2)}_A(x_1,x_2)\langle T(x_1) \rangle\langle T(x_2)\rangle\,,
\eea
where $K^{(1)}_A(x)$, $K^{(2)}_A(x_1,x_2)$ and $\tilde{K}^{(2)}_A(x_1,x_2)$ are some specific kernels. A few comments are in order. First, notice that at this order the entanglement entropy depends on  the specific CFT only through the central charges $a^{*}$ and $C_T$. In fact, for theories that are dual to Einstein gravity, they must be proportional to each other, i.e., $a^{*}\propto C_T$, so the answer depends only on one CFT parameter. Second, (\ref{bc:secondorder}) correctly encodes the first-law of entanglement $\delta S_A=\langle H_A\rangle$ at linear order in $\lambda$ but, in addition, it also contains information about the relative entropy in the excited state $\delta^{(2)}S(\rho_A||\rho_A^{(0)})$, to second order in $\lambda$. Finally, note that equation (\ref{bc:secondorder}) can now be used to specify a canonical boundary condition for $\delta \bm w_\lambda$ in $\partial M$,
\bea
\delta \bm w_\lambda(x)|_{\partial M}&=&4G_N\bigg[K^{(1)}_A(x)\langle T(x)\rangle+\lambda\int K^{(2)}_A(x,x')\langle\mathcal{O}_\alpha(x)\rangle\langle\mathcal{O}_\alpha(x')\rangle\nonumber\\
&&\qquad+\frac{\lambda}{C_T}\int \tilde{K}^{(2)}_A(x,x')\langle T(x) \rangle\langle T(x')\rangle\bigg]\bar{{ \bm \epsilon}}\,,
\eea
and so, it should suffice to uniquely specify the full $(d-1)-$form on $M$ by further requiring bulk locality. This means that at second order in $\lambda$, we need also the one-point functions of all primary operators  $\langle\mathcal{O}_\alpha\rangle$, in addition to stress-energy tensor. From the bulk perspective, \cite{Faulkner:2017tkh} showed that the closedness of the Iyer-Wald form $\bm \chi$, related to $\delta \bm w_\lambda$ through (\ref{w-chi}), encodes now the following data: at first order one recovers (\ref{closeness}), which specialized to an arbitrary slice $\Sigma$ leads to the linearized Einstein's equations $E_{ab}^{(1)}=0$. At second order, one obtains
\be
E_{ab}^{(2)}\equiv (E_{ab}^{(2)})_{\text{grav}}-\frac{1}{2}T_{ab}^{(2)}=0\,,
\ee
where $(E_{ab}^{(2)})_{\text{grav}}$ is the second order Einstein tensor and $T_{ab}^{(2)}$ is the second order stress-energy tensor associated with the bulk fields $\phi_\alpha$ dual to the CFT operators $\mathcal{O}_\alpha$. Thus, \emph{for theories with $a^{*}=C_T$, a CFT state of the form (\ref{cftState}) secretly encodes Einstein's equations (at least up to second order in the perturbation) with matter determined by the CFT one-point functions.} Therefore, on general grounds we can expect that the inversion problem using the canonical bit thread prescription to be well defined at second order in $\lambda$.

One can also go to higher orders in perturbation theory, however, it is clear that one would ultimately need infinitely more CFT data in the full non-linear regime, rendering the problem untractable. To see this, notice that to the $i^{\text{th}}$ order  in  the  perturbation,  the  relative  entropy  $\delta^{(i)}S(\rho_A||\rho_A^{(0)})$ will  generically involve  $i$-point  functions of the primary operators $\mathcal{O}_\alpha$, so at the full non-linear level we would need to specify \emph{all} correlation functions of the theory. Here we have two options to proceed. One could try to pursue the reconstruction problem in a theory with a \emph{know} gravity dual, e.g., $\mathcal{N}=4$ SYM, and with some hard work, one should be able to recover the metric order by order in the perturbation. Alternatively, one could start with a generic CFT and try to constrain the structure of the CFT $i$-point functions such that the calculations match with the gravity side. Indeed, such constraints must start appearing for $i\geq4$, since these correlation functions are not fixed by conformal invariance.

Another possibility would be to consider the full reconstruction problem without resorting to perturbation theory.\footnote{A version of this idea appeared originally in \cite{Freedman:2016zud}.} For example, given a CFT with a known holographic dual, one can start by computing entanglement contours \cite{Chen_2014,Kudler-Flam:2019oru} for several regions $A_i$ in a state of the form (\ref{cftState}). These contours can then be used as boundary conditions in $\partial M$ for a set of closed bulk forms $\bm w^i$ that encode the entanglement pattern of the individual regions. Via the closedness condition, $d\bm w^i=0$, and assuming further input such as bulk locality, one should then be able to reconstruct particular realizations of the set of forms $\bm w^i$ on $M$ that solve the max flow problem in the bulk. If this set is sufficiently dense, then, one could set up the problem of metric reconstruction as a particular optimization problem. More specifically, given a set of such $(d-1)-$ forms $\bm W=\{ \bm w^i\}$, the metric $g_{ab}$ should emerge as the minimal positive definite symmetric $(0,2)$ tensor for which the norm bound constraint
\bea\label{w-norm2}
\frac{1}{(d-1)!}g^{a_1 b_1}\cdots g^{a_{d-1} b_{d-1}} w_{a_1 \cdots a_{d-1}}w_{b_1 \cdots b_{d-1}} \leq 1\,.
\eea
holds for all the elements $\bm w^i$ of the fundamental set $\bm W$. It would be interesting to develop a more precise algorithm based on this optimization problem  and understand how the Einstein's equations would emerge upon its implementation.

\section{Conclusions and outlook}

In this paper we developed a new framework for metric reconstruction based on the bit threads reformulation of entanglement entropy. Our work can be divided roughly into two main parts. In Section \ref{PBT} we explored simple constructions of perturbative thread configurations based on the general methods originally developed in \cite{Agon:2018lwq} but expanded in this paper in various ways. We explored in detail two particular constructions, one that starts by specifying the class of integral curves and a second one that assumes a specific family of level set surfaces. We showed that both methods are efficient and can be easily implemented for the case of perturbative excited states, as we discuss in detail in the concrete example of a local quench in Appendix \ref{app:examples}. However, we realized that both constructions encode the information about the bulk metric in a highly nonlocal way. This implies that these realizations are not particularly useful to tackle the question of metric reconstruction and highlights the necessity of reformulating bit threads in a language that makes background independence manifest.

Motivated by the above results, we started Section \ref{SecForms} by reformulating the bit threads framework in terms of differential forms. We gave general formulas
that translate the relevant equations of the standard description in terms of flows into this new language and studied in detail the case of perturbative excited states. We pointed out that the Iyer-Wald formalism provides us with a canonical choice for the perturbed thread configuration that makes explicit use of bulk locality. More explicitly, we showed that the Iyer-Wald construction yields a particular solution to the max flow problem in the bulk that can be uniquely determined from CFT data, and encodes the Einstein's equations in the bulk through its closedness condition. Assuming that a set of such forms is given, we then showed that the problem of metric reconstruction is equivalent to the inversion of a particular differential operator. We gave explicit inversion formulas for the case of 2d and 3d CFTs, and argued that the problem is also well possed in higher dimensions. Finally, we discussed the generalization of our results to higher orders in the perturbation and its relation to the full Einstein's equations.

There are some open questions related to our work which we think are worth exploring:
\begin{itemize}
  \item \vspace{-0.2cm}\emph{Explicit inversion at higher orders:} In Section \ref{beyond} we have sketched out how to generalize the metric inversion
        problem to higher orders in $\lambda$. We believe that this would be fairly straightforward to second order in the perturbation if one uses \cite{Faulkner:2017tkh} as a starting point, while it would be illuminating to have explicit inversion formulas for the differential operator at this order. Generalizing this story to higher orders should be possible but may require some extra work. To some extent, this study would be even more rewarding since it could yield non-trivial constraints on the space of theories with classical gravity duals, specifically on the structure of their correlation functions. It would also be interesting to work out a more precise algorithm for metric reconstruction at the full non-linear level following the discussion at the end of Section \ref{beyond} and, in particular, understand how the Einstein's equations would emerge in this context.
  \item \vspace{-0.2cm}\emph{More general states and entangling regions:} In our work we have considered perturbations around the vacuum state and focused on spherical regions in the boundary theory. While it is true that in this setting one could in principle recover Einstein's equations systematically (order by order) and hence the bulk geometry, one could relax these two points, with the latter one being arguably the easiest of the two (at least for regions with local modular Hamiltonians). We note that substantial progress on generalizing the former using the Iyer-Wald formalism appeared in \cite{Lewkowycz:2018sgn}. This paper also argued that doing the linear analysis for arbitrary states and shapes of the entangling region is sufficient to capture the full non-linear Einstein's equations in the bulk. It would be interesting to relax these conditions in our method of bulk reconstruction using bit threads, and try to make these statements a bit more precise in our context. It would also be interesting to explicitly start with a state for which the bulk geometry has entanglement shadows, and understand how our approach encodes information about these regions.
  \item \vspace{-0.2cm}\emph{Covariant bulk reconstruction:} In this work we have incorporated time-dependence by combining the maximin prescription introduced in \cite{Wall:2012uf} and the non-covariant formulation of bit threads \cite{Freedman:2016zud}. As explained in Section \ref{pertstatesBT}, this was possible in virtue of various crucial simplifications of the perturbative setting that we consider. However, it should be possible to pose the same question in the fully covariant formulation of bit threads
      \cite{Headrick:toappear}. We believe that in this case, the full modular flow in the bulk should play a role, and could serve as a guide for constructing canonical bit thread configurations in other special cases. Related to this, it would be interesting to ask the question about bulk reconstruction in time-dependent situations that are not easily accessible to the perturbative setting we consider here, e.g., completely far-of-equilibrium states that could lead to black hole formation in the bulk.
  \item \vspace{-0.2cm}\emph{Higher derivative theories:} Finally, we believe it would be worthwhile to explicitly generalize our method of bulk reconstruction to the case of higher derivative theories in the bulk. We point out that the program of \emph{gravitation from entanglement} using extremal surfaces has been worked out in detail for these theories in \cite{Haehl:2017sot}, using the Iyer-Wald formalism both to first and second order in the state deformations around the vacuum. Likewise, the bit thread reformulation of entanglement entropy has already been generalized to the case of higher derivative gravities in \cite{Harper:2018sdd}, incorporating corrections to the local norm bound that depend on the specific theory. It would be interesting to see how our formulas are corrected if we turn on these extra gravity couplings.
\end{itemize}
\vspace{-0.2cm}We hope to come back to some of these points in the near future.

\section*{Acknowledgements}

It is a pleasure to thank Ning Bao, Horacio Casini, Jan de Boer, Matthew Headrick, Martin Ro\v{c}ek, Andrea Russo, Andrew Svesko and Zach Weller-Davies for useful discussions and comments on the manuscript. CAA is supported by the National Science Foundation (NSF) Grant No. PHY-1915093. EC is supported by the NSF Grants No. PHY-1620610 and No. PHY-1820712. JFP is supported by the Simons Foundation through \emph{It from Qubit: Simons Collaboration on Quantum Fields, Gravity, and Information}.

\appendix

\section{Explicit example: local quench state}\label{app:examples}
In this section we will work in detail the concrete example of a local quench using the three methods presented in this paper: the geodesic construction, the level set construction and the canonical construction based on Iyer-Wald. For simplicity, we will focus on the $d=2$ case but a similar analysis can be extended to higher dimensions as well.

Let us consider a three dimensional asymptotically AdS geometry,
\be
ds^2=\frac{1}{z^2}\[(\eta_{\mu\nu}+z^2 H_{\mu\nu}(x^\sigma,z))dx^\mu dx^\nu+dz^2\]\,.
\ee
In a general dimension, the metric perturbation can be expanded as  $H_{\mu\nu}=\frac{16\pi G_N}{d}\sum z^{2n}T^{(n)}_{\mu\nu}$. However, in $d=2$ all $n\geq1$ terms vanish and we have,
\be
H_{\mu\nu}(x^\sigma,z)={8 \pi G_N}\,T_{\mu\nu}(x^\sigma)\,.
\ee
The stress tensor should be traceless and conserved, therefore, its general form is
\be
T_{\mu\nu}(t,x)=f(t-x)\left(
\begin{array}{cc}
	1 & -1 \\
	-1 & 1 \\
\end{array}
\right)+g(t+x)\left(
\begin{array}{cc}
	1 & 1 \\
	1 & 1 \\
\end{array}
\right)\,.
\ee
For general linear perturbations we can perform a Fourier decomposition, and take $f(t+x)$ and $g(t+x)$ to be an appropriate superposition of plane waves. However, for concreteness we will consider an example that is physically motivated: a local quench that arises by the insertion of a local primary operator.  The stress tensor of a local quench is given by the sum of two shock waves \cite{Nozaki:2013wia,Agon:2020fqs},
\bea\label{eq:quench_tensor}
T_{\mu\nu}(t,x)&=&\frac{\lambda \alpha^3}{\pi} \left[\frac{1}{((t-x)^2+\alpha^2)^2}\left(
\begin{array}{cc}
	1 & -1 \\
	-1 & 1 \\
\end{array}
\right)+\frac{1}{((t+x)^2+\alpha^2)^2}\left(
\begin{array}{cc}
	1 & 1 \\
	1 & 1 \\
\end{array}
\right)\right]\,,\nonumber\\
&=&\frac{2 \alpha ^3 \lambda }{\pi}\left(
\begin{array}{cc}
	\displaystyle\frac{(t^2 + x^2 + \alpha^2)^2 + 4 t^2 x^2}{ \left[(x^2-t^2-\alpha^2)^2+4 \alpha ^2 x^2\right]^2} & \displaystyle\frac{-4 t x (t^2 + x^2 + \alpha^2)}{ \left[(x^2-t^2-\alpha^2)^2+4 \alpha ^2 x^2\right]^2} \\
	\displaystyle\frac{-4 t x (t^2 + x^2 +\alpha^2)}{ \left[(x^2-t^2-\alpha^2)^2+4 \alpha ^2 x^2\right]^2} & \displaystyle\frac{(t^2 + x^2 + \alpha^2)^2 + 4 t^2 x^2}{ \left[(x^2-t^2-\alpha^2)^2+4 \alpha ^2 x^2\right]^2} \\
\end{array}
\right)\,,
\eea
where the (small) parameter $\lambda$ gives the total energy inserted and $\alpha$ acts as a UV regulator.

\begin{figure}[t!]
	\centering
	\subfloat[][]{\includegraphics[width=.5\textwidth
		]{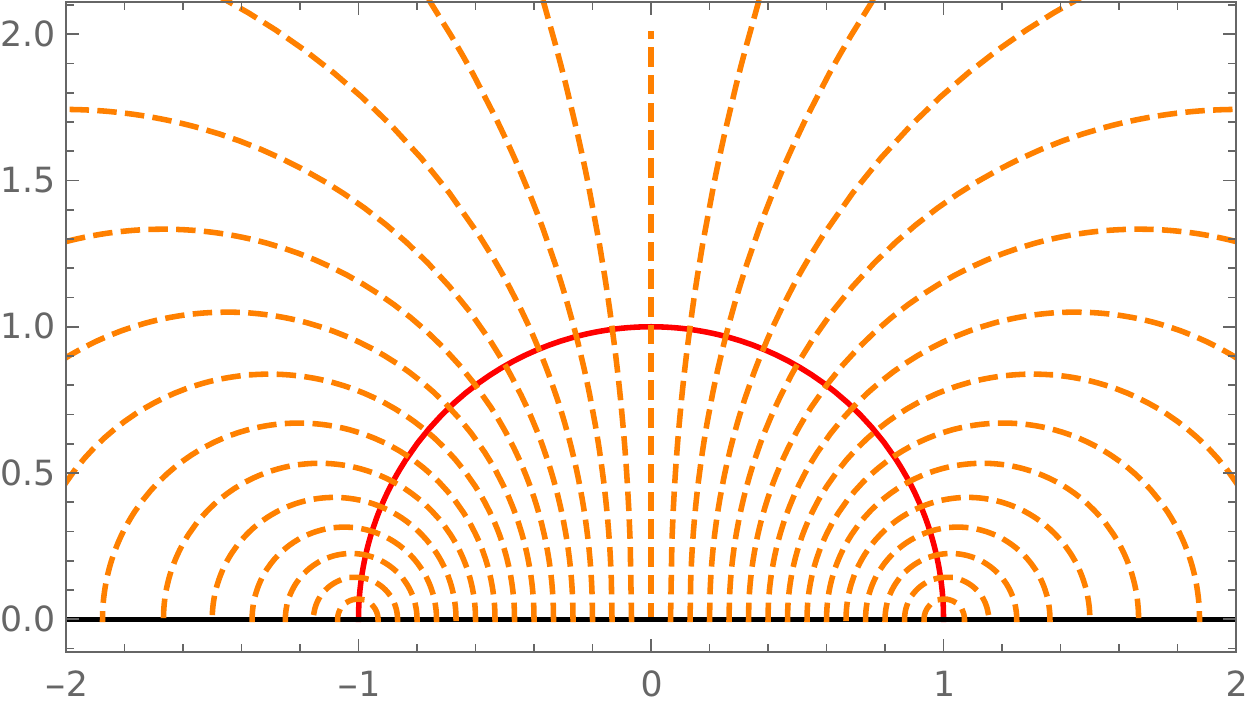}
	}
	\subfloat[][]{	\includegraphics[width=.5\textwidth
		]{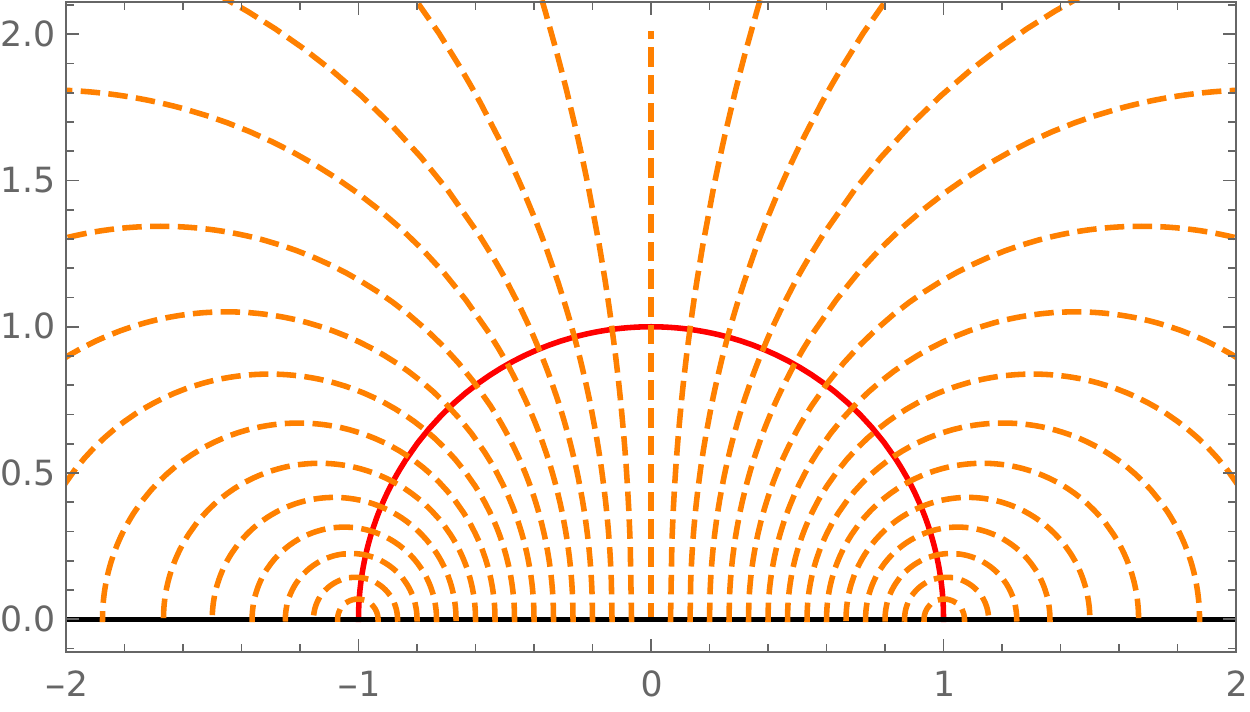}
	}
	
	\subfloat[][]{\includegraphics[width=.5\textwidth
		]{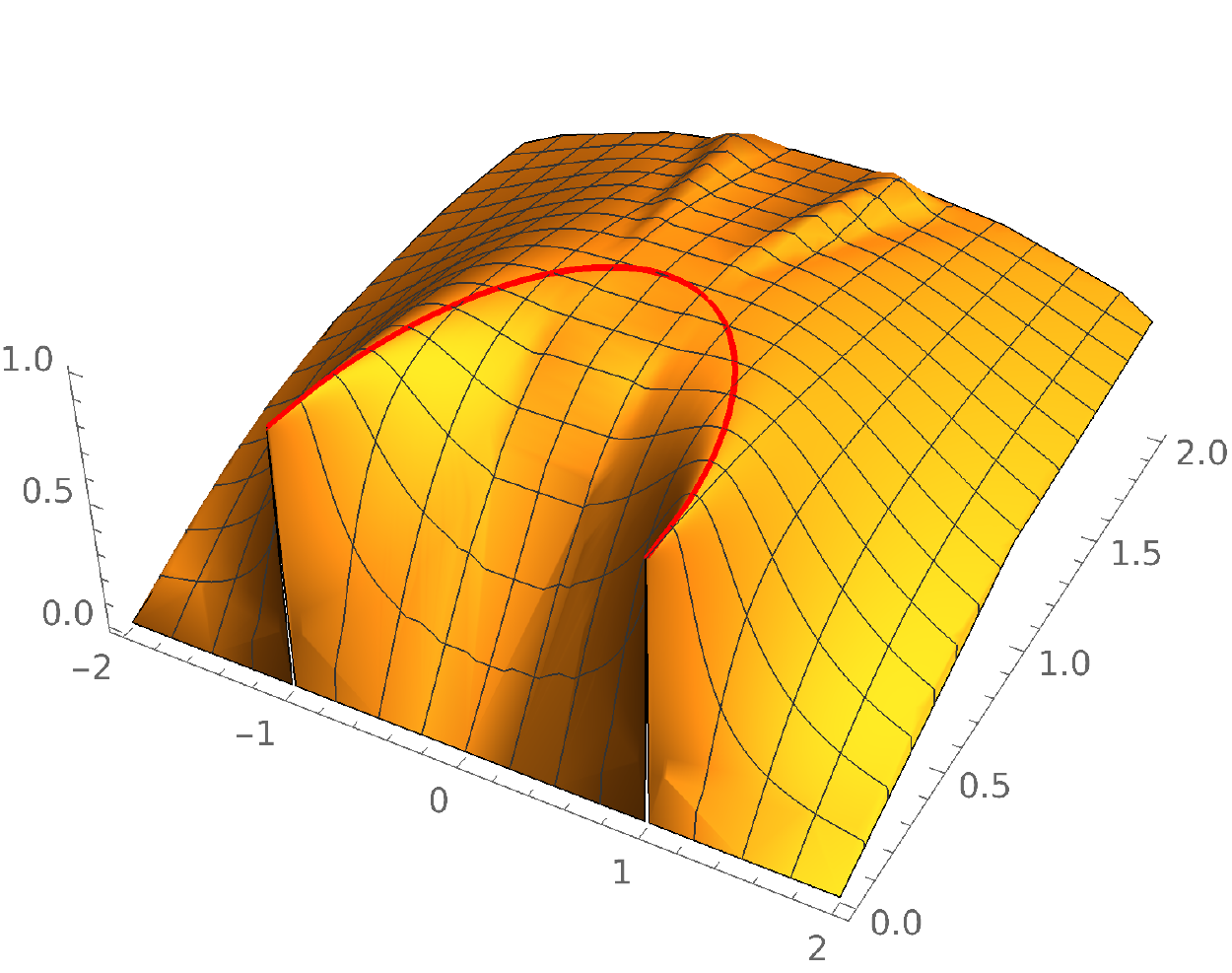}
	}
	\subfloat[][]{	\includegraphics[width=.5\textwidth
		]{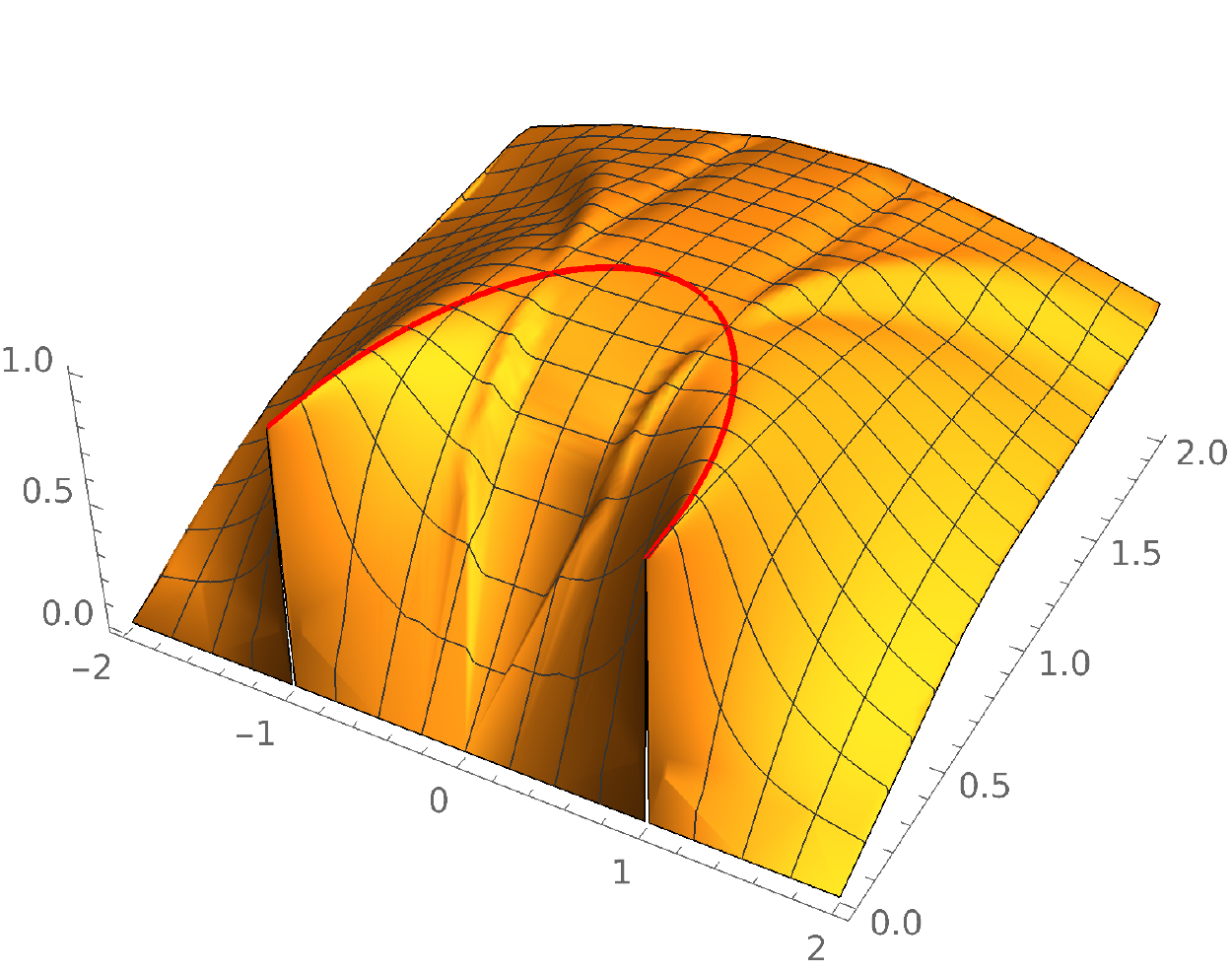}
	}
	\caption{Perturbed vector field obtained using the \emph{unperturbed geodesic} construction. Panels (a) and (b) illustrate the vector lines for  $R=1$,   $\lambda=.1$, $t=.25$ and $t=.5$ respectively. Panels (c) and (d) illustrate the corresponding vector norms. }
	\label{fig_geo}
\end{figure}

As discussed in Section \ref{F1} the unperturbed geodesics provide
a family of integral curves that satisfy the criteria given in \cite{Agon:2018lwq}.
In pure AdS, the geodesics are semicircles anchored at the boundary. These semicircles form a two-parameter family of curves, defined by
\bea\label{geodesic_App}
(x-x_s)^2+z^2=R_s^2\,,
\eea
where $x_s$ is the center of the circle on the $x$-axis and $R_s$ is its radius.
If we denote $(x_m, z_m)$ a point on the minimal surface  $\gamma_A$, we showed in Section \ref{sec:form1-intc} that
\bea
&&R_s(x_m)=\frac{R \sqrt{R^2-x_m^2}}{x_m} \left[1+\frac{\lambda (R^2-x_m^2)^2 H(t,x_m)}{R^2}\right]\,,\label{eq:RsxsApp}\\
&&x_s(x_m)=\frac{R^2}{x_m}\left[1+\frac{\lambda (R^2-x_m^2)^2 H(t,x_m)}{R^2}\right]\,,\label{eq:xsxsApp}
\eea
where $H(t,x)\equiv H_{xx}(t,x)$. Plugging \ref{eq:RsxsApp} and \ref{eq:xsxsApp} in \ref{geodesic_App} we obtain a
family of geodesics orthogonal to $\gamma_A$, parametrized by the point $x_m\in[-R,R]$ on the minimal surface. Likewise, we can use the formula (\ref{magnitudeV}) to obtain the magnitude of the vector field. This calculation can be done following the examples of \cite{Agon:2018lwq}. The final result for the integral curves and magnitude are plotted in Figure \ref{fig_geo}.

\begin{figure}[t!]
	\centering
	\subfloat[][]{\includegraphics[width=.5\textwidth
		]{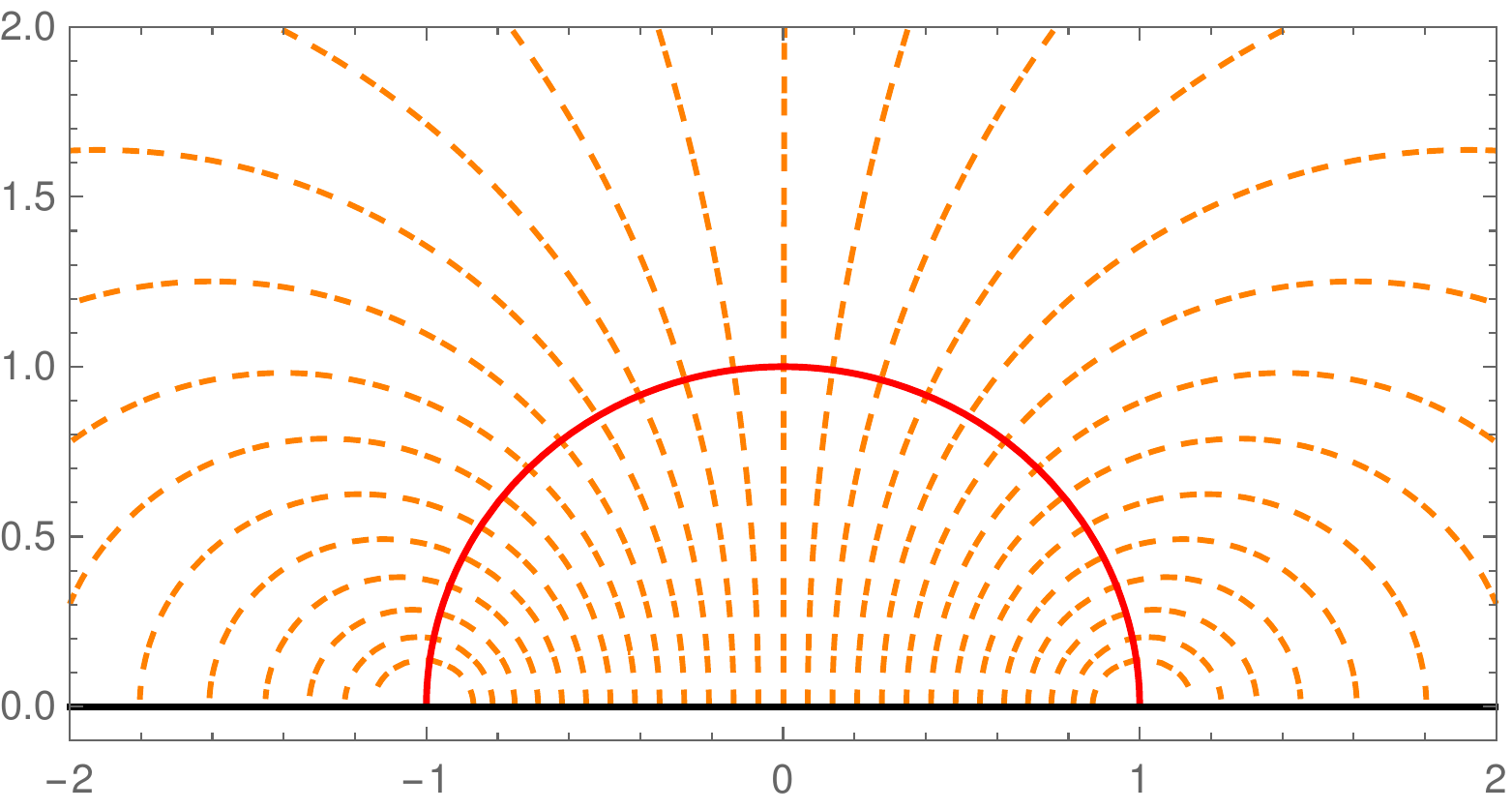}
	}
	\subfloat[][]{	\includegraphics[width=.5\textwidth
		]{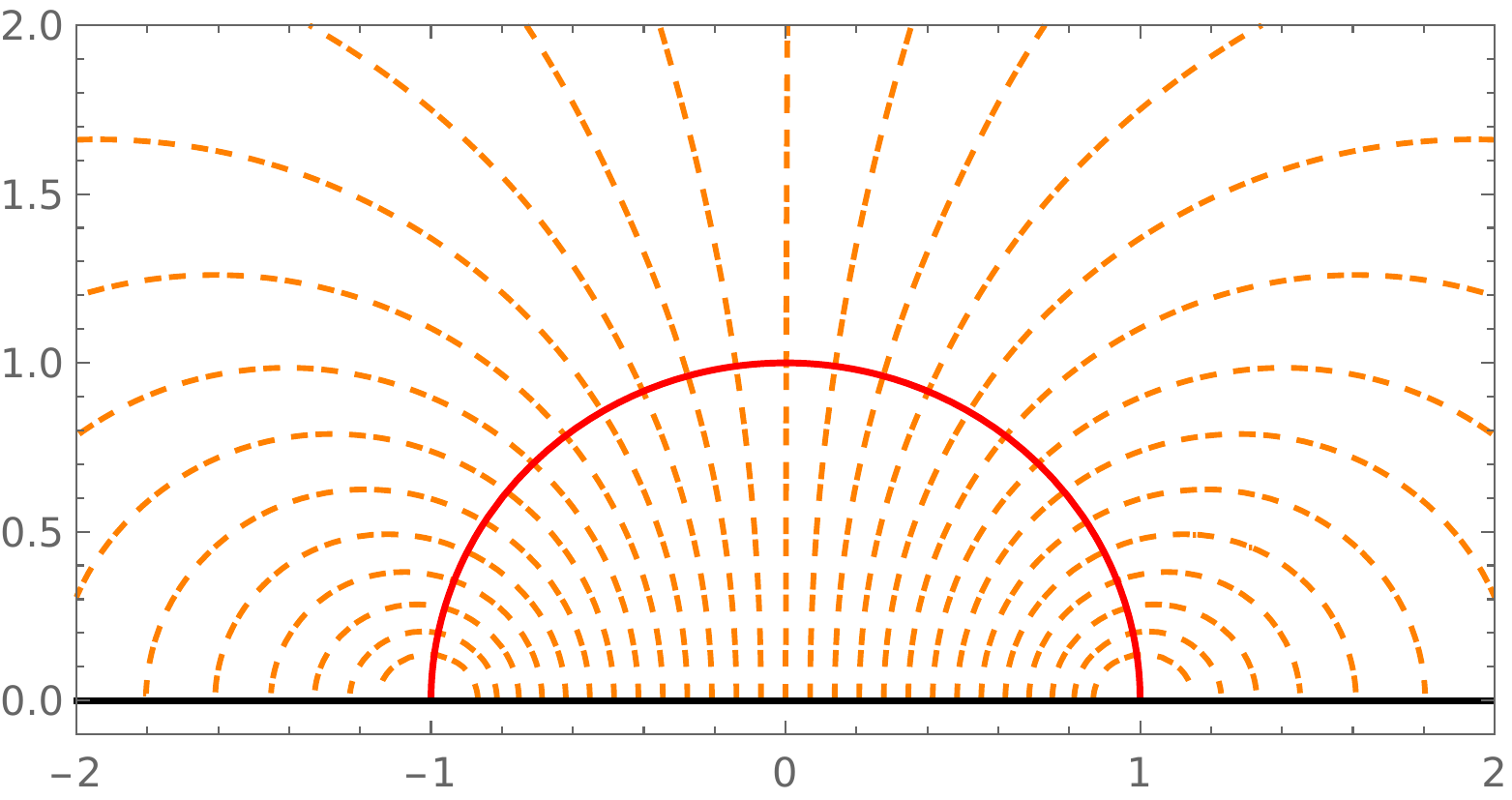}
	}
	
	\subfloat[][]{\includegraphics[width=.5\textwidth
		]{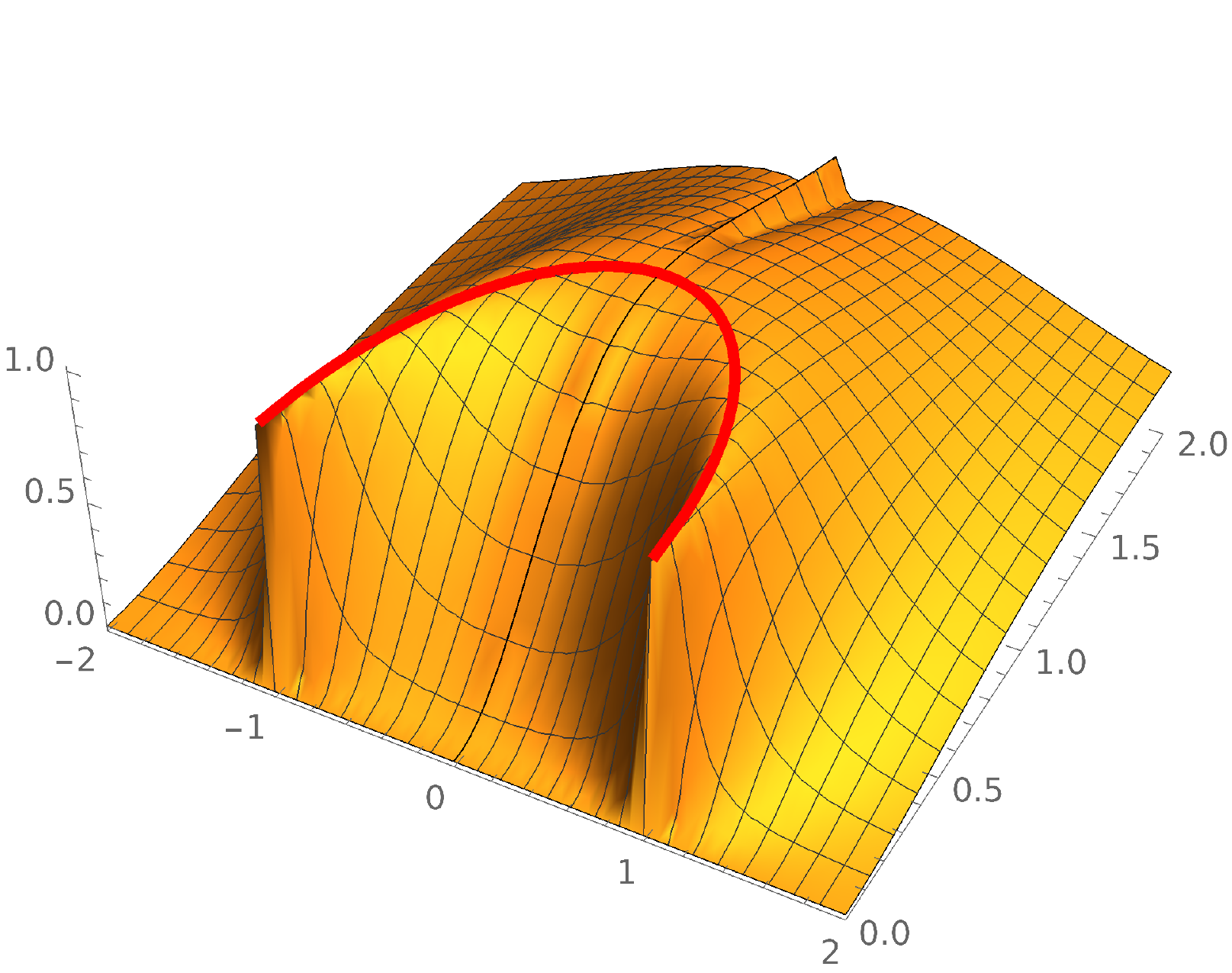}
	}
	\subfloat[][]{	\includegraphics[width=.5\textwidth
		]{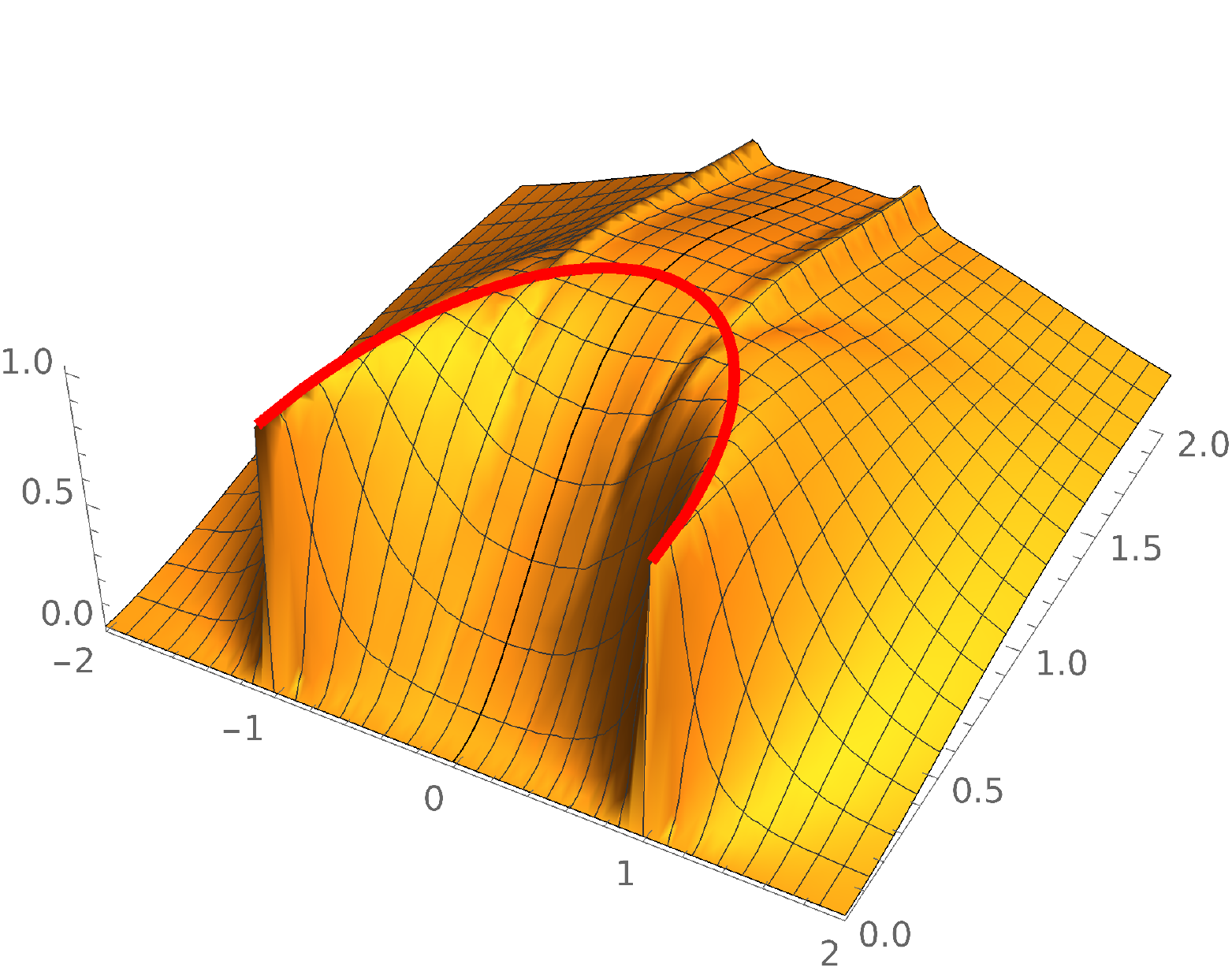}
	}
	\caption{Perturbed vector field obtained  using the\emph{ level set} construction. Panels (a) and (b) illustrate the vector lines for  $R=1$,   $\lambda=.02$, $t=0$ and $t=1/2$ respectively. Panels (c) and (d) illustrate the corresponding vector norms. }
	\label{fig_level_set}
\end{figure}

We can study the same example using the level set construction discussed in Section \ref{lsc}.
In this approach, given a solution to the max flow problem in the unperturbed geometry $v$, the solution for the perturbation of $\delta v$ is
\be
\delta v^a =\Psi v^a-g^{ ab}\delta g_{bc} v^c\,,
\ee
where $\Psi$ is a scalar function that is determined by solving the first order differential equation
\bea
v \cdot \nabla \Psi+\nabla_a (\delta g^{a b} v_b)+\tfrac{1}{2} v\cdot \nabla ( \delta g ) =0\,,
\eea
with boundary condition
\bea
\Psi(\varphi,g_\lambda)|_{\gamma_A}=\frac{1}{2}\delta g_{ab} v^a v^b\,.
\eea
In  Figure \ref{fig_level_set} we present the results obtained using this method.

\begin{figure}[t!]
	\centering
	\subfloat[][]{\includegraphics[width=.5\textwidth
		]{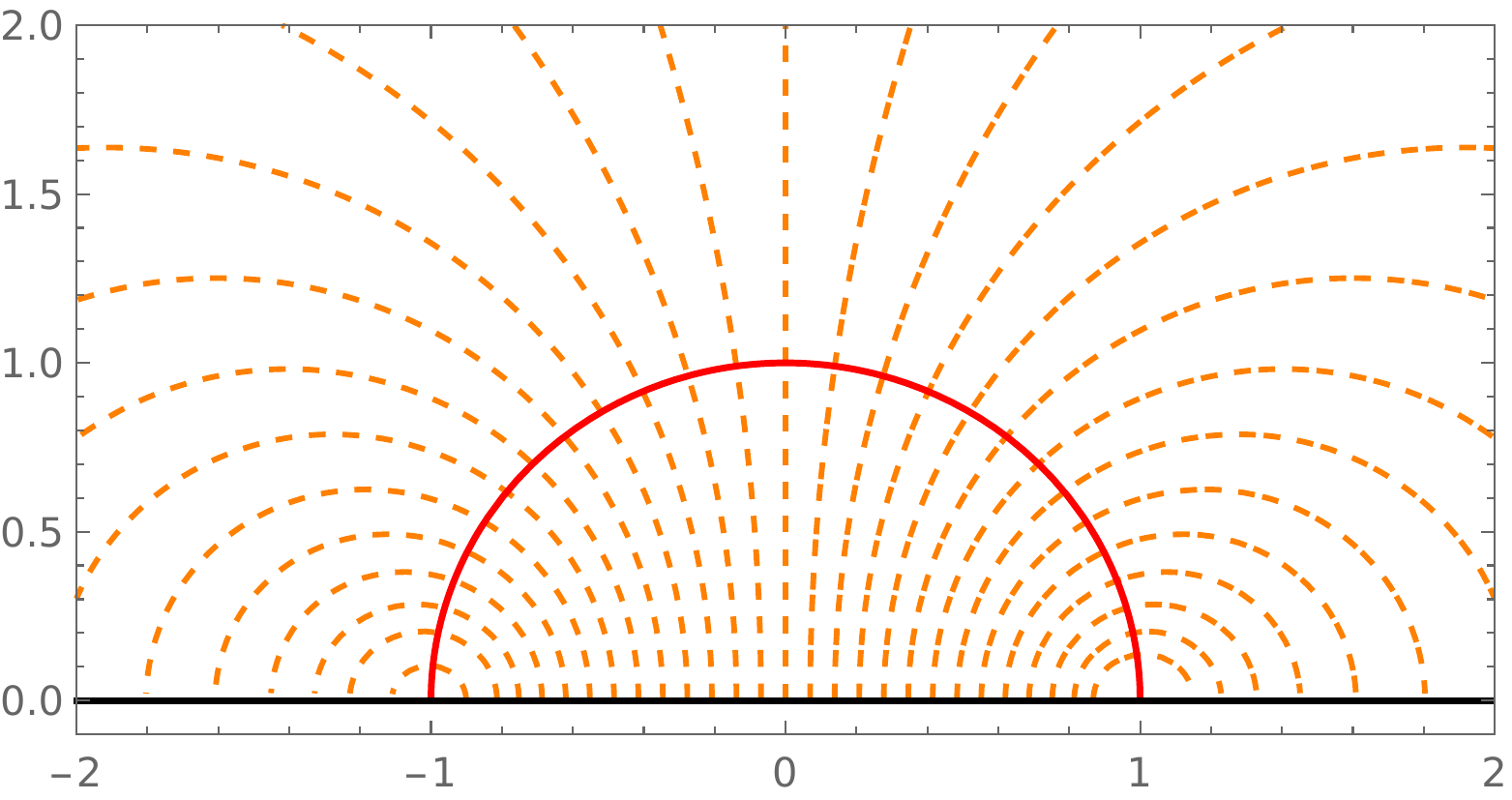}
	}
	\subfloat[][]{	\includegraphics[width=.5\textwidth
		]{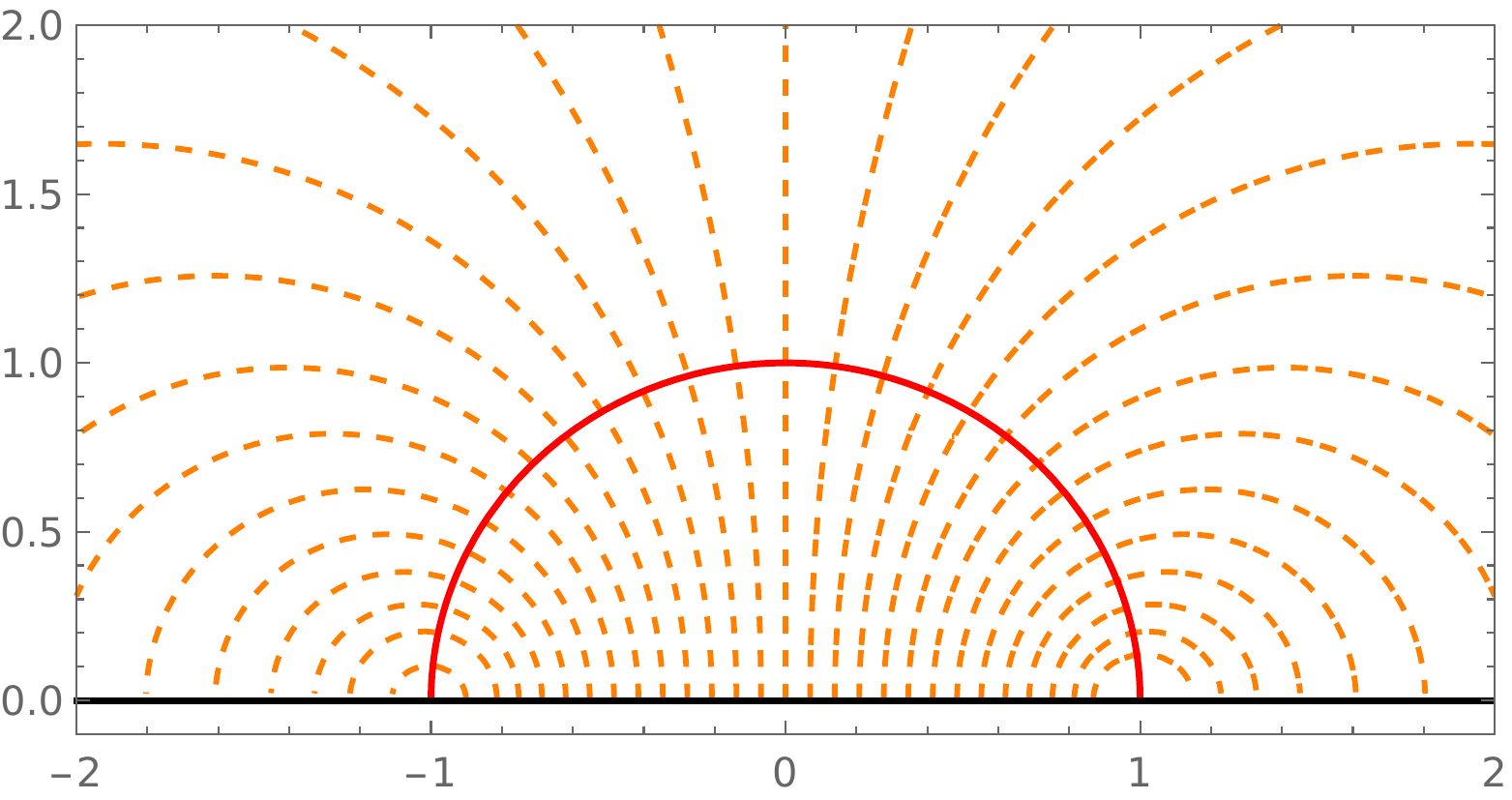}
	}
	
	\subfloat[][]{\includegraphics[width=.5\textwidth
		]{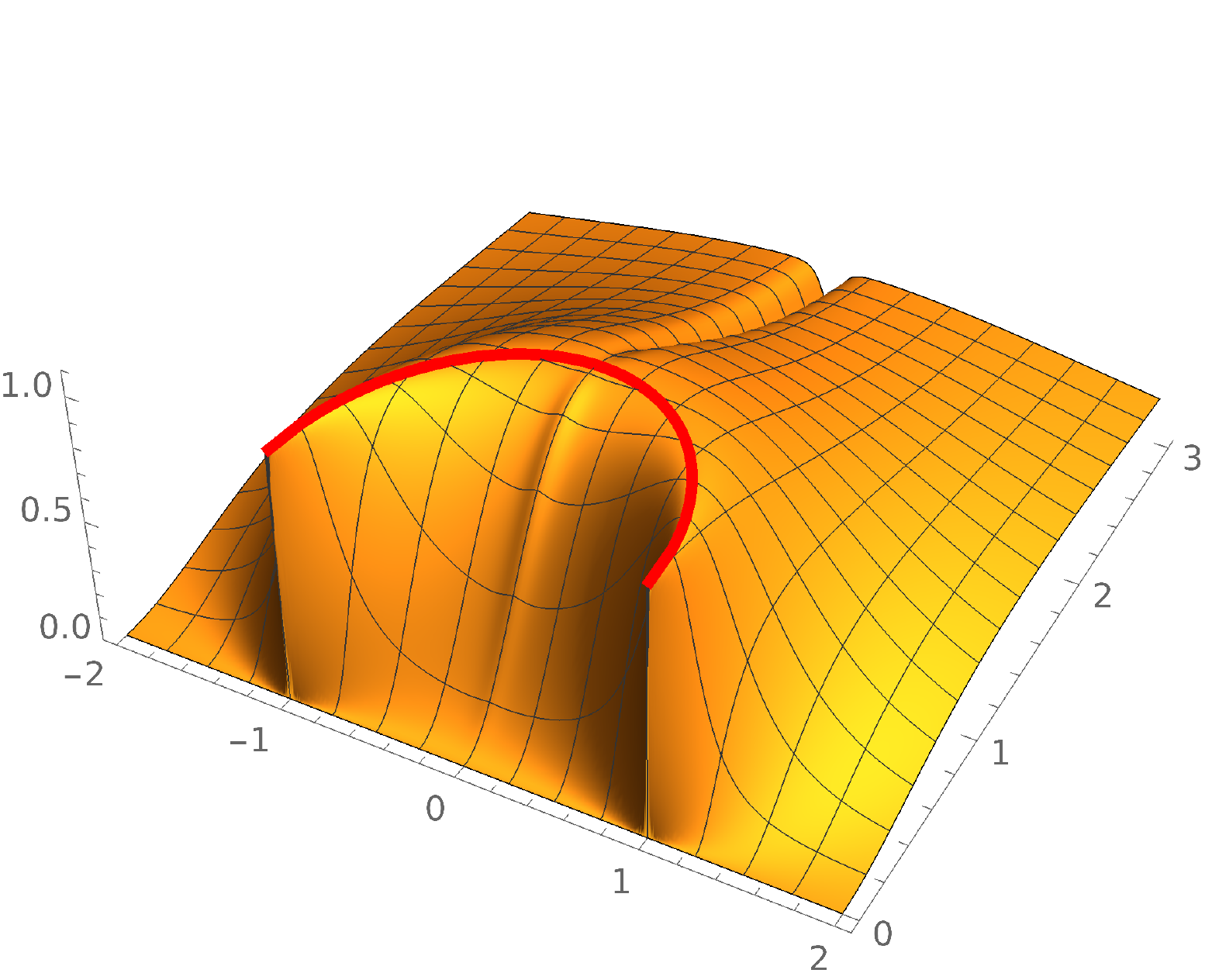}
	}
	\subfloat[][]{	\includegraphics[width=.5\textwidth
		]{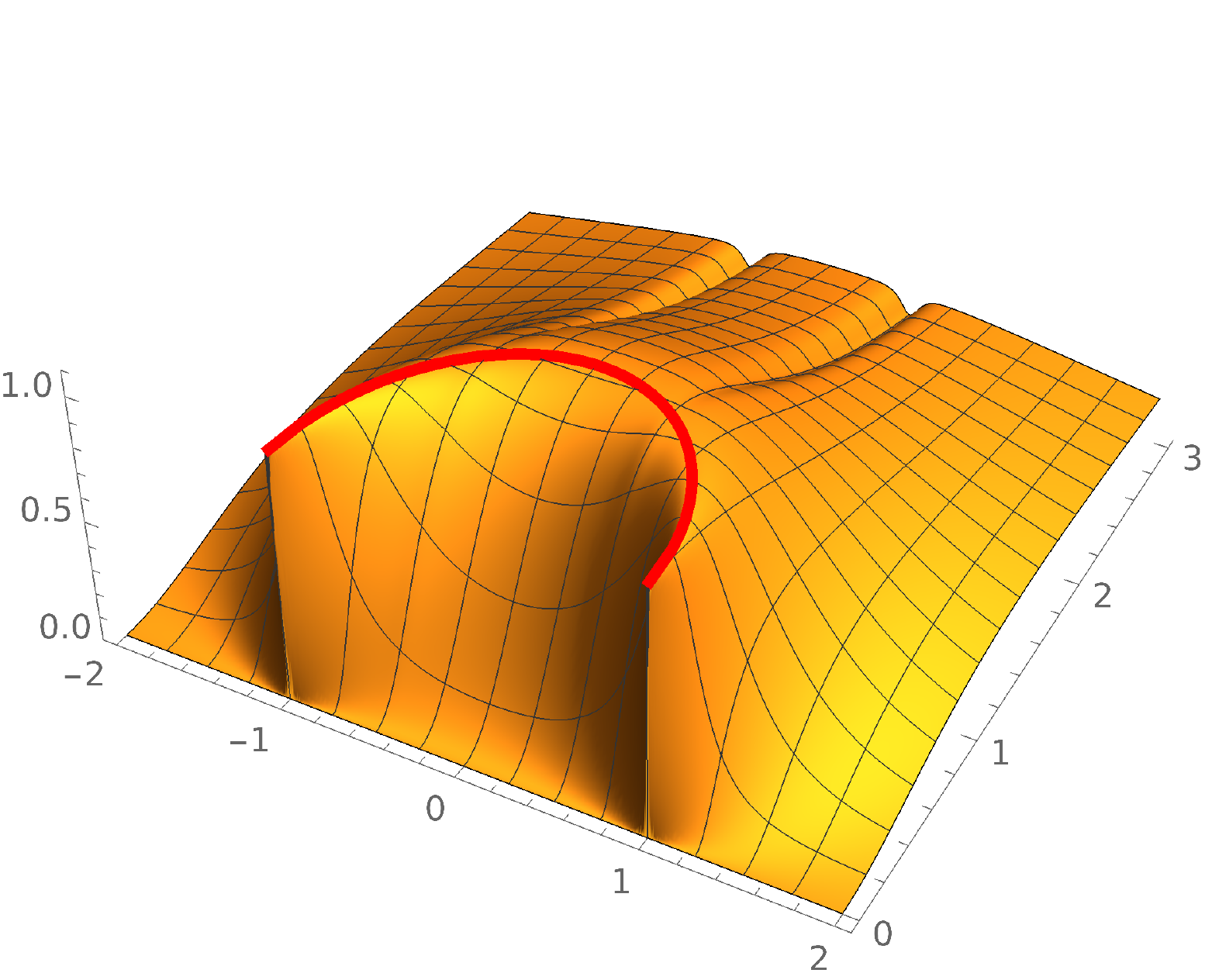}
	}
	\caption{Perturbed vector field obtained  using the\emph{ canonical} bit thread construction that relies on the Iyer-Wald formalism. Panels (a) and (b) illustrate the vector lines for  $R=1$,   $\lambda=.02$, $t=0$ and $t=1/2$ respectively. Panels (c) and (d) illustrate the corresponding vector norms. }
	\label{fig_IW}
\end{figure}

The third method explored in this paper, discussed in Section \ref{5.2},  relies on the Iyer-Wald formalism to define a canonical bit thread perturbation.  This approach uses the language of differential  forms and relates a max flow vector field $v$, to an optimal  closed form $\bm w$. In a background that is perturbatively close to a given geometry, {\it i.e.} $g \rightarrow g+\delta g $, the optimal closed form $\bm w \rightarrow \bm w + \delta \bm w$. Since knowing  $\bm w + \delta \bm w$ determines the max flow $v + \delta v$, the problem now amounts to finding $\delta \bm w$. The Iyer-Wald formalism provides a form, $\bm \chi$, defined in \eqref{chi}, that can be taken as $\delta\bm w$,
\be
\delta \bm w= 4 G_N\, \bm \chi.
\ee
The form $\bm\chi$ is closed when the equation of motions are satisfied. Having $\delta \bm w$ it is straight forward to determine $\delta v$. However, the configuration should satisfy the norm bound (\ref{P-norm-bound}) and that is not guaranteed, in general, in this construction since this condition depends explicitly on the metric. Nevertheless,  the  more detailed analysis performed in Section \ref{5.2}, reveals that up to our order of approximation, the
norm bound is indeed satisfied. To illustrate this point  we plot the  norm for the same perturbation studied with the previous approaches.
The resulting $v$ and its  norm  is plotted in  Figure \ref{fig_IW}  shows that for a perturbative small $\lambda$  the norm bound is indeed satisfied.

\begin{figure}[t!]
	\centering
	\subfloat[][]{\includegraphics[width=.5\textwidth
		]{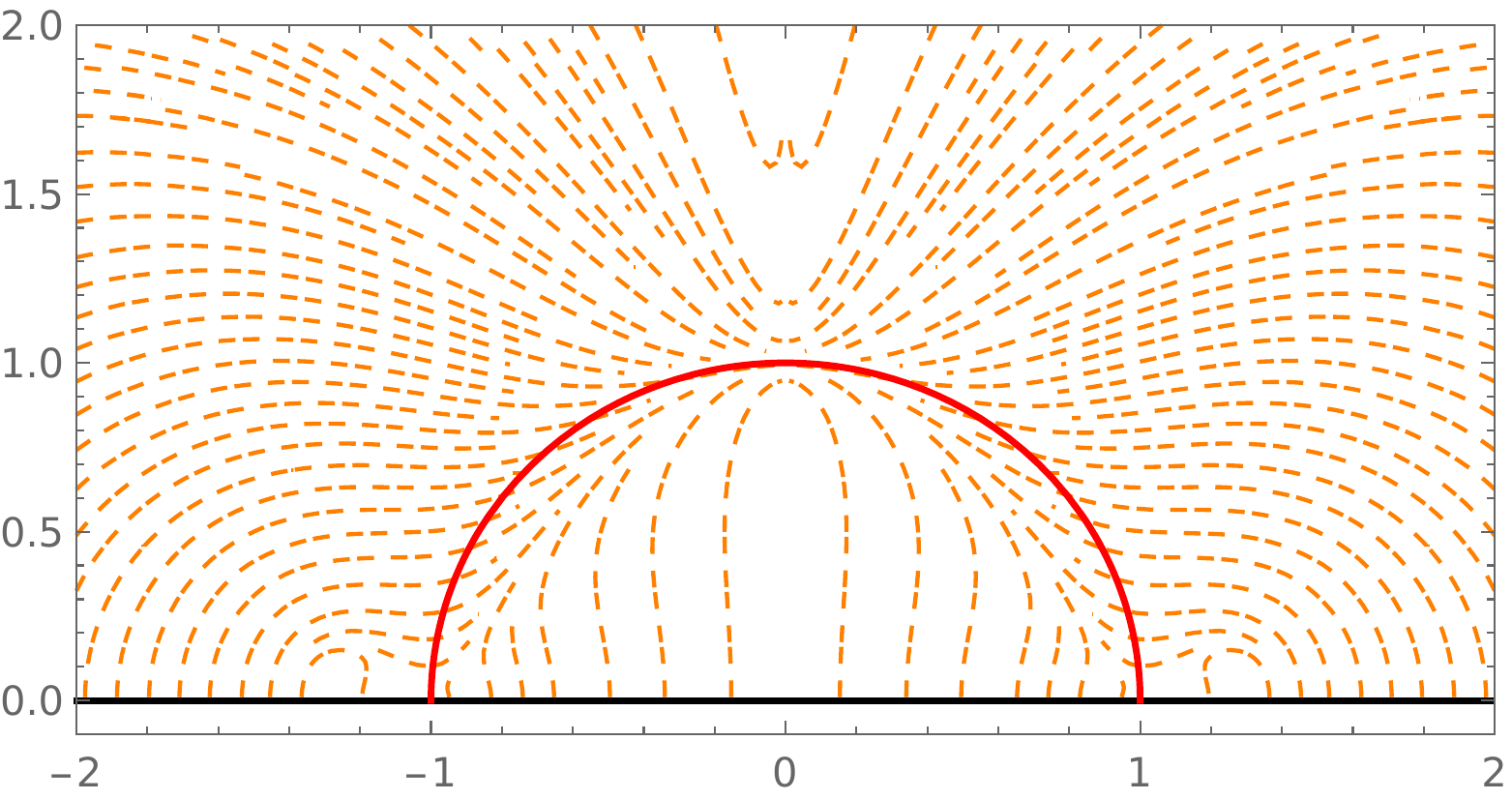}
	}
	\subfloat[][]{	\includegraphics[width=.5\textwidth
		]{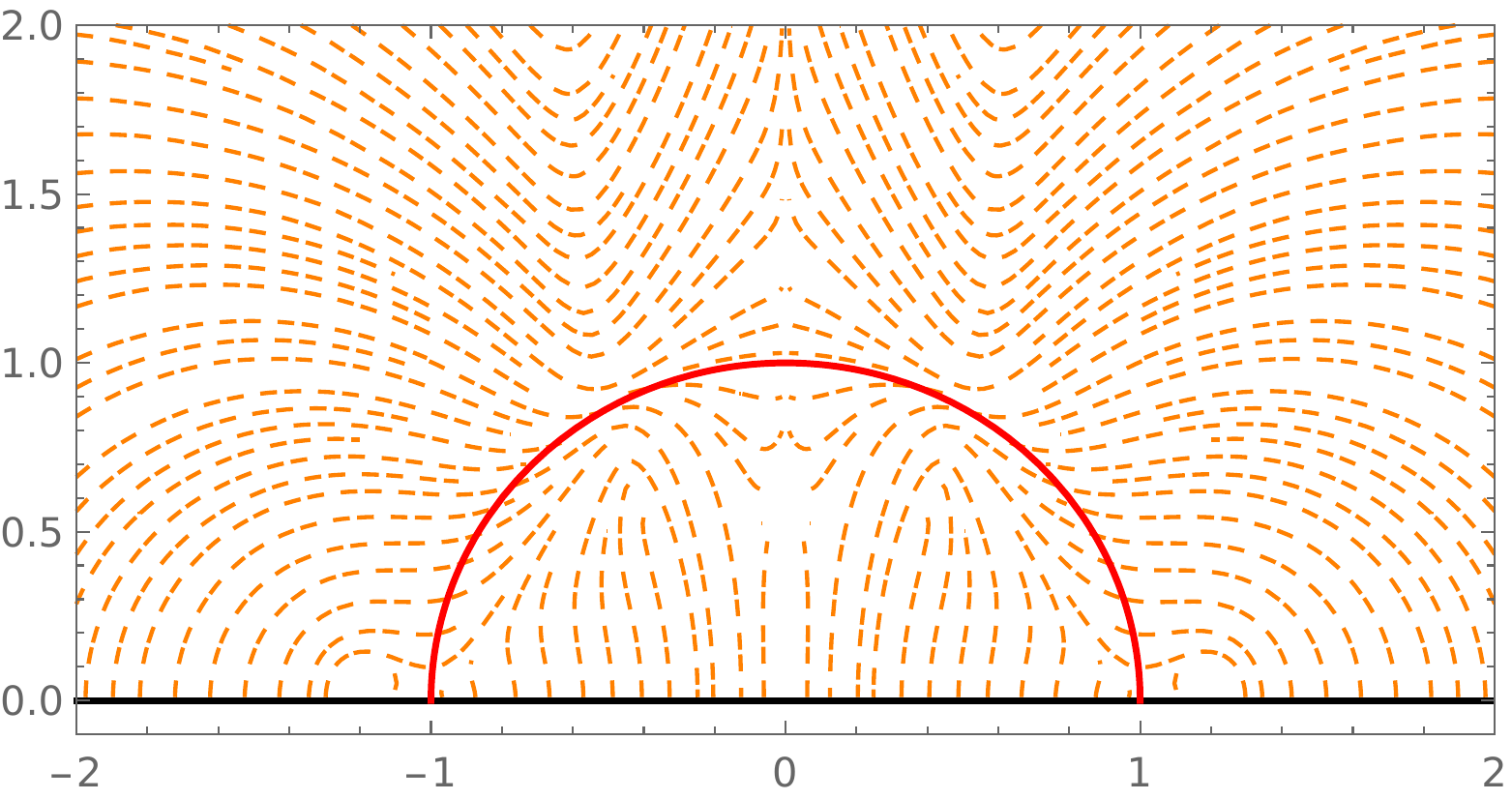}
	}
	
	\subfloat[][]{\includegraphics[width=.5\textwidth
		]{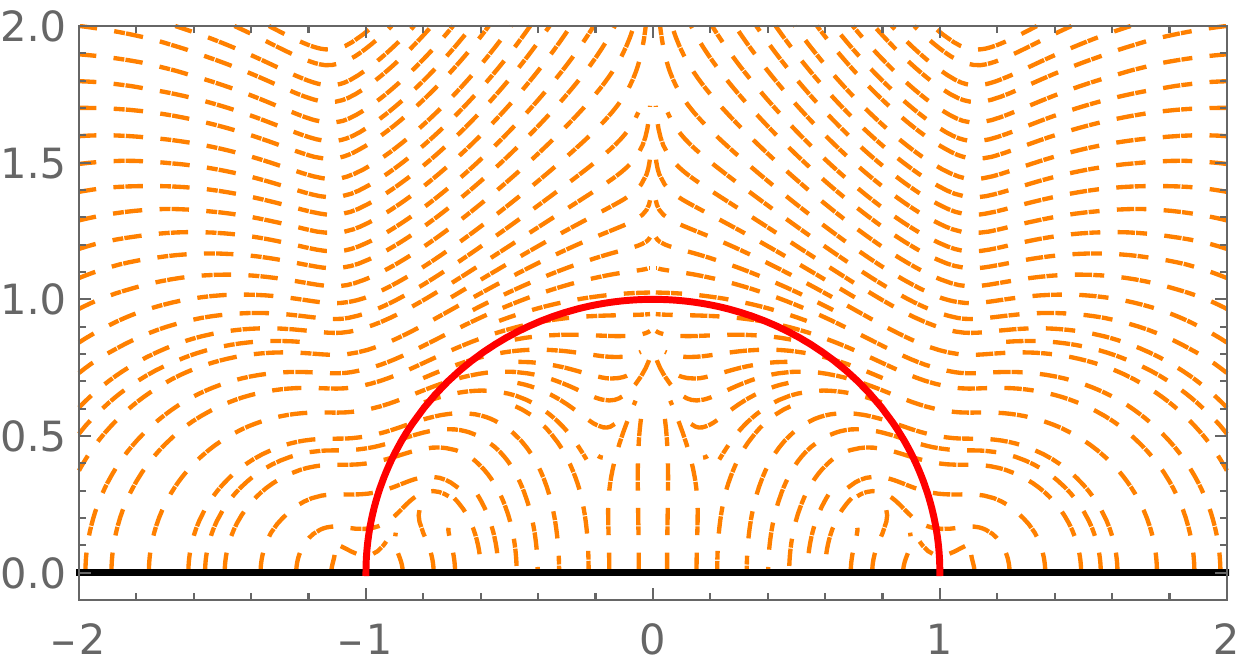}
	}
	\subfloat[][]{	\includegraphics[width=.5\textwidth
		]{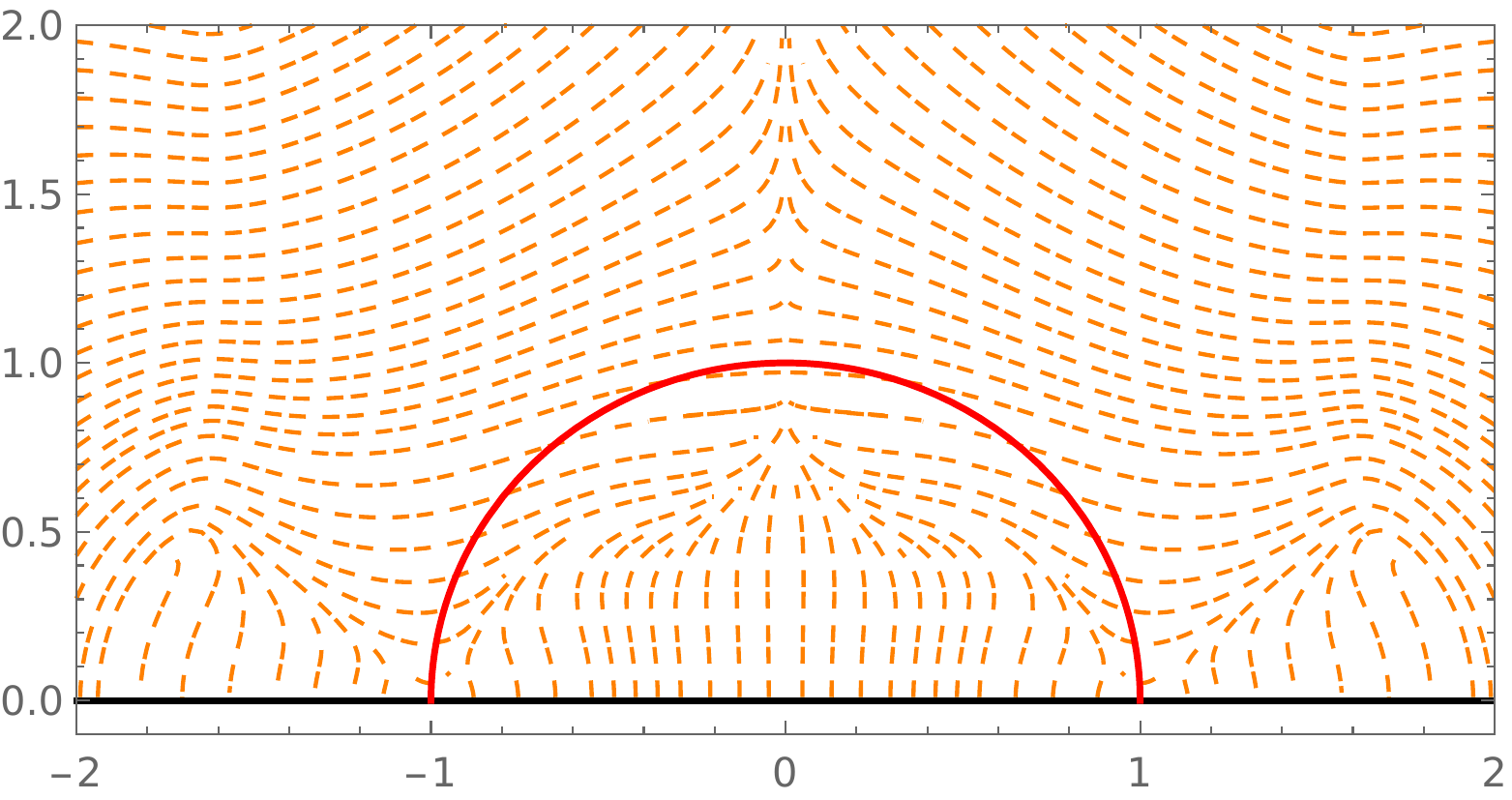}
	}
	\caption{Panels (a), (b), (c) and (d) show $\delta v$ obtained using the level set construction for $t=0,\,.5,\,1,\,1.5$ respectively. We have taken $\lambda=0.02$,\, $R=1$.  }
	\label{fig_pertonly}
\end{figure}

In figures \ref{fig_geo}-\ref{fig_IW} we have presented the perturbed vector field, $v_\lambda =v+ \lambda\, \delta v$, and its  magnitude obtained using the three different methods developed in this paper. It is also interesting to look just at the perturbation $\delta v$ to gain insight into the time-dependence of the local pattern of entanglement induced by the quench. For concreteness we only show results using the level-set method, which are presented in Figure \ref{fig_pertonly}. In \cite{Nozaki:2013wia} it was conjectured that the quench insertion generates an entangled pair of wave packets that move in opposite directions (see Figure 4 of \cite{Nozaki:2013wia} for a pictorial representation). However, it was recently shown in \cite{Agon:2020fqs} that this intuition is only true if one includes the leading $1/N$ corrections coming from the entanglement of bulk fields. More specifically, in this paper it was argued that at the leading order in $G_N$ the two wave packets are effectively unentangled, and only the quantum correlations between the degrees of freedom in each individual packet contribute to the total entanglement entropy. Remarkably, we can reach the same conclusion from the plots in Figure \ref{fig_pertonly}, which exhibit the following features: $i)$ two wave packets moving together with the shocks, i.e., in opposite directions at the speed of light and $ii)$ threads around each wave packet connecting degrees of freedom in their fronts with those in their tails. The fact that we do not see threads connecting the two wave packets implies that they are effectively unentangled at the leading order in $G_N$, in agreement with the result of \cite{Agon:2020fqs}. Moreover, the precise pattern of the threads explains why $S_A$ peaks at $t=R$ (see e.g. Figure 7 of \cite{Agon:2020fqs}): at this time most of the threads connect the degrees of freedom of $A$ with those in its complement (recall that threads connecting points within $A$ do not contribute to the entanglement entropy of the region). It will be very interesting to repeat this analysis for the case of a global quench, and understand how the local pattern of entanglement evolves in time for cases that admit an entanglement tsunami interpretation \cite{Liu:2013iza,Liu:2013qca} (large regions) and cases that do not \cite{Kundu:2016cgh,Lokhande:2017jik} (small regions).

\section{Details of the linear inversion problem\label{app:inversion}}

We start by integrating equation (\ref{eq:inversion}) to obtain
\bea\label{Hii}
\frac{z^d H^i_{\,\, i}}{\xi^t}\bigg|_{z_0}^{z}=\frac{4R^2}{\pi} \, \int_{z_0}^zd\zeta\, \frac{\zeta^{d-1} \delta w_z(\zeta,\vec{x})}{(z_*^2-\zeta^2)^2}\,,\qquad z_*^2\equiv R^2-|\vec{x}-\vec{x}_0|^2\,,
\eea
and hence
\bea \label{Hii-2}
H^i_{\,\, i}(z,\vec{x})=\frac{4R^2 \xi^t(z,\vec{x})}{\pi z^d} \, \int_{z_0}^z d\zeta\, \frac{\zeta^{d-1} \delta w_z(\zeta,\vec{x})}{(z_*^2-\zeta^2)^2} +\frac{z_0^d \xi^t(z,\vec{x})}{z^d \xi^t(z_0,\vec{x})}H^i_{\,\, i}(z_0,\vec{x})\,.
\eea
The second term in the above formula seems to encode the boundary condition since at $z=z_0$ the integral in the first term vanishes and we obtain $H^i_{\,\, i}(z,\vec{x})=H^i_{\,\, i}(z_0,\vec{x})$. However, if we first let $z_0\to 0$, the second term goes away and we seem to naively lose the boundary condition. The presence of the singular term $\sim 1/z^d$ in the integral tell us that we should treat the above limit with some care. Changing the integration variable to $\lambda=\zeta/z$, we obtain
\bea \label{Hii-3}
H^i_{\,\, i} (z,\vec{x})&=&4R\,(z_*^2-z^2) \int_{z_0/z}^1  d\lambda \frac{\lambda^{d-1}\delta w_z(\lambda z,\vec{x})}{[z_*^2-(\lambda z)^2]^2}+\(\frac{z_0}{z}\)^d\frac{ (z_*^2-z^2)}{(z_*^2-z_0^2)}H^i_{\,\, i}(z_0,\vec{x})\,,
\eea
which in the limit $z_0\to 0$ yields
 \bea \label{Hii-4}
H^i_{\,\, i} (z,\vec{x})=4R\,(z_*^2-z^2) \int_{0}^1  d\lambda \frac{\lambda^{d-1}\delta w_z(\lambda z,\vec{x})}{[z_*^2-(\lambda z)^2]^2}\,.
\eea
It can be checked that this equation is now consistent with the boundary condition. Indeed, taking the explicit $z\to 0$ limit  leads to
  \bea \label{Hii-4-0}
H^i_{\,\, i} (0,\vec{x})=\frac{4R \delta w_z(0 ,\vec{x})}{z_*^2}  \int_{0}^1  d\lambda \lambda^{d-1} =\frac{4R \delta w_z(0 ,\vec{x})}{d\hspace{0.5pt}z_*^2}\,,
\eea
which agrees with the $z\to 0$ limit of (\ref{wz}).

Equation (\ref{Hii-4}) is valid for $\vec{x}\in A^c$ and $\forall z$, or for
$\vec{x}\in A$ and $z<z_*$. For $\vec{x}\in A$ and $z\geq z_*$  the integrand of (\ref{Hii-4})
has a double pole at $\lambda_*= z_*/z$ at some point in the range of integration, i.e., $\lambda_*\in[0,1]$. In order to get an expression that
is valid in this region, we first note that the limit $z\to z_*$ (or $\lambda_*\to1$) is in fact finite. To see this, we
integrate up to $1-\epsilon$ and then let $\epsilon\to0$. We further assume that
$\delta w_z(\lambda z, \vec{x})$ is continuous and (at least first) differentiable so that, we can isolate its value at $\lambda\to1$ as follows:
\bea\label{Taylor}
\delta w_z(\lambda z, \vec{x})=\delta w_z(z, \vec{x})-(1-\lambda) z W(\lambda z, \vec{x})\,.
\eea
Here $W(\lambda z, \vec{x})$ is a continuous function which obeys the condition $W(\lambda z, \vec{x})|_{\lambda=1}=\partial_z \delta w_z( z, \vec{x})$. Replacing (\ref{Taylor}) into (\ref{Hii-4}) leads to
 \be\label{Hii-l}
 \!\!\!H^i_{\,\, i} (z,\vec{x})=4R(z_*^2-z^2)\!\left[\delta w_z(z,\vec{x}) \!\int_{0}^{1-\epsilon}\!\!\!\!  d\lambda \frac{\lambda^{d-1}}{[z_*^2-(\lambda z)^2]^2}-z\!\int_{0}^{1-\epsilon}\!\!\!\!d\lambda  \frac{ \lambda^{d-1} (1-\lambda )W(\lambda z,\vec{x}) }{[z_*^2-(\lambda z)^2]^2}\right]\!.
 \ee
The first integral in (\ref{Hii-l}) has a double pole, and evaluates to
 \be
 \int_0^{1-\epsilon}\!\!\!\!d\lambda\frac{\lambda^{d-1}}{[z_*^2-(\lambda z)^2]^2}=\frac{1}{2z_*^2(z_*^2-z^2)}-\frac{d-2}{2\hspace{0.5pt}d\hspace{0.5pt}z_*^4} \,_2F_1\[1,\tfrac{d}{2},\tfrac{d+2}{2},\big(\tfrac{z}{z_*}\big)^2\]+\mathcal{O}(\epsilon)\,.
 \ee
As expected, the result has a single pole at $z\to z_*$; however, the integral is multiplied by a factor $(z_*^2-z^2)$ so the final result is finite. In addition, there is also a
subleading logarithmic divergence that comes from the hypergeometric function, but this term does not contribute in this limit since $(z_*^2-z^2)\log(z_*-z)\to0$ as $z\to z_*$. The second integral in (\ref{Hii-l}) has a single pole and leads to another logarithmic divergence upon integration. However, the same argument is valid here so it ends up not contributing in the $z\to z_*$ limit. Putting all together, we obtain that
  \be\label{bdy-gammaA}
 H^i_{\,\, i} (z_*,\vec{x})=\frac{2R}{z_*^2} \delta w_z(z_*,\vec{x})\,,
 \ee
which is indeed consistent with the $z\to z_*$ limit of (\ref{wz}).

We are now in a position to investigate the region $\vec{x}\in A$ and $z\geq z_*$. Notice that the expression (\ref{Hii-4}) cannot be naively extended to the $z\geq z_*$ region. Instead, we will start from (\ref{Hii-2}) for $z_0$ arbitrarily close to $z_*$ either from above $z_*^+$ (where $z\geq z_*^+$) or from bellow $z_*^-$ (where $0\leq z\leq z_*^-$) in order to avoid crossing the singular point. In either case we can write the following equation for the trace
\bea\label{Hii-5}
\!\!\!\!\!\!\!\!H^i_{\,\, i}(z,\vec{x})&=&\frac{4R(z_*^2-z^2)}{z^d} \, \int_{z_0}^z d\zeta\, \frac{\zeta^{d-1} \delta w_z(\zeta,\vec{x})}{\(z_*^2-\zeta^2\)^2} \nonumber \\
&&+\frac{2 R z_0^{d-2} (z_*^2-z^2)}{z^d}\[\frac{\delta w_z\(z_0,\vec{x}\)}{\(z_*^2-z_0^2\)}-\frac{d}{2z_*^2}\delta w_z(z_*,\vec{x})-\frac{z_0}{4 R}\partial_{z_0}H^{i}_{\,\,i}\(z_0,\vec{x}\) \].
\eea
where the second line above corresponds to the leading term in a series expansion around $z_0\approx z_*$ of the second term in (\ref{Hii-2}). Such expansion is naturally encoded in (\ref{wz}).

In the $z_0\to z_*$ limit we can identify possible divergences coming from the integrand when $\zeta \approx z_0\approx z_*$ and from the first term inside the brackets. However, these divergences turn out to cancel. For instance, the identity
\bea
\int_{z_0}^z \frac{2 \zeta d\zeta}{\(\zeta^2-z_*^2\)^2}=\frac{1}{z_0^2-z_*^2}-\frac{1}{z^2-z_*^2}\,,
\eea
allows us to turn the term with a single pole into an integral which can be combined with the first line of (\ref{Hii-5}). After this replacement is implemented, the $z_0\to z_* $ limit leads to
\bea\label{Hii-7}
\!\!\!\!\!\!\!\!\!\!\!\!\!H^i_{\,\, i}(z,\vec{x})&=&\frac{4R(z_*^2-z^2)}{z^d} \int_{z_*}^z d\zeta\, \frac{\zeta \[\zeta^{d-2} \delta w_z(\zeta,\vec{x})-z_*^{d-2} \delta w_z(z_*,\vec{x})\]}{(z_*^2-\zeta^2)^2}\nonumber \\
&&+\frac{2R z_*^{d-2}}{z^d} \delta w_z\(z_*,\vec{x}\)-\frac{z_*^{d-4}R (z_*^2-z^2)}{z^d}\Big[d\, \delta w_z(z_*,\vec{x})+\frac{z_*^3}{2R}\partial_{z_*}H^{i}_{\,\,i}\(z_*,\vec{x}\) \Big] .
\eea
The second line of the above expression is manifestly finite, while the integrand of the first line has an apparent single pole as we let $\zeta \to z_*$,
\be\label{singlepole}
 \!\!\frac{\zeta^{d-2} \delta w_z(\zeta,\vec{x})-z_*^{d-2} \delta w_z(z_*,\vec{x})}{(z_*^2-\zeta^2)^2}=\frac{z_*^{d-3}[(d-2) \delta w_z(z_*,\vec{x})+z_* \partial_{z_*}\delta w_z(z_*,\vec{x})]}{4z_*^4(\zeta-z_*)}+\text{finite}\,.
\ee
Fortunately, its residue identically vanishes as can be checked from the relation
\bea
(d-2) \delta w_z(z_*,\vec{x}) +z_*\partial_{z_*}\delta w_z(z_*,\vec{x}) =0
\eea
which follows from (\ref{wz}). This makes the expression (\ref{Hii-7}) well defined across the minimal surface and therefore valid $\forall z$.

We note that, in its present form, equation (\ref{Hii-7}) is not fully determined by $\delta w_z(z,\vec{x})$ as it requires knowledge of $\partial_{z_*}H^{i}_{\,\,i}(z_*,\vec{x})$. This can be fixed by considering $z\to 0$ limit of the above expression. After a bit of algebra one finds that finiteness of such limit implies
\bea \label{boundary-partial}
\!\!\!\!\!\!\!\!\!\!\!\!\partial_{z_*}H^{i}_{\,\,i}(z_*,\vec{x}) =-\frac{2R(d-2)}{z_*^3}\delta w_z(z_*,\vec{x})-\frac{8R}{z_*^{d-1}} \int_{0}^{z_*}\!\!\! d\zeta \frac{\zeta \[\zeta^{d-2} \delta w_z(\zeta,\vec{x})-z_*^{d-2} \delta w_z(z_*,\vec{x})\]}{(\zeta^2-z_*^2)^2}.
\eea
Plugging this result back into (\ref{Hii-7}) leads to our final expression for the trace:
\bea\label{Hii-final}
\!\!\!\!\!\!\!\!\!\!H^i_{\,\, i}(z,\vec{x})=\frac{2R z_*^{d-4}}{z^{d-2}}\delta w_z(z_*,\vec{x})+\frac{4R(z_*^2-z^2)}{z^d} \int_{0}^z\!\!\! d\zeta \frac{\zeta \[\zeta^{d-2} \delta w_z(\zeta,\vec{x})-z_*^{d-2} \delta w_z(z_*,\vec{x})\]}{(z_*^2-\zeta^2)^2}.
\eea
Notice that the consistency condition (\ref{boundary-partial}) can be derived from this expression via explicit differentiation at $z=z_*$, and therefore (\ref{Hii-final}) is selfconsistent and finite for all $z$.

Finally, after the change of variables $\zeta\to \lambda z$ one can rewrite (\ref{Hii-final}) as
 \bea\label{Hii-final-2}
\!\!\!\!\!\!\!\!\!\!H^{i}_{\,\,i}(z,\vec{x})=\frac{2R z_{*}^{d-4}\delta w_z(z_{*},\vec{x})}{z^{d-2}}+4R(z_{*}^2-z^2)\! \int_{0}^1\! d\lambda\frac{\lambda [\lambda^{d-2} \delta w_z(\lambda z,\vec{x})-\lambda_{*}^{d-2} \delta w_z(z_{*},\vec{x})]}{[z_{*}^2-(\lambda z)^2]^2}.
\eea
where $\lambda_*=z_*/z$. Equation (\ref{Hii-final-2}) has a close resemblance with (\ref{Hii-4}) and indeed it can be derived from it via various regularizations. For instance, it can be checked that the principle value of the integral (\ref{Hii-4}) yields (\ref{Hii-final-2}). Perhaps a yet simpler way to arrive at (\ref{Hii-final-2}) starting from (\ref{Hii-4}) is to change slightly the integration contour:
 \bea
H^i_{\,\, i} (z,\vec{x})=4R(z_*^2-z^2) \, \int_{i\epsilon}^{1+i\epsilon}\!\!\!  d\lambda \, \frac{ \lambda^{d-1}\delta w_z(\lambda z,\vec{x})}{[z_*^2-(\lambda z)^2]^2}\,,
\eea
for $\epsilon\in \mathbb{R}$ and then letting $\epsilon\to0$. It can be easily checked that this prescription is consistent both for $\vec{x}\in A^c$ and $\vec{x}\in A$ ($\forall z$).

\bibliographystyle{ucsd}
\bibliography{refs-PBT}

\end{document}